\documentclass[pdflatex,sn-basic]{sn-jnl}

\usepackage{graphicx}
\usepackage{multirow}
\usepackage{amsmath,amssymb,amsfonts}
\usepackage{amsthm}
\usepackage{mathrsfs}
\usepackage[title]{appendix}
\usepackage{xcolor}
\usepackage{textcomp}
\usepackage{manyfoot}
\usepackage{booktabs}
\usepackage{tabularx}
\newcolumntype{Y}{>{\raggedright\arraybackslash}X}
\usepackage{bookmark}
\hypersetup{hypertexnames=false}

\theoremstyle{thmstyleone}

\theoremstyle{thmstyletwo}

\theoremstyle{thmstylethree}

\begin{document}

\title[AI Category Ownership Map]{Who Owns the AI Recommendation? A Multi-Industry Empirical Map of Brand Category Ownership Across Large Language Models}

\author*[1,2]{\fnm{Dmitrij} \sur{\.Zatuchin}}\email{dmitrij.zatuchin@eek.ee}

\affil*[1]{\orgdiv{Department of Information Technologies},
  \orgname{Estonian Entrepreneurship University of Applied Sciences (EUAS)},
  \orgaddress{\city{Tallinn}, \country{Estonia}}}
\affil[2]{\orgname{Rankfor.AI},
  \orgaddress{\city{Tallinn}, \country{Estonia}}}

\abstract{This exploratory study measures brand inclusion across five industries, 50 brands and 250 queries, each put five times to GPT-5.2, Gemini 3 Flash and Perplexity sonar-pro in February and September 2026 (3,614 and 3,750 scored answers). Category Inclusion Rate, Recommendation Share, Competitive Vacuum Index and Co-Mention Asymmetry have stated denominators. February inclusion rates sit close together across an industry's sampled brands (mean Gini 0.30), while at least one brand is named in 80\% or more of answers to 204 of 250 queries. Vacuums occur in 7.6\% of queries; provisional open-vocabulary model readings suggest most reflect the sampled brand list. The partially pre-specified September replication shows strong cross-date Recommendation Share correlation (Spearman 0.994), unchanged vacuum prevalence and agreement of 60.8\% against February's 57.2\%. The descriptive size association persists. Fixed margins do not account for all co-mention structure: 31 ordered pairs depart from the September null. Agreement exceeds the query-independent null, and both parametric-pair advantage intervals are positive. All ten clustering seeds yield zero emergent clusters. Matcher validation remains provisional pending author checks. The statistics describe system output and identify no causal mechanism.}


\maketitle

\section{Introduction}\label{sec:intro}

When a buyer asks a language model which payment processor an online store should use, the answer is a short list of named brands: a consideration set assembled by an intermediary and handed to the buyer as text \citep{howard1969buyer, hauser1990consideration, roberts1991consideration}. This paper measures inclusion, the event that a brand is named in such an answer, and its frequency over repeated answers to the same query. No established visibility measure covers that object; generative engine optimisation has treated it as an extension of search visibility \citep{aggarwal2023geo, kumar2026geoscale}. Empirical work on model recommendations has examined one property at a time: demographic bias \citep{zatuchin2026a}, brand-origin bias \citep{kamruzzaman2024brandbias, betello2026passport}, sourcing \citep{jazwinska2025citation}, run-to-run instability \citep{atil2025nondeterminism, schulte2026dontmeasureonce} and one brand's presence across platforms \citep{shin2025hbr}. Which brands are included, how unequally, in whose company and with what cross-system agreement has not been measured across industries with a stated denominator.

Four questions follow, each carrying query type as a design factor, since prompts naming brands require separate reporting from open recommendation queries.

\begin{description}
  \item[RQ1 (Inclusion)] Under a category query, which of the sampled brands do three systems name, and how often over repeated answers?
  \item[RQ2 (Concentration)] How unequal are inclusion rates across the ten sampled brands of an industry, and do the industries order by recommendation concentration as they order in a size benchmark of the sampled brands?
  \item[RQ3 (Co-mention structure)] Within an answer, are two brands named together or apart more often than their inclusion rates and the answer's length imply?
  \item[RQ4 (Cross-system agreement)] How often do the three systems name the same brand most often under a query, read against a Monte Carlo null?
\end{description}

The design is observational with repeated measures. Fifty brands in five industries (SaaS, consulting, fintech, e-commerce, health technology) meet 250 category queries in five types, put to three systems: GPT-5.2 and Gemini 3 Flash, both called with no search or grounding tool, and Perplexity sonar-pro, which retrieves documents at query time. Each query went to each system five times in February 2026 (wave 1) and again in September 2026 (wave 2) with the same identifiers and settings. An empty answer is a missing observation: 136 of the 3,750 wave-1 answers, all from GPT-5.2, are empty, and every denominator counts non-empty answers; wave 2 returned none.

The paper defines four constructs, checks their reliability and association with a size benchmark, maps inclusion, tests co-mention structure against fixed margins, and reports agreement under three denominators.

The study is exploratory. The five hypotheses of the submitted version are reported with their original cut-offs under one interval rule. Three expectations, E1, E2 and E5 (numbered after the hypotheses they replace), were formulated after the wave-1 results were known and fixed in a frozen analysis plan before any wave-2 brand statistic was computed; wave 2 is a partially pre-specified replication. Query-type analysis and clustering are descriptive.

\section{Theoretical background and expectations}\label{sec:theory}

\subsection{The answer as a synthesised consideration set}

Buyers evaluate a small subset of the available brands, retrieved from memory or prompted by the buying situation, and its composition shapes choice before any attribute comparison \citep{howard1969buyer, hauser1990consideration, roberts1991consideration}. A language-model answer reproduces this structure with two differences: an intermediary assembles the set and hands it over as text, and the answer has a finite length budget. Inclusion is therefore the event that matters, binary per answer, measured here as a rate over repeated answers to the same query.

\subsection{Three forces on inclusion, and their signatures}

\emph{Corpus salience.} Parametric memory recalls entities in proportion to their frequency in pre-training text and recalls the long tail poorly \citep{petroni2019language, kandpal2022large, mallen2023trust, dodge2021documenting}; as recommenders, models prefer global to local brands and show popularity bias \citep{kamruzzaman2024brandbias, betello2026passport, lichtenberg2024popularity}. Acting alone, corpus salience would make inclusion follow the corpus, heavier-tailed than the brands' sizes: amplification, a recommendation distribution more concentrated than size \citep{autor2020superstar}.

\emph{Retrieval grounding.} A retrieval-augmented system composes its answer from documents fetched at query time, so its consideration set inherits the ranking of a search layer whose results vary with location and profile \citep{hannak2013personalization, klimansilver2015location}. Document recall and parametric recall can disagree, so a retrieval system should disagree with a parametric system more than two parametric systems disagree.

\emph{The tuning objective.} Instruction-tuned models are optimised on human preference for helpful, complete and even-handed answers \citep{ouyang2022training}, which in a recommendation task rewards naming several plausible options. Acting alone, it would compress inclusion rates toward equality across the brands the system treats as category members: compression, a distribution less concentrated than size.

Sampling non-determinism cuts across all three: identical inputs at low temperature return different answers \citep{atil2025nondeterminism, schulte2026dontmeasureonce}, and a system shown the same candidates in a different order changes its preferred item \citep{ma2026notstable, bito2025positionbias}, so every construct below is defined over repeated answers. None of the three forces is measured here; together they are the account under which the expectations are stated.

\subsection{Neighbouring measures}

\emph{Search visibility} is graded, positional and document-mediated; inclusion in a model answer is binary and direct, and generative engine optimisation treats the two as one problem \citep{aggarwal2023geo}. \emph{Recommender-system output} is ranked for one user from that user's history \citep{celma2010longtail}; the answers studied here carry none. \emph{Share of voice} is an input, a brand's share of category advertising weight \citep{jones1990adspending, binet2013longshort}; inclusion rate is the output of an intermediary. \emph{Brand salience}, or mental availability, is a brand's propensity to be retrieved from a buyer's memory in a buying situation, measured by prompted recall \citep{romaniuk2004salience, romaniuk2016hbg2}; it is the closest relative, a model answer being a prompted retrieval from a text corpus by a system, and the double-jeopardy regularity, that smaller brands have fewer and slightly less loyal buyers \citep{romaniuk2016hbg2}, motivates the size benchmark of E1. Agenda-setting theory says why the distinction matters: intermediaries decide what audiences think about by selecting what appears \citep{mccombs1972agenda, napoli2015socialmedia, nechushtai2019gatekeepers}, and reputation is built from the signals observed through such channels \citep{fombrun1990reputation}.

\subsection{Constructs}\label{sec:constructs}

\emph{Recommendation inclusion} is the event that a brand is named at least once in an answer to a category query: binary per answer, with no rank, endorsement or valence. The guarded matcher of Section~\ref{sec:extraction} detects it; an empty answer is missing and never scored.

\emph{Inclusion rate} (Category Inclusion Rate, CIR; the Category Ownership Index of the submitted version) for a brand and a query is the proportion of non-empty answers to the query in which the event occurs, in $[0, 1]$; a system's own rate divides by that system's non-empty answers and is undefined when it has none. It is the system's salience for a brand and says nothing about any buyer's awareness, consideration or choice.

\emph{Recommendation share} (RS) for a brand and an industry is its mean inclusion rate over the industry's queries, with a bootstrap interval over queries.

\emph{Category-level concentration} is the inequality of recommendation shares across an industry's sampled brands: a Gini coefficient, the top-three share and a Herfindahl-Hirschman index of normalised shares \citep{rhoades1993herfindahl}, each with a bootstrap interval over queries; it is a property of ten sampled brands under the sampled queries, and no property of the industry.

\emph{Competitive Vacuum Index} (CVI) for a query is one minus the highest inclusion rate any sampled brand reaches under it. A query is a vacuum when CVI exceeds 0.50, every sampled brand being named in fewer than half of the non-empty answers, and has high maximum inclusion when the highest rate is at least 0.80, a label for repeated inclusion that several brands can meet on one query. A vacuum can mean thin inclusion or brands named outside the sample, so every vacuum query is also read without the sample restriction.

\emph{Co-mention asymmetry} (CMA; the Displacement Score of the submitted version) between brands $A$ and $B$ under a query is $P(A \mid \neg B) - P(A \mid B)$ over the query's non-empty answers, defined only when $B$ is present in some and absent from others; an undefined cell stays undefined. It is an association between two brands' presence in the same answer; displacement would need $B$'s presence to cause $A$'s absence, which a design with no manipulation and no temporal order cannot show.

\emph{Platform agreement} is the proportion of queries on which the systems queried name the same brand most often, each system's top brand computed over its own non-empty answers; it is a property of the systems and versions queried on the dates queried.

The emergent-category clustering of the submitted version (a topic model over answer texts, reported there as H4) is re-run as a descriptive check and predicts nothing.

\subsection{Expectations}\label{sec:expectations}

Section~\ref{sec:results} reports the five hypotheses of the submitted version with their original cut-offs. The expectations below are what the account says the data should show; each names its reference and what would count against it.

\paragraph{E1, concentration relative to size.} The submitted version read amplification against compression from a Gini level against a cut of 0.60, an author judgement with no unique top-three-share or HHI equivalent: a ten-brand vector with one share of 0.70 and nine of 1/30 has Gini 0.60 and HHI 5,000, and a ten-brand Zipf vector with exponent 1 has Gini about 0.42. No distribution is fitted \citep{clauset2009powerlaw, newman2005power}: ten purposively selected brand values do not supply a defensible sample for tail-distribution inference. E1 expects the orderings to agree: within each industry, the brands' ordering by RS follows their ordering in a size benchmark of the sampled brands (rank correlation, bootstrap interval over brands), with the benchmark's top three among the recommendation top three; across the five industries, the ordering by recommendation concentration follows the ordering by benchmark concentration. A Gini on ten bounded rates can reach 0.90, so its level beside a benchmark Gini on revenues is an empirical comparison. With five purposively chosen industries no test is applied, and E1 is reported descriptively in both waves.

\paragraph{E2, co-mention structure.} Pooled over answers of varying length, a long answer names $A$ and $B$ together and a short one names one or neither, so co-mention dominates and the brand named in most answers travels with everyone; at a fixed list length a slot taken by $B$ is unavailable to $A$, which pushes the asymmetry positive (two names from ten brands: $2/9 - 1/9 = +0.11$). Both follow from the margins. For one query with $N$ non-empty answers,
\begin{equation}
  \mathrm{CMA}(A,B) = \frac{n_A}{N - n_B} - \frac{n_{AB}\,N}{n_B (N - n_B)},
\end{equation}
where $n_A$, $n_B$ and $n_{AB}$ count the answers naming $A$, $B$ and both; summed over $A$, $n_{AB}$ gives $S_B - n_B$, with $S_B$ the total list length of the answers naming $B$, whose sum over brands the list lengths fix. A query's mean CMA is therefore nearly fixed by its margins, and the null produces a negative RS to mean-CMA rank correlation in these data. E2 is the claim that co-mention structure is accounted for by marginal inclusion rates and answer length, with no pair-specific departure beyond those margins. Its reference is a within-query fixed-margins null: curveball trades on each query's answer-by-brand presence matrix keep every answer's list length and every brand's count (2,000 draws per query, seed 20260910), and a query that admits no trade contributes its observed matrix to every draw. E2-a$'$ (primary) is supported when the pooled rank correlation between RS and mean CMA as brand $A$, on within-industry ranks with RS held fixed, lies inside its null 2.5th to 97.5th percentile band (tolerance $10^{-12}$), and not supported outside it. E2-b$'$ is supported when the pooled asymmetric-to-co-mentioned record ratio at $|\mathrm{CMA}| > 0.10$ lies inside its band and no ordered pair departs, a pair departing when its pooled CMA differs from the null after Benjamini-Hochberg correction at $q = 0.05$ with its sign reproduced in both split halves of the iterations, $\{1, 3, 5\}$ and $\{2, 4\}$; it is not supported when the ratio lies outside and at least one pair departs, and inconclusive otherwise. E2 is supported when both components are supported, not supported when either is not, and inconclusive with missing or non-finite evidence; the primary verdict uses all 250 queries and the companion the 200 non-comparison queries, each set filtered before any computation. On wave 1 E2 is not supported on either set (Section~\ref{sec:res-cma}); the rules followed inspection of the earlier movable-query analysis; these final all-query values are replication references. The frozen rules, the sign of the correlation and list-length strata against an industry-pooled null, are reported as descriptive statistics.

\paragraph{Query type.} The five query types are modelled explicitly and no ordering is predicted: the reporting is the odds that a sampled brand is named under each type relative to direct recommendation, from a pooled logistic model, and the vacuum rate, leader and concentration by type. Comparison queries name individual brands or brand groups, so the non-comparison estimates exclude them.

\paragraph{E5, platform agreement.} Two systems answered from parametric memory and one retrieves documents at query time (Section~\ref{sec:collection}); E5 predicts that the two parametric systems agree with each other on the most-named brand more than either agrees with the retrieval system. The reference is a Monte Carlo null in which each system's top brand is redrawn, query by query, from that system's own distribution of top brands over the industry's queries (2,000 draws). Missing or non-finite required evidence makes E5 inconclusive (evidence unavailable). E5 is supported when the lower Wilson bound of full agreement among the queries every system answered exceeds the null's 97.5th percentile and both pair differences (the parametric pair minus each parametric-retrieval pair) have paired-bootstrap intervals entirely above 0; not supported when the upper bound lies below the null's 2.5th percentile or either difference interval lies entirely below 0; inconclusive otherwise. Split-half self-agreement of each system is a descriptive reliability reference and is not a bound on cross-system agreement.

Brand origin and demographic framing can affect recommendations \citep{kamruzzaman2024brandbias, betello2026passport, xu2025genderrace}; neither receives a separate test here (Section~\ref{sec:limitations}).

\paragraph{Status of the expectations.} The five hypotheses of the submitted version were set by the authors before the analysis, and none was registered. Their directions came from Section~\ref{sec:theory} (corpus salience for H1, slot competition for H2, long-tail recall failure for H3, retrieval grounding for H5), and their cut-offs were judgements: a Gini of 0.60 with no unique share equivalent, a 3:1 ratio at which asymmetric records would clearly outnumber co-mentioned ones, vacuum and emergent-cluster shares of 20 and 15 per cent with no literature anchor, and 50 per cent agreement, below which the systems would disagree on most queries. E1, E2 and E5 were formulated after the wave-1 results were known and fixed in a frozen analysis plan (Section~\ref{sec:res-crosswave}). The consideration-set model is an account of the corrected wave-1 pattern: wave 2 tests the stability of the map and the three expectations, and the forces yield no brand-level prediction to test.

\section{Method}\label{sec:method}\label{sec:methods}

\subsection{Design}\label{sec:design}

The observational design crosses five industries, ten brands per industry, 50 queries per industry in five types, three systems and five iterations: 3,750 calls per wave. Wave 1 ran on 23 and 24 February 2026, wave 2 on 9 and 10 September 2026 with the same identifiers, settings and queries. Of the 3,750 wave-1 answers, 136 are empty and treated as missing, so 3,614 are scored; the submitted statement that all 3,750 were usable is withdrawn. Inclusion statistics are computed per industry over its ten brands; mentions of a brand in another industry's answers (632 rows) stay out of them.

\subsection{Industries and brands}\label{sec:brands}

The five industries and 50 brands (Table~\ref{tab:brands}) were chosen purposively: five industries spanning B2B and B2C buying, and in each ten brands taken as the category's revenue leaders plus regional challengers with a Nordic or Baltic footprint; no sampling frame was drawn. This revision adds a size benchmark of the sampled brands: the best comparable annual figure each brand publishes, in USD at ECB 2025 average rates, one public source per brand, each with a quality flag (29 are audited filings); Temu, Oracle Health and Bolt Market have no comparable figure (Supplementary Section S4). Adding all fifteen candidates a size-based frame would have listed places six among the three largest measured brands of their industry. An alias dictionary of 77 entries covers the 50 brands (Supplementary Section S1).

\begin{table}[tbp]
\caption{The sampled brands per industry, in the order of the size benchmark (figures, sources and quality flags in Supplementary Table S2), with headquarters countries.}\label{tab:brands}
\centering
\footnotesize
\setlength{\tabcolsep}{3pt}
\begin{tabularx}{\textwidth}{@{}>{\raggedright\arraybackslash}p{2.0cm}Y>{\raggedright\arraybackslash}p{2.6cm}@{}}
\toprule
\textbf{Industry} & \textbf{Brands} & \textbf{Headquarters} \\
\midrule
SaaS & Salesforce, HubSpot, Monday.com, Asana, Miro, Notion, Semrush, ClickUp, Pipedrive, Ahrefs & US, IL, SG, EE \\
Consulting & Deloitte, Accenture, PwC, EY, KPMG, McKinsey, BCG, Bain, Roland Berger, Implement Consulting & US, UK, IE, NL, DE, DK \\
Fintech & PayPal, Block, Stripe, Revolut, Klarna, Adyen, Wise, N26, Plaid, Montonio & US, UK, NL, SE, DE, EE \\
E-commerce & Amazon, H\&M, Zalando, Shopify, ASOS, Allegro, Vinted, Verkkokauppa.com, Temu, Bolt Market & US, CA, DE, SE, UK, LT, PL, CN, EE, FI \\
Health technology & GE HealthCare, Philips Healthcare, Epic Systems, Veeva Systems, Teladoc, CompuGroup Medical, MEDITECH, Doctolib, Kry, Oracle Health & US, DE, FR, SE, NL \\
\bottomrule
\end{tabularx}
\end{table}

\subsection{Category queries}\label{sec:queries}

Fifty queries per industry, 250 in all, were written by the research team in five types with fixed index ranges, the same in every industry (Table~\ref{tab:queries}), to span the segments of each industry. The submitted version called the set 250 brand-free queries. That was false: 49 comparison queries (indices 15 to 24) name individual sampled brands and one compares Big 4 with MBB. Six further queries name a sampled brand (SaaS 29, consulting 28, fintech 28, e-commerce 12 and 28, health technology 28), making 55 prompts with individual sampled-brand names; a second group-only prompt asks about the Big 4. Every statistic is therefore reported on all 250 queries and on the 200 outside the comparison block, the non-comparison set (the analysis plan's brand-free companion; six name individual sampled brands, while the other 194 include Big 4, SAP and Slack prompts), and query type enters every model as a factor. The external corpus examined, 47,019 labelled queries, did not provide suitable consumer supplier-selection queries covering all five categories, so the study keeps the author-written prompt set.

\begin{table}[tbp]
\caption{The five query types: index ranges (the same in every industry), $n$ per industry and in total, prompts that name an individual sampled brand, and empty GPT-5.2 answers in wave 1.}\label{tab:queries}
\centering
\footnotesize
\setlength{\tabcolsep}{3pt}
\begin{tabular}{@{}p{2.2cm}p{1.1cm}p{1.1cm}p{3.6cm}p{2.2cm}p{1.3cm}@{}}
\toprule
\textbf{Type} & \textbf{Indices} & \textbf{$n$} & \textbf{Template} & \textbf{Names a sampled brand} & \textbf{Empty, wave 1} \\
\midrule
Direct recommendation & 0 to 14 & 15 / 75 & ``What is the best [category] for [use case]?'' & 1 of 75 (one brand) & 5 \\
Comparison & 15 to 24 & 10 / 50 & ``[Brand A] vs [Brand B] for [use case]'' & 49 of 50; one group-only & 83 \\
Category exploration & 25 to 34 & 10 / 50 & ``What options exist for [category] in [market]?'' & 5 of 50 (one brand) & 17 \\
Problem-solution & 35 to 44 & 10 / 50 & ``I need to [solve problem]. What should I use?'' & none & 31 \\
Trust & 45 to 49 & 5 / 25 & ``Which [category] company is most trusted?'' & none & 0 \\
\bottomrule
\end{tabular}
\end{table}

\subsection{Systems and collection}\label{sec:collection}

Three systems were called through their vendors' APIs under one configuration (Table~\ref{tab:systems}): temperature 0.3, an output cap of 1,024 tokens and one system prompt. GPT-5.2 and Gemini 3 Flash were called with no search or grounding tool, so both answered from parametric memory; Perplexity sonar-pro retrieves on every call. The submitted version called Gemini web-grounded; it was not. Wave 1 ran from a Google Colab notebook runtime whose region was not recorded and stored the answer text only. Wave 2 ran from 9 September 2026, 22:30 UTC, to 10 September, 11:46 UTC, from a workstation on one residential connection in northern Europe, and also stores the returned identifier and the token counts; the identity of the systems between waves rests on the requested identifiers (Supplementary Section S3). Wave 2 is complete: 3,750 answers, five for each of the 750 query-system cells, with no transport error and no empty body.

\begin{table}[tbp]
\caption{The three systems as called in both waves (one system prompt, temperature 0.3); wave 1 recorded no returned identifier and no token counts; wave 2 returned 1,250 answers per system with no transport error.}\label{tab:systems}
\centering
\footnotesize
\setlength{\tabcolsep}{3pt}
\begin{tabular}{@{}p{3.2cm}p{3.0cm}p{3.2cm}p{3.0cm}@{}}
\toprule
 & \textbf{GPT-5.2 (OpenAI)} & \textbf{Gemini 3 Flash (Google)} & \textbf{Perplexity sonar-pro} \\
\midrule
Model requested / returned (wave 2) & \texttt{gpt-5.2} / \texttt{gpt-5.2-}\allowbreak\texttt{2025-12-11} & \texttt{gemini-3-}\allowbreak\texttt{flash-preview} / the same alias & \texttt{sonar-pro} / the same alias \\
Output cap & 1,024 completion tokens & 1,024 output tokens, thinking included & 1,024 tokens \\
Retrieval & none & none & built in, every call \\
Empty answers, wave 1 / wave 2 & 136 / 0 & 0 / 0 & 0 / 0 \\
At the 1,024-token cap, share of answers (wave 2) & 10.08\% (126/1,250) & 0\% & 0\% \\
\bottomrule
\end{tabular}
\end{table}

The 136 empty wave-1 answers were returned with no error flag, every one from GPT-5.2 (10.9\% of its answers); 43 queries carry at least one, and on 13 all five GPT-5.2 answers are empty (Supplementary Section S3). Wave 1 stored no token usage or finish reason, so the cause cannot be determined; GPT-5.2 returned no empty answer in wave 2 (recorded reasoning tokens: zero in all 1,250 answers).

Three facts about the answer budget bear on any wave comparison. An answer whose completion tokens equal the cap may end mid-list, and the matcher scores the available text; Deviation 2 of the analysis plan, written when 27 of the first 425 wave-2 GPT-5.2 answers had ended at the cap, flags every such answer and repeats every cross-wave statistic without them. Gemini's thinking tokens count against its cap in both waves, so part of each 1,024-token budget is spent before any visible text. Perplexity sonar-pro uses a low retrieval-context setting in wave 2 that wave 1 did not record; answer-length comparisons appear in Section~\ref{sec:res-crosswave} and Supplementary Section S12.

\subsection{Brand mention extraction}\label{sec:extraction}

The submitted version matched each alias as a case-insensitive substring, so ``ey'' matched inside ``they'', ``key'' and ``journey'' and ``implement'' matched the verb. The revision replaces it with a guarded matcher intended to identify brand references: word boundaries for every alias, capitalised or all-caps form for twenty-two ordinary-word aliases, a context rule for eight, one record per brand and answer, and no run on an empty answer (Supplementary Section S1). The change removes about a quarter of the recorded mentions (14,005 presence rows against 10,463); the largest movement is EY, from 2,912 answers to 437.

A gold sample of 400 items (300 precision hits, 100 whole answers) was adjudicated independently by two language models under one written rubric. Both labelled all 300 precision hits true: precision is 300 of 300 (Wilson 95\% interval 98.7 to 100.0 per cent), with kappa undefined because neither used a second label. Recall is 278 of 282 (98.6 per cent; 96.4 to 99.4) under the rubric's rule that an AWS mention names an Amazon service, applied to every item after the first adjudication had left it out of two answers. The four misses are AWS written without the string Amazon; no ``aws'' alias is added, since an answer naming AWS as infrastructure does not recommend retail Amazon, the sampled brand. The figures remain provisional: the author's hand check of a stratified 78-row subset is pending, and no completed human validation is claimed.

The matcher applies written string rules and reads no meaning, and it is frozen for both waves. Markdown emphasis placed directly around an alias hides the neighbouring words a context guard reads, so ``H\&R \texttt{**Block**}'' passes the guard that rejects ``H\&R Block''; in wave 1, 458 of the 2,850 accepted hits of the eight guarded aliases carry such markers, and none would be rejected once the markers are removed. Twenty constructed adversarial sentences (6 pass, 14 fail) are released separately and are not pooled into precision or recall (Supplementary Section S1). Wave 2 is scored by the same matcher; one language model accepted all 100 hits in a fresh precision sample (Wilson 95\% interval 96.3 to 100.0 per cent); its 25-row author hand check is pending.

\subsection{Constructs and measures}\label{sec:measures}

This subsection fixes the operational definitions, every denominator counting non-empty answers. Let $N_q$ be the number of non-empty answers to query $q$ (10 to 15 in wave 1), $N_{q,s}$ the number of them from system $s$ (0 to 5), and $n(b,q)$ the number naming brand $b$.

\emph{Category Inclusion Rate} (CIR):
\begin{equation}
  \mathrm{CIR}(b,q) = \frac{n(b,q)}{N_q}, \qquad \mathrm{CIR}_s(b,q) = \frac{n_s(b,q)}{N_{q,s}},
\end{equation}
the second undefined when $N_{q,s} = 0$; both lie in $[0, 1]$.

\emph{Recommendation Share} (RS) for brand $b$ in industry $k$ is the mean CIR over the industry's query set, on all 50 and on the 40 non-comparison queries, with a BCa interval over queries; a rank-weighted variant weighting each naming answer by $1 / \text{mention rank}$ is reported beside it. \emph{Concentration} is the Gini of the ten RS values (maximum 0.90 at $n = 10$), the HHI of normalised RS (1,000 to 10,000 for ten brands) and the top-three share, each with a BCa interval over queries.

\emph{Competitive Vacuum Index} (CVI) for query $q$ is $1 - \max_b \mathrm{CIR}(b,q)$ over the industry's ten brands: a \emph{vacuum} when $\mathrm{CVI} > 0.50$ (a brand in exactly half of the answers leaves the query outside), and \emph{high maximum inclusion} when $\max_b \mathrm{CIR}(b,q) \geq 0.80$, reported with the number of brands at that level and the count under the strict rule $> 0.80$.

\emph{Co-Mention Asymmetry} (CMA) for ordered brands $A$, $B$ under query $q$ is $P(A \mid \neg B) - P(A \mid B)$ over the non-empty answers to $q$, defined only when $B$ is present in some and absent from others; an undefined cell stays undefined (the submitted code set it to zero, Section~\ref{sec:corrections}). The pooled CMA of an ordered pair is its mean over the queries where it is defined; records (pair, query) and pooled pairs above 0.10 are asymmetric and below $-0.10$ co-mentioned, the submitted cut-offs, with 0.05, 0.15 and 0.20 as a sensitivity.

\emph{Cross-system agreement.} The top brand of a (query, system) cell is the sampled brand named in most of the system's non-empty answers. Ties are broken by the lowest mean ordinal first-mention rank among the answers naming each tied brand; a remaining tie is broken by the original industry brand-list order deposited with the data. A system has no top brand when it returned no non-empty answer or when no answer names a sampled brand. Full agreement is reported on every query (250), on the queries where each system returned at least one non-empty answer (237), the denominator of the E5 rule, and on those where all three have a top brand (211); pairwise and majority (two of three) agreement follow the same rule. The Monte Carlo null of Section~\ref{sec:expectations} is the reference; split-half self-agreement is reported beside it as a description.

\emph{Two estimands.} The empty answers are not missing at random: all 136 are GPT-5.2's, 83 fall on comparison prompts, and 13 queries hold no GPT-5.2 answer. The conditional estimand, inclusion among the answers a system returns, divides by non-empty answers and carries every table. The user-facing estimand counts an empty answer as an answer naming no brand (denominator fifteen, the submitted convention) and is printed beside the headline rows; the two differ exactly where GPT-5.2 returned no answer (Supplementary Section S7).

\subsection{Statistical procedures}\label{sec:procedures}

Every bootstrap, permutation and within-query null uses seed 20260910 with 2,000 draws; the industry-pooled list-length null uses 100 matrices per stratum and the clustering ten seeds. Intervals for RS, Gini, HHI and top-three share are BCa bootstraps over queries resampled within industry, the bias correction counting ties with half weight; where the bias correction reaches its clipping limit, the interval has zero width while the resamples vary, or a resample has a zero denominator, the percentile interval of the full bootstrap distribution is reported with that status. Rates carry Wilson 95\% intervals. Every query-type estimate is reported: its subset holds 5 to 15 queries, so BCa and percentile intervals are given together with degeneracy diagnostics (Supplementary Section S6), and interpretation at that size is cautious. One rule gives every verdict: an interval entirely on the predicted side of its reference is supported, entirely on the other side not supported, and one that includes the reference inconclusive; the submitted hypotheses are evaluated under it with their original cut-offs, E2 and E5 by the rules of Section~\ref{sec:expectations}.

Each pooled CMA pair carries a two-sided permutation $p$ under a null stratified within system. The 450 ordered-pair tests are one dependent family: nominal counts are reported with Benjamini-Hochberg, Benjamini-Yekutieli and Holm adjustments, figures mark nominal $p$ only, and pair results are exploratory. In the within-query null of E2, 130 of the 250 queries admit no trade and enter as point masses; the other 120 show 2 to 2,000 distinct matrices in the 2,000 sampled draws, and the computation restricted to them is a sensitivity. A secondary industry-pooled null groups answers by the number of sampled brands they name (1 to 2, 3 to 4, 5 or more) and draws 100 curveball matrices per stratum. Query type and system enter a binomial logistic regression on the answer $\times$ brand rows, pooled with brand fixed effects, reference levels direct queries and GPT-5.2, standard errors clustered by query, empty answers dropped. Reliability compares iterations $\{1, 3, 5\}$ with $\{2, 4\}$, and brand-set sensitivity removes each brand in turn. Criterion validity is the Spearman correlation between RS and the size benchmark, with a percentile bootstrap over brands, per industry and pooled on within-industry ranks; the industry orderings by recommendation and benchmark concentration are tabulated with no test. For every vacuum query an adjudicating language model read all non-empty answers without the sample restriction and counted every brand named; the authors' check of those readings is pending. The replication follows the frozen analysis plan (Section~\ref{sec:res-crosswave}); Supplementary Section S2 specifies every procedure.

\subsection{Corrections to the submitted analysis}\label{sec:corrections}

Reproduction of the submitted analysis from its own data found four defects, recorded in full in Supplementary Section S5: the method text misdescribed the instrument, the public deposit disagrees with itself and the per-industry values of two submitted tables do not reproduce; alias matching by substring; the claim of 250 brand-free queries; and empty answers scored as answers. The submitted pipeline, re-run on its own mention table, reproduces the printed mean Gini (0.28; 0.284 recomputed), the 41.6 per cent agreement and all 20 printed vacuums. Of the 157 classified quantities in Tables 2 to 7 of the submitted version, 12 flip, 13 otherwise change and 132 are unchanged. Table~\ref{tab:corrections} decomposes the two headline changes. The step sizes of the agreement change depend on their order, because the own-industry restriction and the guarded matcher remove largely the same false top brands, substring hits such as EY in other industries' answers; the Shapley values average each change over all 24 orders, and both sequential orders are in the supplement.

\begin{table}[tbp]
\caption{Agreement corrections: Shapley contributions average 24 orders; historical intermediate denominators vary. Increments are percentage points. Both orders appear in Supplementary Section S5.}\label{tab:corrections}
\centering
\footnotesize
\setlength{\tabcolsep}{3pt}
\begin{tabular}{@{}lccc@{}}
\toprule
Change & Shapley & Increment & Agreement \\
\midrule
Submitted pipeline & & & 41.6\% (104 of 250) \\
R: own-industry top brand & $+9.63$ & $+18.9$ & 60.5\% (147 of 243) \\
T: first-mention rank and brand order & $+2.42$ & $+2.1$ & 62.6\% (152 of 243) \\
M: guarded matcher & $+9.20$ & $+2.7$ & 65.3\% (156 of 239) \\
S: strict no-top rule & $-5.65$ & $-8.1$ & 57.2\% (143 of 250) \\
Missing convention & 0.0 & 0.0 & 57.2\% (143 of 250) \\
\midrule
Vacuum count & \multicolumn{3}{l}{20 submitted; 21 guarded, fixed-15; 19 conditional} \\
\bottomrule
\end{tabular}
\end{table}

\section{Results}\label{sec:results}

Wave 1 holds 3,750 answers, of which the 136 empty GPT-5.2 answers are missing; every rate below divides by the non-empty answers to the query and is the conditional estimand. The user-facing estimand differs only where GPT-5.2 returned no answer: mean Gini 0.303 against 0.302, vacuum rate 7.6 against 8.4 per cent, high-maximum-inclusion rate 81.6 against 76.0 per cent, full agreement 57.2 per cent under both. Intervals are BCa bootstraps over queries, or Wilson intervals for rates.

\subsection{Inclusion rates and leaders}\label{sec:res-inclusion}

\begin{figure}[t]
\centering
\includegraphics[width=\textwidth]{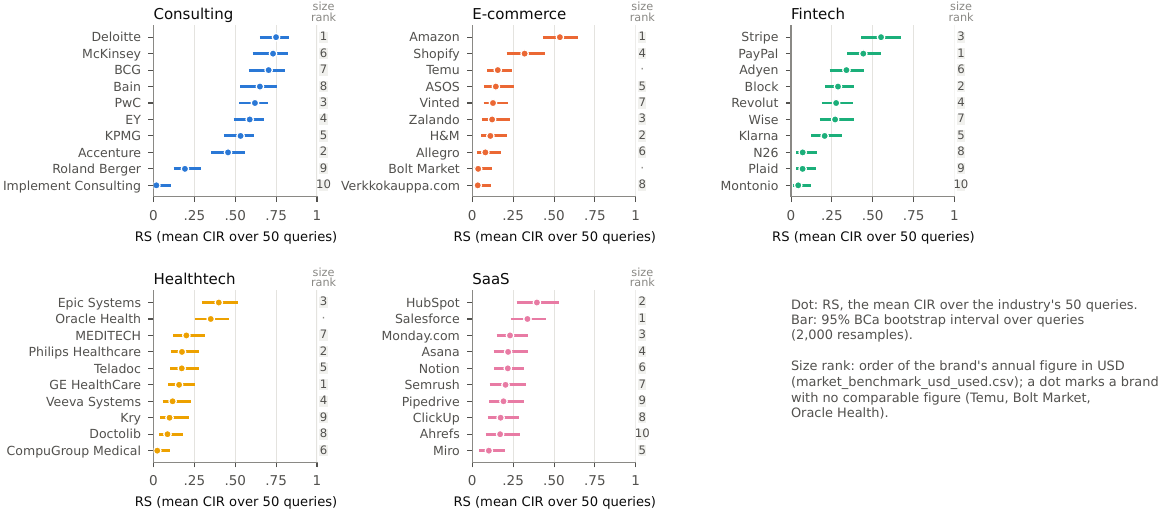}
\caption{Recommendation Share of all fifty brands with BCa 95\% intervals, and each brand's rank in the size benchmark of the sampled brands (grey column; a dot marks a brand without a comparable figure).}
\label{fig:rs}
\end{figure}

Supplementary Figure S2 maps the Category Inclusion Rate of every brand under every query. Of the 2,500 (query, brand) cells, 1,433 are zero and 472 are at or above 0.80. Table~\ref{tab:rs} gives Recommendation Share for the five leading brands per industry and Figure~\ref{fig:rs} all fifty.

\begin{table}[tbp]
\caption{Recommendation Share of the five leading brands per industry, all 50 queries, with BCa 95\% intervals; high inclusion = queries with CIR $\geq 0.80$; absent = CIR $= 0$; rank-weighted RS weights each naming answer by the reciprocal of its mention position; size rank = rank in the size benchmark of the sampled brands (none: no comparable figure).}\label{tab:rs}
\centering
\footnotesize
\setlength{\tabcolsep}{2pt}
\begin{tabular}{@{}llcccccc@{}}
\toprule
Industry & Brand & RS & 95\% CI & High inc. & Absent & RS (rank-weighted) & Size rank \\
\midrule
Consulting & Deloitte & 0.749 & [0.65, 0.83] & 31 & 4 & 0.333 & 1 \\
 & McKinsey & 0.731 & [0.61, 0.82] & 32 & 7 & 0.557 & 6 \\
 & BCG & 0.704 & [0.58, 0.81] & 32 & 8 & 0.310 & 7 \\
 & Bain & 0.651 & [0.53, 0.75] & 25 & 8 & 0.251 & 8 \\
 & PwC & 0.620 & [0.52, 0.70] & 21 & 5 & 0.188 & 3 \\
\midrule
E-commerce & Amazon & 0.536 & [0.43, 0.65] & 18 & 6 & 0.416 & 1 \\
 & Shopify & 0.320 & [0.21, 0.44] & 15 & 28 & 0.271 & 4 \\
 & Temu & 0.156 & [0.09, 0.24] & 2 & 33 & 0.094 & none \\
 & ASOS & 0.144 & [0.07, 0.25] & 6 & 38 & 0.098 & 5 \\
 & Vinted & 0.127 & [0.07, 0.22] & 3 & 34 & 0.086 & 7 \\
\midrule
Fintech & Stripe & 0.551 & [0.43, 0.67] & 22 & 16 & 0.434 & 3 \\
 & PayPal & 0.442 & [0.35, 0.55] & 14 & 14 & 0.216 & 1 \\
 & Adyen & 0.339 & [0.24, 0.44] & 11 & 23 & 0.145 & 6 \\
 & Block & 0.287 & [0.21, 0.38] & 8 & 21 & 0.115 & 2 \\
 & Revolut & 0.276 & [0.19, 0.38] & 6 & 24 & 0.165 & 4 \\
\midrule
Health technology & Epic Systems & 0.399 & [0.30, 0.51] & 12 & 16 & 0.347 & 3 \\
 & Oracle Health & 0.350 & [0.25, 0.46] & 10 & 18 & 0.171 & none \\
 & MEDITECH & 0.201 & [0.12, 0.31] & 8 & 33 & 0.082 & 7 \\
 & Philips Healthcare & 0.174 & [0.11, 0.28] & 5 & 33 & 0.090 & 2 \\
 & Teladoc & 0.173 & [0.10, 0.28] & 6 & 34 & 0.142 & 5 \\
\midrule
SaaS & HubSpot & 0.395 & [0.27, 0.53] & 17 & 25 & 0.314 & 2 \\
 & Salesforce & 0.337 & [0.24, 0.45] & 12 & 19 & 0.217 & 1 \\
 & Monday.com & 0.231 & [0.15, 0.34] & 7 & 27 & 0.114 & 3 \\
 & Asana & 0.219 & [0.13, 0.34] & 9 & 32 & 0.165 & 4 \\
 & Notion & 0.217 & [0.13, 0.32] & 4 & 30 & 0.133 & 6 \\
\bottomrule
\end{tabular}
\end{table}

The consulting leader changes from the submitted version. The substring matcher had credited EY with a share of 0.955 and dominance in 47 of 50 queries; the guarded matcher gives EY 0.588 [0.49, 0.68], rank 6 of 10, and the top three is Deloitte, McKinsey and BCG. Outside consulting the leaders and top-three sets are those of the submitted version. The rank-weighted CIR keeps the leader in four industries and puts McKinsey ahead of Deloitte in consulting, two brands whose intervals overlap.

\subsection{Concentration}\label{sec:res-concentration}

Table~\ref{tab:concentration} gives three concentration statistics of the ten RS values per industry. The mean Gini over the five industries is 0.30 [0.26, 0.32] on all queries and 0.33 [0.29, 0.35] on the non-comparison set. E-commerce has the highest value on all three statistics (Gini 0.42 [0.32, 0.49]; 0.46 non-comparison). SaaS has the lowest Gini and HHI (0.19 [0.10, 0.25]; 1,128), while consulting has the lowest top-three share (41.7 per cent). Concentration does not rest on one brand: dropping any one brand moves the Gini by at most 0.12, dropping the leader by $-0.10$ (e-commerce) to $+0.01$ (consulting), and single removals can swap adjacent industries (Supplementary Section S7).

\begin{sidewaystableorg}[tbp]
\caption{Concentration of Recommendation Share per industry: Gini, HHI (normalised shares, 1,000 to 10,000 for ten brands) and top-three share with BCa 95\% intervals; LOO range = spread of the Gini when one brand is dropped; without leader = Gini of the remaining nine; benchmark top-3 = top-three share of the size benchmark of the sampled brands.}\label{tab:concentration}
\centering
\footnotesize
\setlength{\tabcolsep}{4pt}
\resizebox{\linewidth}{!}{%
\begin{tabular}{@{}lcccccccc@{}}
\toprule
 & \multicolumn{3}{c}{All 50 queries} & \multicolumn{2}{c}{Non-comparison (40)} & \multicolumn{2}{c}{Brand-set sensitivity} & Benchmark \\
Industry & Gini [95\% CI] & HHI [95\% CI] & Top-3 [95\% CI] & Gini [95\% CI] & Top-3 & LOO range & Without leader & top-3 \\
\midrule
Consulting & 0.23 [0.21, 0.24] & 1192 [1150, 1214] & 41.7\% [40.1, 43.4] & 0.23 [0.21, 0.25] & 41.6\% & 0.10 & 0.24 & 59.8\% \\
E-commerce & 0.42 [0.32, 0.49] & 1757 [1418, 2139] & 60.8\% [51.6, 69.3] & 0.46 [0.35, 0.54] & 64.3\% & 0.12 & 0.33 & 96.9\% \\
Fintech & 0.34 [0.28, 0.40] & 1376 [1252, 1516] & 52.1\% [47.0, 58.7] & 0.36 [0.29, 0.42] & 53.0\% & 0.07 & 0.32 & 77.9\% \\
Health technology & 0.33 [0.22, 0.41] & 1389 [1183, 1633] & 53.5\% [44.4, 61.9] & 0.41 [0.29, 0.49] & 58.9\% & 0.08 & 0.30 & 83.8\% \\
SaaS & 0.19 [0.10, 0.25] & 1128 [1035, 1255] & 43.0\% [35.8, 49.5] & 0.21 [0.11, 0.29] & 45.0\% & 0.05 & 0.15 & 93.6\% \\
\midrule
MEAN & 0.30 [0.26, 0.32] & 1368 & 50.2\% & 0.33 [0.29, 0.35] & 52.6\% & & & \\
\bottomrule
\end{tabular}%
}
\end{sidewaystableorg}

\begin{figure}[t]
\centering
\includegraphics[width=\textwidth]{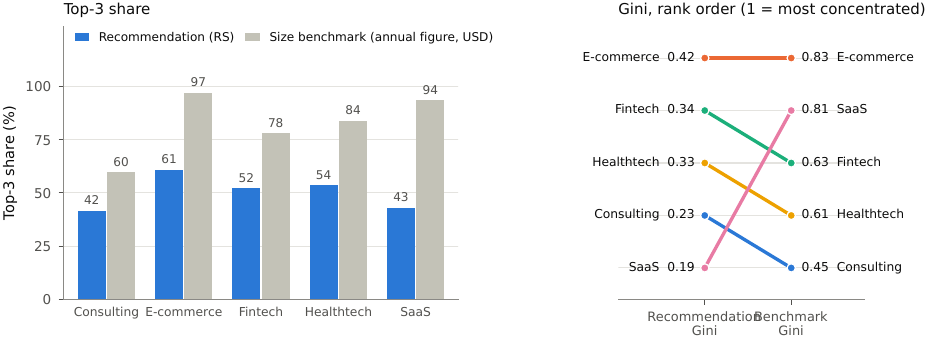}
\caption{Recommendation concentration beside the size benchmark of the sampled brands: top-three share of RS and of the annual figures (left), and each industry's rank by recommendation and by benchmark Gini (right).}
\label{fig:market}
\end{figure}

Figure~\ref{fig:market} sets recommendation concentration beside the size benchmark. Bounded rates can reach a Gini of 0.90 (one brand in every answer, nine in none), so the difference in level is an empirical result: the recommendation Gini is below the benchmark Gini in every industry (0.19 against 0.81 in SaaS, 0.23 against 0.45 in consulting), and so is the top-three share. The orderings are descriptions of five purposively chosen industries, and no test is applied: by recommendation Gini e-commerce, fintech, health technology, consulting, SaaS; by benchmark Gini e-commerce, SaaS, fintech, health technology, consulting. E-commerce leads both, and SaaS is where they part.

Within industries, the rank correlation between RS and the size benchmark is 0.418 [$-0.36$, 0.94] in consulting (10 brands), 0.548 [$-0.33$, 1.00] in e-commerce (8), 0.855 [0.39, 1.00] in fintech (10), 0.467 [$-0.32$, 0.84] in health technology (9) and 0.794 [0.21, 1.00] in SaaS (10); pooled on within-industry ranks over the 47 brands with a figure it is 0.635 ($p < 0.001$). The evidence is precise enough to distinguish positive association from zero in fintech and SaaS and remains imprecise in the other three, and the figures behind it mix audited revenue, recurring-revenue estimates and trade sources. The benchmark's top three share one to three members with the recommendation top three in every industry, and the RS leader is the size leader in two of five.

\subsection{Vacuums}\label{sec:res-vacuums}

\begin{figure}[t]
\centering
\includegraphics[width=0.6\textwidth]{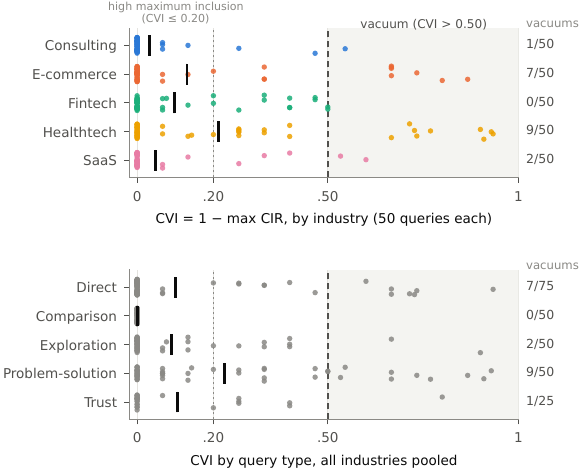}
\caption{Competitive Vacuum Index by industry and by query type, one point per query; the tick is the group mean, the shaded region beyond 0.50 marks vacuums and the dotted line at 0.20 the high-maximum-inclusion boundary.}
\label{fig:cvi}
\end{figure}

A query is a vacuum when every sampled brand is named in fewer than half of its non-empty answers (CVI $> 0.50$; Figure~\ref{fig:cvi}). Vacuums occur in 19 of 250 queries, 7.6 per cent [4.9, 11.6] (user-facing: 21, 8.4 per cent), and in 19 of the 200 non-comparison queries, 9.5 per cent [6.2, 14.4]: health technology 9 of 50 (18.0 per cent [9.8, 30.8]), e-commerce 7 (14.0 per cent [7.0, 26.2]), SaaS 2, consulting 1, fintech 0 [0.0, 7.1]; by type, problem-solution 9 of 50 (18.0 per cent), direct 7 of 75, exploration 2, trust 1, and comparison 0: no comparison query was a vacuum in wave 1, these prompts name brands or groups that may encourage repetition. At least one sampled brand is named in 80 per cent or more of the non-empty answers to 204 of 250 queries, 81.6 per cent [76.3, 85.9] (consulting 94.0, SaaS 90.0, e-commerce 80.0, fintech 80.0, health technology 64.0; strict rule 200; user-facing 190). Two or more brands reach that level on 137 of the 204 and exactly one on 67, so the statistic measures repeated inclusion and shows neither exclusivity nor concentration within a query.

Table~\ref{tab:vacuums} lists the 19 vacuum queries. Read without the sample restriction (Supplementary Section S10; author check pending), the most-named brand is outside the sample in 18 of the 19, and in 15 of those it is named in more than half of the non-empty answers: Instacart, Doxy.me and Follow Up Boss in every answer to the quick-commerce, telehealth and real-estate CRM queries. Three queries remain vacuums under the same strict rule after the open-vocabulary reading: SAP implementation (Deloitte in 5 of 11 non-empty answers), life-sciences technology and remote patient monitoring; on the Nordic digital-health query Microsoft Azure is named in exactly 7 of 14, on the boundary. The 7.6 per cent vacuum rate is therefore an upper bound on category-level vacuums; the rest measure the distance between the 50-brand sample and the sub-category the query targets.

\begin{table}[tbp]
\caption{The 19 vacuum queries (CVI $> 0.50$) with the best sampled brand by the matcher; type D direct, E exploration, P problem-solution, T trust; the most-named brand without the sample restriction is in Supplementary Table S9.}\label{tab:vacuums}
\centering
\footnotesize
\setlength{\tabcolsep}{2pt}
\begin{tabularx}{\textwidth}{@{}lcYcc>{\raggedright\arraybackslash}p{3.2cm}@{}}
\toprule
Industry & Idx & Query & Type & CVI & Best sampled brand (CIR) \\
\midrule
Health technology & 11 & Best EHR for small medical practices & D & 0.93 & Epic Systems (0.07) \\
Health technology & 44 & Launch a digital health service in the Nordics & P & 0.93 & Kry (0.07) \\
Health technology & 40 & Monitor patients remotely & P & 0.91 & Epic Systems (0.09) \\
Health technology & 31 & Which health tech startups are emerging & E & 0.90 & Teladoc (0.10) \\
E-commerce & 44 & Buy luxury fashion online & P & 0.87 & Amazon (0.13) \\
E-commerce & 47 & Fashion retailer with the best reputation & T & 0.80 & H\&M (0.20) \\
Health technology & 43 & Radiology AI software & P & 0.77 & Epic Systems (0.23) \\
E-commerce & 6 & Marketplace for electronics in the Nordics & D & 0.73 & Amazon (0.27) \\
Health technology & 39 & Online appointment booking for patients & P & 0.73 & Epic Systems (0.27) \\
Health technology & 13 & Health tech for pharmaceutical companies & D & 0.73 & Veeva Systems (0.27) \\
Health technology & 5 & Health technology for life sciences & D & 0.71 & Veeva Systems (0.29) \\
E-commerce & 5 & Best quick commerce delivery app & D & 0.67 & Amazon (0.33) \\
E-commerce & 10 & Marketplace for sustainable shopping & D & 0.67 & Amazon (0.33) \\
E-commerce & 30 & Quick commerce services available & E & 0.67 & Amazon (0.33) \\
E-commerce & 42 & Find sustainable products online & P & 0.67 & Amazon (0.33) \\
Health technology & 36 & Offer telehealth to my patients & P & 0.67 & Teladoc (0.33) \\
SaaS & 13 & Best CRM for real estate agents & D & 0.60 & HubSpot (0.40) \\
Consulting & 37 & Consultant for an SAP implementation & P & 0.55 & Deloitte (0.45) \\
SaaS & 44 & Build landing pages without a developer & P & 0.53 & HubSpot (0.47) \\
\bottomrule
\end{tabularx}
\end{table}

\subsection{Co-mention structure}\label{sec:res-cma}

\begin{figure}[t]
\centering
\includegraphics[width=\textwidth]{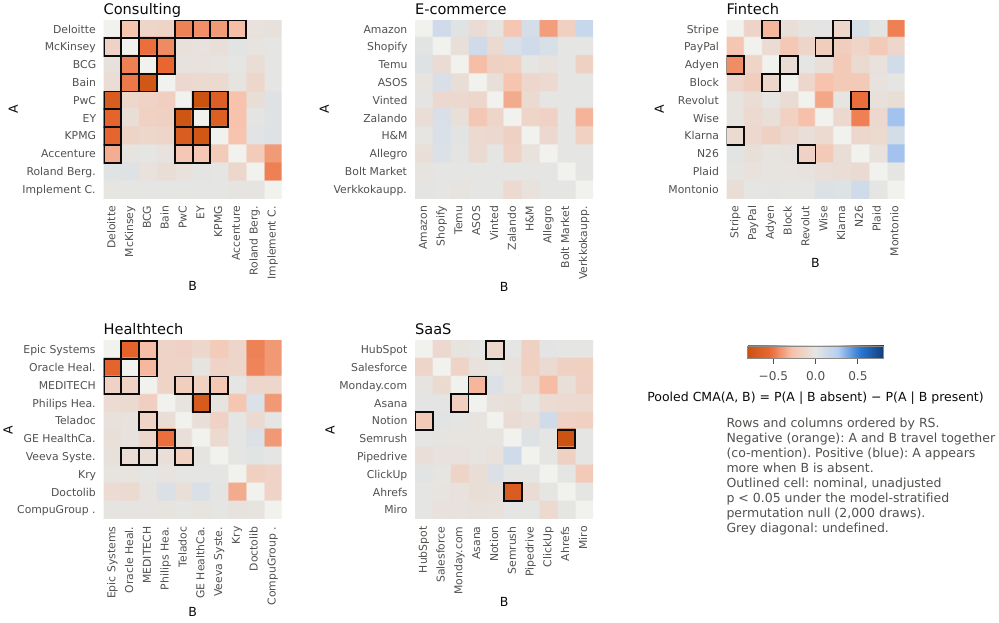}
\caption{Co-Mention Asymmetry between every ordered pair of brands, ordered by RS; orange (negative): $A$ is named more often when $B$ is also named; blue (positive): the reverse; an outline marks nominal, unadjusted $p < 0.05$ under the system-stratified permutation null.}
\label{fig:cma}
\end{figure}

Table~\ref{tab:cma} and Figure~\ref{fig:cma} give records (pair, query) and pooled pairs beyond $|\mathrm{CMA}| > 0.10$. Pooled over all answers, co-mention dominates: the records ratio is 0.28:1 [0.21, 0.36] (487 asymmetric against 1,769 co-mentioned; user-facing 0.23:1), and every industry is below 1:1 at every threshold from 0.05 to 0.20 (Supplementary Section S9), e-commerce at 0.58:1 the highest. Of the 450 ordered-pair tests under the system-stratified null, 53 have nominal $p < 0.05$ (49 of them with $|\mathrm{CMA}| > 0.10$), all with negative CMA; Benjamini-Hochberg retains 26, and Benjamini-Yekutieli and Holm retain none, since with 2,000 permutations the smallest attainable $p$ exceeds Holm's first threshold. The strongest pairs are co-mention within tiers: PwC and EY ($-0.80$), Semrush and Ahrefs ($-0.77$), Bain and BCG ($-0.73$). The submitted ratios above 1 came from setting an undefined $P(A \mid B)$ to zero: on the corrected mention table that substitution still gives 1.49:1 pooled.

\begin{sidewaystableorg}[tbp]
\caption{Co-Mention Asymmetry at $|\mathrm{CMA}| > 0.10$, all 50 queries. Upper block: records are (pair, query) combinations; ratio = asymmetric : co-mentioned with a BCa 95\% interval; significant pairs have nominal, unadjusted $p < 0.05$ under the system-stratified permutation null. Lower block: E2 against the within-query fixed-margins null (2,000 draws; band = 2.5th to 97.5th percentile), all queries and the non-comparison set.}\label{tab:cma}
\centering
\footnotesize
\setlength{\tabcolsep}{4pt}
\resizebox{\linewidth}{!}{%
\begin{tabular}{@{}lcccccc@{}}
\toprule
Industry & Records asym. & Records co-m. & Records ratio [95\% CI] & Pairs ratio & Sig. pairs (asym. / co-m.) & User-facing ratio \\
\midrule
Consulting & 193 & 881 & 0.22:1 [0.13, 0.36] & 0.00:1 & 0 / 24 & 0.20:1 \\
E-commerce & 67 & 116 & 0.58:1 [0.30, 1.03] & 0.17:1 & 0 / 0 & 0.57:1 \\
Fintech & 115 & 361 & 0.32:1 [0.20, 0.56] & 0.11:1 & 0 / 7 & 0.25:1 \\
Health technology & 45 & 260 & 0.17:1 [0.08, 0.32] & 0.00:1 & 0 / 12 & 0.15:1 \\
SaaS & 67 & 151 & 0.44:1 [0.24, 0.72] & 0.10:1 & 0 / 6 & 0.30:1 \\
\midrule
Pooled & 487 & 1769 & 0.28:1 [0.21, 0.36] & 0.06:1 & 0 / 49 & 0.23:1 \\
\midrule
Query set & Mean CMA [null band] & Ratio at 0.10 [null band] & RS and mean CMA, $\rho$ [null band] & Pairs BH-significant & Pairs departing & E2 (reference) \\
\midrule
All (250) & $-0.1171$ [$-0.1190$, $-0.1166$] & 0.275 [0.176, 0.201] & $-0.525$ [$-0.513$, $-0.362$] & 58 & 43 & not supported \\
Non-comparison (200) & $-0.1197$ [$-0.1217$, $-0.1191$] & 0.279 [0.177, 0.203] & $-0.515$ [$-0.474$, $-0.321$] & 58 & 43 & not supported \\
\bottomrule
\end{tabular}%
}
\end{sidewaystableorg}

Against the within-query fixed-margins null of Section~\ref{sec:expectations}, which keeps each answer's list length and each brand's count within its query, the mean CMA lies inside a band 0.0025 wide, as the margins nearly force, and carries almost no evidence (Table~\ref{tab:cma}, lower block). The tails carry it. The ratio at 0.10 is 0.275 against [0.176, 0.201], above the band pooled and in consulting, fintech and SaaS and inside it in e-commerce and health technology. Of the 450 ordered pairs, 58 differ from their null after Benjamini-Hochberg at $q = 0.05$ and 43 of them reproduce their sign in both split halves: 31 are named together more often than their margins imply and 12 less often (Section~\ref{sec:discussion}). RS and mean CMA as brand $A$ are rank-correlated at $-0.525$, more negative than the null band $[-0.513, -0.362]$ that the margins alone produce. Under the rules of Section~\ref{sec:expectations} E2 is not supported on both query sets; the rules followed inspection of the earlier 120-query result; the corrected all-query verdicts are references for wave 2. Every result here is an association in answer text: a departing pair is named together, or apart, more often than its query's margins imply, and nothing identifies one brand's presence as the cause of another's absence.

\subsection{Cross-system agreement}\label{sec:res-agreement}

\begin{figure}[t]
\centering
\includegraphics[width=0.6\textwidth]{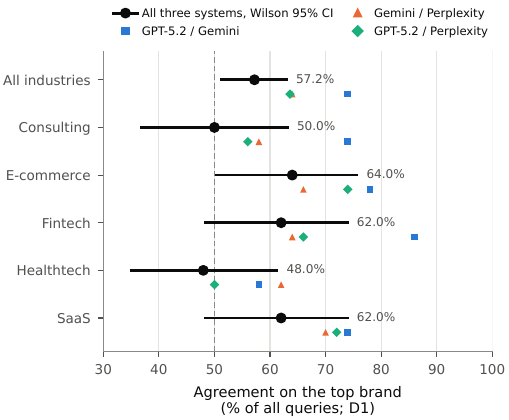}
\caption{Share of all queries on which the three systems have the same top brand, with Wilson 95\% intervals, by industry and overall; small markers give the pairwise rates; the dashed line is the submitted H5 cut-off of 50 per cent. Among the queries every system answered, the overall rate is 60.3 per cent (143 of 237).}
\label{fig:agreement}
\end{figure}

Table~\ref{tab:agreement} reports agreement under the three denominators (Figure~\ref{fig:agreement}, every query). The three systems name the same top brand in 143 of 250 queries, 57.2 per cent [51.0, 63.2], a query on which a system has no top brand counting as disagreement; in 143 of the 237 queries on which every system returned at least one non-empty answer, 60.3 per cent [54.0, 66.4]; and in 67.8 per cent of the 211 on which all three have a top brand. Each rate is the same under both estimands; the first two differ by their denominator. Pairwise agreement is 74.0 per cent between GPT-5.2 and Gemini 3 Flash, the two parametric systems, and 64.0 and 63.6 per cent between each of them and Perplexity sonar-pro; a majority of two agrees on 87.2 per cent of queries.

\begin{sidewaystableorg}[tbp]
\caption{Cross-system agreement on the top brand per query: every query (no top brand counts as disagreement; Wilson 95\% interval), answered (every system returned an answer) and complete (all three have a top brand); pairwise and majority over every query; MC null = each system redrawing its top brand from its own distribution (mean and 95\% band); self-agreement = the lowest of the three systems' split-half rates, a description.}\label{tab:agreement}
\centering
\footnotesize
\setlength{\tabcolsep}{3pt}
\resizebox{\linewidth}{!}{%
\begin{tabular}{@{}lcccccccccc@{}}
\toprule
Group & $n$ & Every query [95\% CI] & Answered & Complete & GPT and Gemini & Gemini and Pplx & GPT and Pplx & Majority & MC null [95\% band] & Self-agreement, lowest \\
\midrule
Consulting & 50 & 50.0\% [36.6, 63.4] & 50.0\% & 52.1\% & 74.0\% & 58.0\% & 56.0\% & 88.0\% & 10.9\% [4.0, 20.0] & 76.0\% \\
E-commerce & 50 & 64.0\% [50.1, 75.9] & 64.0\% & 76.2\% & 78.0\% & 66.0\% & 74.0\% & 90.0\% & 5.7\% [0.0, 14.0] & 78.0\% \\
Fintech & 50 & 62.0\% [48.2, 74.1] & 66.0\% & 70.5\% & 86.0\% & 64.0\% & 66.0\% & 92.0\% & 6.0\% [0.0, 14.0] & 74.0\% \\
Health technology & 50 & 48.0\% [34.8, 61.5] & 55.8\% & 75.0\% & 58.0\% & 62.0\% & 50.0\% & 74.0\% & 4.2\% [0.0, 10.0] & 56.0\% \\
SaaS & 50 & 62.0\% [48.2, 74.1] & 66.0\% & 68.9\% & 74.0\% & 70.0\% & 72.0\% & 92.0\% & 2.6\% [0.0, 8.0] & 86.0\% \\
\midrule
All industries & 250 & 57.2\% [51.0, 63.2] & 60.3\% & 67.8\% & 74.0\% & 64.0\% & 63.6\% & 87.2\% & 5.9\% [3.2, 8.8] & 77.6\% \\
All, non-comparison & 200 & 56.0\% [49.1, 62.7] & 57.1\% & 65.9\% & 73.5\% & 62.0\% & 63.0\% & 86.5\% & 7.9\% [4.5, 11.5] & 75.0\% \\
\bottomrule
\end{tabular}%
}
\end{sidewaystableorg}

Three systems each naming their usual brands with no regard to the query would agree on 5.9 per cent of queries [3.2, 8.8]. Observed agreement lies above the null's upper percentile in every industry, and among the 237 answered queries its lower Wilson bound, 54.0 per cent, far exceeds the null's 9.3 per cent. Split-half self-agreement, 78.4 per cent for GPT-5.2, 88.0 for Gemini and 77.6 for Perplexity, describes each system's reliability between halves of three and two answers and is not a bound on cross-system agreement. A per-query independence reference (65.7 per cent) equals 1 on every agreeing query and tracks the rate it references, so it provides no null model. Under the answered-query pair denominators, the parametric pair agrees more than either parametric-retrieval pair: differences are 0.141 [0.083, 0.199] and 0.110 [0.047, 0.171]. Both E5 components are supported in February and September (Supplementary Section S12).

\subsection{Query type and system effects}\label{sec:res-glm}

Table~\ref{tab:glm} reports the pooled logistic model of brand presence per non-empty answer (250 query clusters; 36,140 answer-by-brand rows). Relative to direct queries, comparison prompts lower the odds that a sampled brand is named (odds ratio 0.73 [0.56, 0.95]), exploration prompts raise them (1.70 [1.35, 2.13]), problem-solution prompts lower them (0.75 [0.58, 0.97]) and trust prompts show no clear difference (0.99 [0.75, 1.30]). Gemini 3 Flash and Perplexity sonar-pro name a sampled brand less often than GPT-5.2 (0.64 [0.57, 0.71] and 0.66 [0.59, 0.75]); under the user-facing estimand, where the empties counted against GPT-5.2, the two terms were 0.76 and 0.79. The per-industry models agree in direction on the exploration and system terms in all five industries (Supplementary Section S8). Full cross-system agreement is 42.0 per cent [29.4, 55.8] on problem-solution queries against 60.0 to 64.0 on the other types.

\begin{table}[tbp]
\caption{Pooled logistic model of brand presence per non-empty answer: odds ratios with 95\% intervals relative to direct queries and to GPT-5.2, brand fixed effects, standard errors clustered by query; the user-facing column scores an empty answer as naming nobody.}\label{tab:glm}
\centering
\footnotesize
\setlength{\tabcolsep}{3pt}
\begin{tabular}{@{}lccc@{}}
\toprule
Term & Odds ratio [95\% CI] & $p$ & User-facing [95\% CI] \\
\midrule
Comparison vs direct & 0.73 [0.56, 0.95] & 0.021 & 0.61 [0.47, 0.78] \\
Exploration vs direct & 1.70 [1.35, 2.13] & $< 0.001$ & 1.63 [1.30, 2.04] \\
Problem-solution vs direct & 0.75 [0.58, 0.97] & 0.030 & 0.70 [0.53, 0.92] \\
Trust vs direct & 0.99 [0.75, 1.30] & 0.921 & 0.99 [0.75, 1.31] \\
Gemini 3 Flash vs GPT-5.2 & 0.64 [0.57, 0.71] & $< 0.001$ & 0.76 [0.67, 0.85] \\
Perplexity sonar-pro vs GPT-5.2 & 0.66 [0.59, 0.75] & $< 0.001$ & 0.79 [0.70, 0.89] \\
\bottomrule
\end{tabular}
\end{table}

\subsection{Stability and sensitivity}\label{sec:res-stability}

Recommendation Share is stable across draws: computed on iterations $\{1, 3, 5\}$ and on $\{2, 4\}$, the fifty brand shares correlate at $r = 0.998$ (Spearman 0.995), the per-industry Gini differs by at most 0.013, the vacuum classification agrees with Cohen's $\kappa = 0.81$, brand sets across iterations have Jaccard similarity 0.79 to 0.84, and pooled pair CMA has split-half sign agreement of 84.0 per cent (Supplementary Table S5). Dropping the 13 queries with no GPT-5.2 answer or all 43 queries with an empty answer moves the mean Gini by at most 0.01 and full agreement to 60.3 and 60.4 per cent (Supplementary Figure S3). Every hypothesis verdict of Section~\ref{sec:res-hypotheses} is the same under both estimands and on both drop sets.

\subsection{The submitted hypotheses revisited}\label{sec:res-hypotheses}

None of the submitted hypotheses is supported under the common interval rule in either estimand. H1, H2, H3 and H5 are not supported overall; H3 and H5 have inconclusive industry rows. On the non-comparison set H5 is inconclusive and H4 is not separately clustered. H4's ten-seed re-run finds zero emergent clusters at every seed after excluding the 136 empty answers that formed the submitted three emergent clusters (4.2 per cent); the corrected range lies below the submitted 15 per cent cut. Supplementary Section S11 reports every scope, cut-off and verdict.

\subsection{Cross-wave replication}\label{sec:res-crosswave}

The plan was frozen after 950 answers and their length and token diagnostics were available, before brand statistics; its six deviations also preceded brand statistics. Deviations 5 and 6 revised E2 after inspection of wave-1 results. This is a partially pre-specified replication, with chronology and rules in Supplementary Section S2 and hashes in Data availability.

Recommendation Share is stable between dates (Spearman 0.994, Pearson 0.996), with all industry leaders unchanged; ASOS replaces Temu in the e-commerce top three. Mean Gini changes by $-0.006$ [$-0.014$, 0.005], the vacuum rate remains 7.6 per cent, and agreement rises from 57.2 to 60.8 per cent, a 3.6-point change [$-1.6$, 8.8]. These intervals include zero; they do not establish equivalence. Perplexity has lower brand-set overlap (0.672) than OpenAI (0.886) and Gemini (0.882).

E1 remains descriptive: size and RS correlate at $\rho=0.616$ [0.315, 0.790]. E2-a$'$ is supported on all queries ($-0.455$ within [$-0.627$, $-0.428$]), but E2-b$'$ and the composite are not: the ratio 0.220 exceeds [0.144, 0.166], and 31 of 441 tested ordered pairs depart. E2-a$'$ is not supported on the non-comparison companion. E5 is supported: both pair-difference intervals are positive (0.120 [0.064, 0.184] and 0.120 [0.060, 0.180]), and answered-query agreement's lower Wilson bound, 54.6 per cent, exceeds the null's upper percentile, 8.8 per cent. E2 and E5 composite verdicts persist in the companion and at-cap arms (Supplementary Section S12). All five historical pooled hypotheses are not supported; H4 has zero emergent clusters at every seed, with one unanchored cluster under the stricter benchmark at seed 0.

The system records constrain interpretation. OpenAI's missing February answers reappear, and 126 September answers reach its cap. Perplexity answers shorten from 2,343 to 1,551 characters (paired change $-793$ [$-846$, $-741$]); its retrieval-context setting and unobserved backend changes prevent a causal attribution. The at-cap sensitivity retains the composite findings, with all four frozen populations reported beside the primary results. Comparison and exploration odds ratios are 0.78 [0.61, 1.00] and 2.00 [1.56, 2.55]; problem-solution is 0.88 [0.68, 1.13], whose interval includes one.

\begin{table}[tbp]
\caption{February against September 2026, all 250 queries; S supported, NS not supported. Rate changes are percentage points; intervals are paired percentile query bootstraps. Answered/complete denominators are 237/211 in February and 250/223 in September. O/G/P denote OpenAI/Gemini/Perplexity. Supplementary Section S12 gives all four frozen populations and the at-cap sensitivity, expectation intervals, lengths, leaders and regressions.}\label{tab:crosswave}
\centering\footnotesize\setlength{\tabcolsep}{3pt}
\begin{tabularx}{\textwidth}{@{}>{\raggedright\arraybackslash}p{4.2cm}Y>{\raggedright\arraybackslash}p{2.6cm}>{\raggedright\arraybackslash}p{2.8cm}@{}}
\toprule
Statistic & February & September & Change [95\% CI] \\
\midrule
RS cross-wave Spearman / Pearson &  & 0.994 / 0.996 &  \\
Mean industry Gini & 0.303 & 0.297 & -0.006 [-0.014, +0.005] \\
Vacuum rate & 7.6\% & 7.6\% & +0.0 [-2.4, +2.4] \\
High maximum inclusion & 81.6\% & 82.4\% & +0.8 [-2.4, +4.0] \\
Full agreement, every query & 57.2\% & 60.8\% & +3.6 [-1.6, +8.8] \\
Full agreement, answered / complete & 60.3\% / 67.8\% & 60.8\% / 68.2\% &  \\
Same pooled top brand between waves &  & 88.8\% &  \\
Same top per system (O / G / P) &  & 86.4\% / 91.6\% / 73.6\% &  \\
Brand-set Jaccard (O / G / P) &  & 0.886 / 0.882 / 0.672 &  \\
E1 size--RS $\rho$, descriptive & 0.635 & 0.616 &  \\
E2-a$'$ / E2-b$'$ / composite & NS / NS / NS & S / NS / NS &  \\
E5-a / E5-b / composite & S / S / S & S / S / S &  \\
Submitted H1 to H5, primary & All NS & All NS &  \\
\bottomrule
\end{tabularx}
\end{table}

\section{Discussion}\label{sec:discussion}

The corrected map has two levels, and the consideration-set model, written after the wave-1 results, accounts for both. Across the ten sampled brands of an industry, inclusion rates sit close together: a mean Gini of 0.30, a top-three share of 50 per cent, and in four of five industries a leader with RS below 0.56. Within a query the systems repeat themselves: at least one sampled brand is named in 80 per cent or more of the non-empty answers to 204 of 250 queries, and on 137 of those two or more brands are. The tuning objective accounts for the several members named; corpus salience accounts for which members a query calls up, so a brand is named in nearly every answer to the query that describes it and in none to the query that does not, and averaging over queries yields the flat distribution.

Recommendation concentration is below benchmark concentration in every industry. Consulting has both the lowest recommendation top-three share (41.7 per cent) and the lowest benchmark top-three share (59.8 per cent). SaaS departs: its sampled brands are the second most concentrated by size (Salesforce holds 85 per cent of their revenue), yet its recommendation Gini is the lowest (0.19), and HubSpot, at less than a tenth of Salesforce's revenue, leads it. The tuning-objective account fits the direction; in SaaS size orders the brands (Spearman 0.794) and says nothing about the level. Health technology shows where the sample runs out before salience does: its 18.0 per cent vacuum rate reads, without the sample restriction, as the systems naming Doxy.me, SimplePractice, Aidoc and Tebra in nearly every answer. The three vacuums that survive an open reading lack a brand named in half the non-empty answers.

The co-mention structure follows the answer budget in its mean and departs from it in its tails. Mean asymmetry is nearly fixed by inclusion rates and list lengths, and the observed mean sits inside that null's narrow band. The 43 departing pairs have a pattern: 31 are named together more often than their margins imply and 12 less often. Illustrative examples are Adyen with Stripe for more co-mention and KPMG with BCG for less.  The pattern is consistent with grouping by tier beyond the fixed margins; tier membership was not a pre-specified covariate. A concentrated industry need not show asymmetric exclusion of its weaker members: consulting combines relatively low concentration with the strongest co-mention. The submitted displacement ratios measured a coding convention.

Cross-system agreement exceeds the query-independent marginal null. Three systems naming their usual brands with no regard to the query would agree on 5.9 per cent of queries; they agree on 57.2 per cent. Each system's split-half self-agreement, 77.6 to 88.0 per cent, describes its reliability over halves of three and two answers and sets no limit on agreement between systems. The two systems called without retrieval agree with each other (74.0 per cent) more than either agrees with the retrieval-grounded one (64.0 and 63.6 per cent), consistent with a shared parametric layer, and both pair-difference intervals are positive in each wave under the answered-query denominators. On the systems and dates studied, the three part on the top brand in about four queries in ten.

Split-half stability ($r = 0.998$; vacuum $\kappa = 0.81$) supports reproducibility within this collection \citep{atil2025nondeterminism, schulte2026dontmeasureonce}; it does not establish stability across dates. The association with query type calls for reporting each type separately.

Recommendation Share relates to size moderately and unevenly (Section~\ref{sec:res-concentration}): it follows size in rank more than in level, as a system-side salience construct would under double jeopardy \citep{romaniuk2016hbg2}, and the RS leader is the size leader in two industries of five.

For the theory of intermediaries, a language model answering a category query sets an agenda of a few named brands \citep{mccombs1972agenda}, drawn from a memory whose recall follows corpus frequency \citep{petroni2019language, kandpal2022large, mallen2023trust} and shaped by a preference for even-handed lists \citep{ouyang2022training}. Mental availability \citep{romaniuk2004salience} has a system-side analogue that behaves like its consumer-side original in rank and unlike it in range. Generative engine optimisation treats inclusion as a search problem \citep{aggarwal2023geo}; the map shows why it is a different one: a brand is either in the set or outside it, entry is decided query by query, and there is no position to climb.

\section{Implications for practice}\label{sec:implications}

An inclusion-rate audit tells a marketing team four things and no more. It gives the brand's Category Inclusion Rate under every query it cares about, on every system it audits, with an interval, since a single answer is one draw from a propensity. It gives the brand's Recommendation Share against the sampled competitors and the concentration of the set, so a team can see whether the systems name many members of its category (consulting, SaaS) or few (e-commerce). It gives the vacuum queries, each with the open-vocabulary reading that says whether the systems name nobody or a specialist the brand list did not carry, the common case. It gives the co-mention structure: which pairs are named together, and which more often than their inclusion rates and answer lengths imply.

Four terms the submitted version used interchangeably describe four different things, and an audit report should keep them apart. AI visibility is any mention in any answer. Recommendation inclusion is a brand named in an answer to a category query, with rates separated by whether the prompt names a sampled brand. Category concentration is a statistic of the sampled set, comparable across industries by rank. Competitive advantage is a consumer outcome that needs consumer data and a design that manipulates exposure; an audit of system output observes none of it.

What the audit cannot tell a team is what to do about a number. The study manipulates no content, observes no consumer and measures no change in any system after any intervention, so it supports no statement about the return on content investment, structured data, citations or any other action a vendor might sell as generative engine optimisation. Treating a vacuum as an uncontested opportunity would have sent a fashion retailer after a luxury-fashion query that Net-a-Porter already fills. A co-mention pair is an association and licenses no offensive or defensive strategy. A share that is high on one system and low on another is a fact about two systems on one date, and the four-in-ten disagreement rate says that a team auditing one system should audit the others. The audit's value is descriptive and repeated: the same queries and systems every wave, with intervals, so that a change is separable from a draw.

\section{Limitations}\label{sec:limitations}

Three systems were queried, and inclusion is a property of a system version on a date. GPT-5.2, Gemini 3 Flash and Perplexity sonar-pro differ in more than retrieval at once, so the agreement figures describe these three systems and isolate no architectural cause. Wave 1 recorded no returned model string, token counts or client region, and ran from a Google Colab notebook runtime; wave 2 ran from a workstation on one residential connection in northern Europe, so the waves are not matched on client environment.

The brands and the queries were chosen purposively; a size-based frame would add fifteen brands, every concentration statistic is a property of the sampled set, and most vacuums are the distance between the sample and a sub-category. The 250 queries were written by the authors, 55 name individual sampled brands and two name only brand groups. The external corpus we examined did not provide suitable consumer supplier-selection queries covering all five categories.

Wave 1 holds 136 empty GPT-5.2 answers, 10.9 per cent of that system's answers and 33 per cent of its comparison answers, which the submitted version scored as answers naming nobody. They are not missing at random, so the paper reports the conditional and the user-facing estimand; the pooled submitted-hypothesis verdicts and the E2 composite remain the same under both, and wave 2 supplies GPT-5.2 answers for the 13 queries that had none.

The five hypotheses were set before analysis and not registered. E1, E2 and E5 were formulated after the wave-1 results were known, the plan that fixes them was frozen after 950 wave-2 answers had been collected, and its E2 rules were replaced after inspecting the earlier wave-1 null, so the wave-1 verdicts are references and wave 2 is a partially pre-specified replication. The February gold sample was adjudicated by two language models, the September precision sample by one, and the open-vocabulary readings by one; until the author hand checks are recorded, those figures are provisional. The stored wave-1 answer embeddings have no recorded encoder revision. Re-encoding 292 answers gave a minimum cosine similarity of 0.972; the cause remains unresolved and limits direct cross-wave clustering comparisons (Supplementary Section S11).

Demographic framing of the query, which prior work found to change the number of brands named \citep{zatuchin2026a} and which brands are named \citep{xu2025genderrace}, is not varied here; conditioning inclusion on it needs a separate factorial collection with demographic framing varied explicitly. Brand origin is carried through the sampled set and tested by no separate expectation.

\section{Conclusion}\label{sec:conclusion}

Brand inclusion in three language-model systems across five industries has two levels, rates close together across an industry's sampled brands and a few members named in nearly every answer to a query, and the consideration-set model accounts for both. Vacuums are mostly the distance between a brand list and a sub-category the systems know better than the list does. Fixed margins nearly determine mean co-mention asymmetry, while conditional pair associations depart from the null. The three systems agree above the query-independent marginal null, and the two parametric systems agree most, consistent with a shared parametric layer. None of the submitted version's hypotheses is supported under the interval rule, and the one reported as supported rested on an extraction artefact. The constructs are audit statistics of system output and claim no consumer outcome or effect of content. The September replication retains the positive descriptive size--RS association (E1), rejects full fixed-margin compatibility (E2), and supports the specified agreement ordering and null comparison (E5). High cross-date RS correlation accompanies query-level changes; neither stability nor agreement identifies an architectural cause.

\bmhead{Data availability}

Published version 2.0 (DOI \href{https://doi.org/10.5281/zenodo.22693819}{10.5281/zenodo.22693819}) contains the corrected February outputs, guarded matcher and specification, model adjudications, size benchmark, open-vocabulary readings, clustering code and pinned dependencies, both raw collections, and the scrubbed notebook. The actual September analyses and completed author checks require a subsequent version, not yet published. The earlier release's inclusion-rate file used a one-system denominator inconsistent with its share and vacuum files; version 2.0 replaces those outputs. The original plan was frozen on 2026-09-10T01:15:54Z with 950 answers collected, at SHA-256 \texttt{73d8425c\allowbreak 8388c563\allowbreak 9034d6ca\allowbreak 6a6704bc\allowbreak 42a756e8\allowbreak 046265a9\allowbreak 6647ace0\allowbreak 721dfa1f}. Six deviations were appended before September brand statistics (whole-file SHA-256 \texttt{34dd9092\allowbreak 13f08a07\allowbreak a46628f6\allowbreak a314d932\allowbreak b8e75592\allowbreak 3d92d5c4\allowbreak 73633092\allowbreak 0ac8fa7d}); the published plan is redacted and the original was retained for validation.

\bmhead{Competing interests}

D.~\.Zatuchin is affiliated with Rankfor.AI, which develops AI brand intelligence tools. The research was conducted independently; the company had no influence on study design, methodology, analysis, or conclusions.

\bmhead{Ethics approval}

The study queried commercial language-model interfaces and involved no human participants; no ethics approval was required.

\bmhead{Use of generative AI}

The three language-model systems are the objects of study. A further language model, following a written rubric, adjudicated the 400-item gold sample and read the answers to the 19 vacuum queries without the sample restriction; a second language model adjudicated the same gold sample independently (Sections~\ref{sec:extraction} and~\ref{sec:procedures}). One language model also adjudicated the fresh 100-item September precision sample under the same rubric. The author checks are pending. Language-model editing assistance was used on the prose; the authors retain responsibility for its final review before submission.

\setlength{\bibsep}{0.9em}
\bibliography{references-revised}

\begin{thebibliography}{36}
\providecommand{\natexlab}[1]{#1}
\providecommand{\url}[1]{{#1}}
\providecommand{\urlprefix}{URL }
\providecommand{\doi}[1]{\url{https://doi.org/#1}}
\providecommand{\eprint}[2][]{\url{#2}}
 \bibcommenthead

\bibitem[{Aggarwal et~al.(2024)Aggarwal, Murahari, Rajpurohit, Kalyan,
  Narasimhan, and Deshpande}]{aggarwal2023geo}
Aggarwal P, Murahari V, Rajpurohit T, et~al (2024) {GEO}: Generative engine
  optimization. In: Proceedings of the 30th ACM SIGKDD Conference on Knowledge
  Discovery and Data Mining (KDD '24). ACM, pp 5--16,
  \doi{10.1145/3637528.3671900}, preprint: arXiv:2311.09735

\bibitem[{At{\i}l et~al.(2025)At{\i}l, Aykent, Chittams, Fu, Passonneau,
  Radcliffe, Rajagopal, Sloan, Tudrej, T{\"u}re, Wu, Xu, and
  Baldwin}]{atil2025nondeterminism}
At{\i}l B, Aykent S, Chittams A, et~al (2025) Non-determinism of
  ``deterministic'' {LLM} system settings in hosted environments. In:
  Proceedings of the 5th Workshop on Evaluation and Comparison of NLP Systems
  (Eval4NLP). Association for Computational Linguistics, pp 135--148,
  \doi{10.18653/v1/2025.eval4nlp-1.12}, preprint: arXiv:2408.04667,
  ``Non-Determinism of ``Deterministic'' LLM Settings''

\bibitem[{Autor et~al.(2020)Autor, Dorn, Katz, Patterson, and
  Van~Reenen}]{autor2020superstar}
Autor D, Dorn D, Katz LF, et~al (2020) The fall of the labor share and the rise
  of superstar firms. The Quarterly Journal of Economics 135(2):645--709.
  \doi{10.1093/qje/qjaa004}

\bibitem[{Betello et~al.(2026)Betello, Missault, Czinczoll, and
  Beliuha}]{betello2026passport}
Betello F, Missault P, Czinczoll T, et~al (2026) Your {LLM} has a passport:
  Investigating brand origin geographic bias in e-commerce rankings. In:
  Proceedings of the 2026 ACM Conference on Fairness, Accountability, and
  Transparency (FAccT '26). ACM, pp 8089--8106, \doi{10.1145/3805689.3806438}

\bibitem[{Binet and Field(2013)}]{binet2013longshort}
Binet L, Field P (2013) The Long and the Short of It: Balancing Short and
  Long-Term Marketing Strategies. Institute of Practitioners in Advertising,
  London

\bibitem[{Bito et~al.(2025)Bito, Ren, and He}]{bito2025positionbias}
Bito E, Ren Y, He E (2025) Evaluating position bias in large language model
  recommendations.
  {\href{https://arxiv.org/abs/2508.02020}{{arXiv:2508.02020}}}

\bibitem[{Celma(2010)}]{celma2010longtail}
Celma {\`O} (2010) Music Recommendation and Discovery: The Long Tail, Long
  Fail, and Long Play in the Digital Music Space. Springer, Berlin, Heidelberg,
  \doi{10.1007/978-3-642-13287-2}

\bibitem[{Clauset et~al.(2009)Clauset, Shalizi, and
  Newman}]{clauset2009powerlaw}
Clauset A, Shalizi CR, Newman MEJ (2009) Power-law distributions in empirical
  data. SIAM Review 51(4):661--703. \doi{10.1137/070710111}

\bibitem[{Dodge et~al.(2021)Dodge, Sap, Marasovi{\'c}, Agnew, Ilharco,
  Groeneveld, Mitchell, and Gardner}]{dodge2021documenting}
Dodge J, Sap M, Marasovi{\'c} A, et~al (2021) Documenting large webtext
  corpora: A case study on the {Colossal Clean Crawled Corpus}. In: Proceedings
  of the 2021 Conference on Empirical Methods in Natural Language Processing
  (EMNLP). Association for Computational Linguistics, pp 1286--1305,
  \doi{10.18653/v1/2021.emnlp-main.98}

\bibitem[{Dubois et~al.(2025)Dubois, Dawson, and Jaiswal}]{shin2025hbr}
Dubois D, Dawson J, Jaiswal A (2025) Forget what you know about search.
  optimize your brand for {LLMs}. Harvard Business Review
  \url{https://hbr.org/2025/06/forget-what-you-know-about-seo-heres-how-to-optimize-your-brand-for-llms}

\bibitem[{Fombrun and Shanley(1990)}]{fombrun1990reputation}
Fombrun C, Shanley M (1990) What's in a name? reputation building and corporate
  strategy. Academy of Management Journal 33(2):233--258. \doi{10.2307/256324}

\bibitem[{Hannak et~al.(2013)Hannak, Sapiezynski, Molavi~Kakhki, Krishnamurthy,
  Lazer, Mislove, and Wilson}]{hannak2013personalization}
Hannak A, Sapiezynski P, Molavi~Kakhki A, et~al (2013) Measuring
  personalization of web search. In: Proceedings of the 22nd International
  Conference on World Wide Web (WWW '13). ACM, pp 527--538,
  \doi{10.1145/2488388.2488435}

\bibitem[{Hauser and Wernerfelt(1990)}]{hauser1990consideration}
Hauser JR, Wernerfelt B (1990) An evaluation cost model of consideration sets.
  Journal of Consumer Research 16(4):393--408. \doi{10.1086/209225}

\bibitem[{Howard and Sheth(1969)}]{howard1969buyer}
Howard JA, Sheth JN (1969) The Theory of Buyer Behavior. John Wiley \& Sons,
  New York

\bibitem[{Ja{\'z}wi{\'n}ska and Chandrasekar(2025)}]{jazwinska2025citation}
Ja{\'z}wi{\'n}ska K, Chandrasekar A (2025) {AI} search has a citation problem.
  Columbia Journalism Review, Tow Center for Digital Journalism, 6 March 2025.
  \url{https://www.cjr.org/tow_center/we-compared-eight-ai-search-engines-theyre-all-bad-at-citing-news.php}

\bibitem[{Jones(1990)}]{jones1990adspending}
Jones JP (1990) Ad spending: Maintaining market share. Harvard Business Review
  \url{https://hbr.org/1990/01/ad-spending-maintaining-market-share}

\bibitem[{Kamruzzaman et~al.(2024)Kamruzzaman, Nguyen, and
  Kim}]{kamruzzaman2024brandbias}
Kamruzzaman M, Nguyen HM, Kim GL (2024) ``{Global} is good, local is bad?'':
  Understanding brand bias in {LLMs}. In: Proceedings of the 2024 Conference on
  Empirical Methods in Natural Language Processing. Association for
  Computational Linguistics, pp 12695--12702,
  \doi{10.18653/v1/2024.emnlp-main.707}

\bibitem[{Kandpal et~al.(2023)Kandpal, Deng, Roberts, Wallace, and
  Raffel}]{kandpal2022large}
Kandpal N, Deng H, Roberts A, et~al (2023) Large language models struggle to
  learn long-tail knowledge. In: Proceedings of the 40th International
  Conference on Machine Learning (ICML), Proceedings of Machine Learning
  Research, vol 202. PMLR, pp 15696--15707

\bibitem[{Kliman-Silver et~al.(2015)Kliman-Silver, Hannak, Lazer, Wilson, and
  Mislove}]{klimansilver2015location}
Kliman-Silver C, Hannak A, Lazer D, et~al (2015) Location, location, location:
  The impact of geolocation on web search personalization. In: Proceedings of
  the 2015 Internet Measurement Conference (IMC '15). ACM, pp 121--127,
  \doi{10.1145/2815675.2815714}

\bibitem[{Kumar(2026)}]{kumar2026geoscale}
Kumar P (2026) Generative engine optimization at scale: Measuring brand
  visibility across {AI} search engines.
  {\href{https://arxiv.org/abs/2606.20065}{{arXiv:2606.20065}}}

\bibitem[{Lichtenberg et~al.(2024)Lichtenberg, Buchholz, and
  Schw{\"o}bel}]{lichtenberg2024popularity}
Lichtenberg JM, Buchholz A, Schw{\"o}bel P (2024) Large language models as
  recommender systems: A study of popularity bias. Accepted at the Gen-IR@SIGIR
  2024 workshop, {\href{https://arxiv.org/abs/2406.01285}{{arXiv:2406.01285}}}

\bibitem[{Ma et~al.(2026)Ma, Cheng, Zheng, Zhu, and Xiong}]{ma2026notstable}
Ma T, Cheng Y, Zheng Z, et~al (2026) Large language models are not stable
  recommender systems: A position bias perspective. In: Knowledge Science,
  Engineering and Management (KSEM 2025), Lecture Notes in Computer Science,
  vol 15919. Springer Nature Singapore, p 415--429,
  \doi{10.1007/978-981-95-3001-4_31}, preprint: arXiv:2312.15746 (2023), Ma,
  Cheng, Zhu and Xiong, ``Large Language Models are Not Stable Recommender
  Systems''

\bibitem[{Mallen et~al.(2023)Mallen, Asai, Zhong, Das, Khashabi, and
  Hajishirzi}]{mallen2023trust}
Mallen A, Asai A, Zhong V, et~al (2023) When not to trust language models:
  Investigating effectiveness of parametric and non-parametric memories. In:
  Proceedings of the 61st Annual Meeting of the Association for Computational
  Linguistics (Volume 1: Long Papers). Association for Computational
  Linguistics, pp 9802--9822, \doi{10.18653/v1/2023.acl-long.546}

\bibitem[{McCombs and Shaw(1972)}]{mccombs1972agenda}
McCombs ME, Shaw DL (1972) The agenda-setting function of mass media. Public
  Opinion Quarterly 36(2):176--187. \doi{10.1086/267990}

\bibitem[{Napoli(2015)}]{napoli2015socialmedia}
Napoli PM (2015) Social media and the public interest: Governance of news
  platforms in the realm of individual and algorithmic gatekeepers.
  Telecommunications Policy 39(9):751--760. \doi{10.1016/j.telpol.2014.12.003}

\bibitem[{Nechushtai and Lewis(2019)}]{nechushtai2019gatekeepers}
Nechushtai E, Lewis SC (2019) What kind of news gatekeepers do we want machines
  to be? {Filter} bubbles, fragmentation, and the normative dimensions of
  algorithmic recommendations. Computers in Human Behavior 90:298--307.
  \doi{10.1016/j.chb.2018.07.043}

\bibitem[{Newman(2005)}]{newman2005power}
Newman MEJ (2005) Power laws, {Pareto} distributions and {Zipf}'s law.
  Contemporary Physics 46(5):323--351. \doi{10.1080/00107510500052444}

\bibitem[{Ouyang et~al.(2022)Ouyang, Wu, Jiang, Almeida, Wainwright, Mishkin,
  Zhang, Agarwal, Slama, Ray, Schulman, Hilton, Kelton, Miller, Simens, Askell,
  Welinder, Christiano, Leike, and Lowe}]{ouyang2022training}
Ouyang L, Wu J, Jiang X, et~al (2022) Training language models to follow
  instructions with human feedback. In: Advances in Neural Information
  Processing Systems, pp 27730--27744

\bibitem[{Petroni et~al.(2019)Petroni, Rockt{\"a}schel, Riedel, Lewis, Bakhtin,
  Wu, and Miller}]{petroni2019language}
Petroni F, Rockt{\"a}schel T, Riedel S, et~al (2019) Language models as
  knowledge bases? In: Proceedings of the 2019 Conference on Empirical Methods
  in Natural Language Processing and the 9th International Joint Conference on
  Natural Language Processing (EMNLP-IJCNLP). Association for Computational
  Linguistics, pp 2463--2473, \doi{10.18653/v1/D19-1250}

\bibitem[{Rhoades(1993)}]{rhoades1993herfindahl}
Rhoades SA (1993) The {Herfindahl}-{Hirschman} index. Federal Reserve Bulletin
  79(3):188--189

\bibitem[{Roberts and Lattin(1991)}]{roberts1991consideration}
Roberts JH, Lattin JM (1991) Development and testing of a model of
  consideration set composition. Journal of Marketing Research 28(4):429--440.
  \doi{10.1177/002224379102800405}

\bibitem[{Romaniuk and Sharp(2004)}]{romaniuk2004salience}
Romaniuk J, Sharp B (2004) Conceptualizing and measuring brand salience.
  Marketing Theory 4(4):327--342. \doi{10.1177/1470593104047643}

\bibitem[{Romaniuk and Sharp(2016)}]{romaniuk2016hbg2}
Romaniuk J, Sharp B (2016) How Brands Grow: Part 2. Emerging Markets, Services,
  Durables, New and Luxury Brands. Oxford University Press, Melbourne

\bibitem[{Schulte et~al.(2026)Schulte, Bleeker, and
  Kaufmann}]{schulte2026dontmeasureonce}
Schulte J, Bleeker M, Kaufmann P (2026) Don't measure once: Measuring
  visibility in {AI} search ({GEO}).
  {\href{https://arxiv.org/abs/2604.07585}{{arXiv:2604.07585}}}

\bibitem[{Xu et~al.(2025)Xu, Potka, and Thomo}]{xu2025genderrace}
Xu K, Potka S, Thomo A (2025) Gender and race bias in consumer product
  recommendations by large language models. In: Advanced Information Networking
  and Applications (AINA 2025), Lecture Notes on Data Engineering and
  Communications Technologies, vol 246. Springer Nature Switzerland, p
  245--258, \doi{10.1007/978-3-031-87766-7_22}

\bibitem[{\.Zatuchin(2026)}]{zatuchin2026a}
\.Zatuchin D (2026) Gender bias in large language model brand recommendations:
  A three-study analysis of prompt-induced disparities across seasonal and
  recipient contexts. Research Square preprint, version 1,
  \doi{10.21203/rs.3.rs-8883056/v1}

\end{thebibliography}


\begin{thebibliography}{1}
\providecommand{\natexlab}[1]{#1}
\providecommand{\url}[1]{{#1}}
\providecommand{\urlprefix}{URL }
\providecommand{\doi}[1]{\url{https://doi.org/#1}}
\providecommand{\eprint}[2][]{\url{#2}}
 \bibcommenthead

\bibitem[{Rhoades(1993)}]{rhoades1993herfindahl}
Rhoades SA (1993) The {Herfindahl}-{Hirschman} index. Federal Reserve Bulletin
  79(3):188--189

\end{thebibliography}

\end{document}


\title[Supplementary material: AI Category Ownership Map]{Supplementary material for: Who Owns the AI Recommendation? A Multi-Industry Empirical Map of Brand Category Ownership Across Large Language Models}

\author*[1,2]{\fnm{Dmitrij} \sur{\.Zatuchin}}\email{dmitrij.zatuchin@eek.ee}

\affil*[1]{\orgdiv{Department of Information Technologies},
  \orgname{Estonian Entrepreneurship University of Applied Sciences (EUAS)},
  \orgaddress{\city{Tallinn}, \country{Estonia}}}
\affil[2]{\orgname{Rankfor.AI},
  \orgaddress{\city{Tallinn}, \country{Estonia}}}

\abstract{This supplement gives the guarded matcher and its adjudications (S1), statistical procedures (S2), collection configuration (S3), the size benchmark and inclusion heatmap (S4), the correction record and agreement decomposition (S5), every query-type concentration estimate and interval (S6), stability and scoring conventions (S7), industry regressions (S8), co-mention checks (S9), open-vocabulary readings (S10), the submitted hypotheses and clustering re-run (S11), and the actual September replication and cross-wave results (S12). Results use the conditional estimand unless labelled fixed-15. Author checks remain pending. Published version 2.0 contains the February results and both raw collections; the September outputs require a subsequent deposit version.}

\keywords{Large language models, Consideration sets, Brand inclusion, Recommendation concentration, Cross-system agreement}

\maketitle

\section{The guarded alias matcher}\label{ssec:matcher}

Section 3.5 of the main text states the matcher's principle, its gold figures and the rule on the ``AWS'' token; this section gives the rule set and the alias classes. The 77 aliases fall in three classes.

\begin{enumerate}
  \item \emph{Word boundary.} Every alias matches only at a word boundary; hyphens, punctuation and markup count as boundaries, so ``EY-Parthenon'' matches and ``they'' does not.
  \item \emph{Proper names.} Aliases that are proper names and nothing else match case-insensitively.
  \item \emph{Ordinary words.} Twenty-two aliases that are ordinary words match only in capitalised or all-caps form: allegro, asana, bain, block, bolt food, cash app, click up, epic, h and m, implement, kry, livi, miro, monday, notion, philips, plaid, square, stripe, temu, vinted, wise. ``ey'' and ``cgm'' match only as the all-caps token, and ``Ernst \& Young'' also counts as EY.
\end{enumerate}

Eight aliases carry a context rule that accepts or rejects a hit from the surrounding text. Block and Square need a fintech query or a nearby payments term; Implement needs ``Implement Consulting'' or a consulting query; Monday needs ``.com'' or a product term outside day-of-week contexts; CGM needs CompuGroup or an electronic-health-record term and no glucose term; ``H\&R Block'', ``Epic Games'', ``Wise choice'' and ``Bain Capital'' are rejected. Overlapping hits of one brand collapse to the longest alias, a brand is recorded at most once per answer (as in the submitted code), and the matcher is never run on an empty answer.

The change removes about a quarter of the recorded mentions: 14,005 (answer, brand) presence rows under substring matching, 10,635 under word boundaries alone and 10,463 under the guarded matcher, of which 9,831 are own-industry rows that enter the inclusion statistics and 632 are guarded off-industry presence rows (Shopify in SaaS answers, Salesforce in consulting and health technology answers, Amazon in SaaS answers, Epic in consulting answers). The largest movements, in answers hit, are EY (2,912 to 437; 2,196 of the substring hits were in other industries' answers), Implement Consulting (735 to 13), Block (416 to 249), Revolut (241 to 202, inside ``revolution''), Wise (231 to 199, inside ``otherwise'') and CompuGroup Medical (34 to 13, the ``CGM'' alias having matched glucose monitors).

\emph{Gold sample.} Seed 20260910 selected 300 precision hits (120 characters of context, stratified by aliases and brands) and 100 whole answers for recall, stratified by industry and system. Two language models independently applied one rubric: a precision hit refers to the sampled company, product, service or platform; a recall label lists every sampled brand named. Both labelled all 300 precision hits true (Wilson 95\% interval 98.7 to 100.0 per cent), so precision kappa is undefined. Recall is 278 of 282 brand-answer pairs (98.6 per cent; Wilson 96.4 to 99.4), with four misses and no spurious brand in the recall sample; eight recall answers name no sampled brand. The rubric treats AWS as an Amazon service. Two first-round rows initially omitted it; those original labels remain recorded alongside the corrected labels and the independent second adjudication. All four misses name AWS without Amazon. No AWS alias is added: the frozen sampled retail-brand matcher does not equate infrastructure mentions with retail recommendations. The authors' stratified 78-row hand check (70 precision, 8 recall) is pending. Wave-2 precision uses a fresh 100-item sample (seed 20260911) with a 25-row author check: one language model labelled 100/100 TRUE (Wilson 95\% interval 96.3 to 100.0 per cent); all 25 author labels remain blank. With no FALSE or UNSURE item, the author worksheet is a uniform draw without replacement of 25 precision items using seed 20260911; this worksheet-selection seed was chosen at preparation because the plan did not specify it.

\emph{Limits and adversarial cases.} The matcher applies string and local-context rules. Constructed negatives expose ordinary-word, calendar, imperative, domain-only and acronym errors. Markdown emphasis can hide a guard's neighbouring words: ``H\&R \texttt{**Block**}'' bypasses the rejection of ``H\&R Block''. Among 2,850 accepted wave-1 hits of the eight guarded aliases, 458 carry emphasis markers; stripping those markers reverses no decision. These are hit counts, separate from the probability sample. Twenty released adversarial sentences produce 6 passes and 14 failures; four explicitly constructed emphasis bypasses demonstrate a limitation even though the observed-corpus check found no reversal. An ambiguous early CGM hit can also change first-mention rank. These results neither validate universal precision nor establish semantic entity linking. The same frozen matcher is used in both waves.

\emph{The ``Bolt Food'' alias.} Bolt Food is the delivery app of the group whose grocery arm is the sampled Bolt Market. The alias contributes 12 of the brand's 27 presence rows; with it Bolt Market's RS is 0.036 on all queries (rank 9) and without it 0.020 (rank 10); the e-commerce Gini moves from 0.421 to 0.433 on all queries and from 0.457 to 0.474 on the brand-free set; the vacuum count is 7 either way. The alias is kept because the sampled brand is the Bolt offer as consumers meet it.

\section{Statistical procedures in detail}\label{ssec:procedures}

\emph{Resampling.} Seed 20260910 and 2,000 draws are used for bootstraps and permutations. Query resampling is stratified by industry; the mean over industries resamples each industry independently. BCa bias correction gives ties half weight, with query jackknife acceleration. A clipped bias correction, zero-width BCa interval despite varying replicates, or any non-finite replicate triggers a full-distribution percentile fallback with its reason recorded. Infinite ratios caused by zero denominators stay in the distribution; they are never discarded to force finite intervals. Rates carry Wilson 95\% intervals. Every query-type estimate and its interval is reported even when the interval does not bracket its estimate (Section~\ref{ssec:intervals}); the 5 to 15 queries per type warrant caution. Conditional Gini's retained percentile column uses linear interpolation; new fallback and fixed-15, HHI and top-three percentile columns use the inverted empirical CDF. Thus a fallback need not equal the retained Gini percentile column.

\emph{Verdict rule.} An interval entirely on the predicted side of its reference is supported, entirely on the other side is not supported, and an interval that includes the reference is inconclusive. The five submitted hypotheses are evaluated under that rule, pooled and per industry, with their original cut-offs (H1, Gini above 0.60; H2, an asymmetric-to-co-mentioned records ratio above 3:1 at $|\mathrm{CMA}| > 0.10$; H3, vacuums in more than 20\% of queries; H4, more than 15\% emergent clusters, with a range over ten clustering seeds in place of an interval (Section~\ref{ssec:hypotheses}); H5, full agreement below 50\%). E1 is descriptive. E2 and E5 follow the frozen analysis plan with six deviations, as stated in Section 2.5 of the main text; missing or non-finite required evidence gives an inconclusive (evidence unavailable) verdict.

\emph{Permutation null for CMA.} Each pooled CMA pair carries a two-sided permutation $p$ under a null stratified within system: brand $A$'s presence is shuffled within each system's non-empty answers to a query, so every system keeps its own count of $A$ and only the within-system pairing of $A$ and $B$ is broken. The exchangeable null that shuffles across all non-empty answers, which mixes systems that differ in list length and CIR, is kept in the deposit for comparison; under it 150 pairs are significant (7 asymmetric, 143 co-mentioned) against 53 (0 and 53) under the stratified null. The 450 ordered pairs form one dependent exploratory family. Missing-convention counts are 53 nominal, 26 Benjamini-Hochberg, 0 Benjamini-Yekutieli and 0 Holm; fixed-15 counts are 84, 50, 34 and 0. With 2,000 permutations the minimum attainable p-value, 1/2001, exceeds Holm's first threshold, 0.05/450. These adjustments are distinct from the within-query E2 family and its split-half requirement.

\emph{Within-query null.} For each query, curveball trades preserve every non-empty answer's sampled-brand count and every brand's query count. Unchanged trades are retained; row and column sums are checked. The primary design includes all 250 queries: 130 have fixed matrices, including 90 with defined CMA records (1,575 of the 6,516 defined records). The remaining 120 queries show 2 to 2,000 distinct matrices in 2,000 sampled draws; these counts do not enumerate every reachable matrix. Queries whose CMA is undefined contribute no records; they stay in the design. The frozen companion includes 200 non-comparison queries, six of which name individual sampled brands; the other 194 include Big 4, SAP and Slack prompts. Filtering precedes every RS, CMA and null calculation. Each query contributes 2,000 draws, with 20 trades per answer. A 5/20/60-trade check retains the mean, ratio and rho band positions in both conventions and sets; this is a stability diagnostic, without a claim of complete mixing. Restricting to the 120 movable queries is a labelled sensitivity (Section~\ref{ssec:cma}).

\emph{Secondary list-length control.} Answers are pooled by industry and by sampled-brand count (1 to 2, 3 to 4, 5 or more); zero-brand answers lie outside these strata, which makes their inputs identical under the two conventions. One hundred curveball matrices per stratum preserve answer lengths and brand counts. This secondary pooled analysis does not replace the within-query E2 test or isolate length as the explanation for the primary co-mention pattern.

\begin{figure}[t]
\centering
\includegraphics[width=0.6\textwidth]{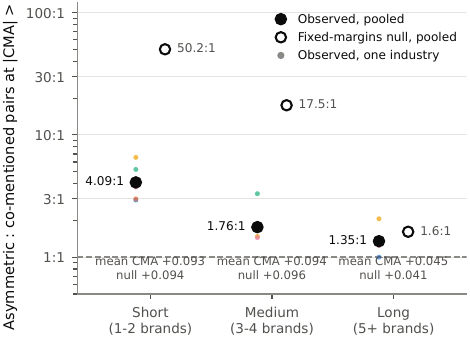}
\caption{List-length control for CMA. Filled dots: observed ratio of asymmetric to co-mentioned ordered pairs at $|\mathrm{CMA}| > 0.10$ within answers naming one to two, three to four, and five or more sampled brands (pooled over industries: 1,807, 919 and 567 answers); hollow dots: the same ratio under a fixed-margins null that keeps every answer's list length and every brand's count (100 matrices per stratum); small coloured dots: the observed ratio per industry. Text below each stratum: observed mean CMA against the null mean.}
\label{fig:s-listlength}
\end{figure}

\emph{Query-type regression.} Query type and system enter a binomial regression with logit link on the answer $\times$ brand rows of each industry and pooled: the probability that a brand is named as a function of brand fixed effects, query type and system, with reference levels the industry's RS leader, direct queries and GPT-5.2, standard errors clustered by query (50 clusters per industry, 250 pooled), rows of empty answers dropped (36,140 pooled rows). The pooled model carries the estimates; per-industry rows are shown for direction (Section~\ref{ssec:glm}), and a term identified from fewer than ten clusters (every per-industry trust term rests on five) is flagged because the clustered sandwich variance is downward biased with so few clusters.

\emph{Reliability.} Within a wave the iterations split into $\{1, 3, 5\}$ (up to nine non-empty answers per query) and $\{2, 4\}$ (up to six): Pearson and Spearman correlation of RS over brands, the Gini of each half, Cohen's kappa of the vacuum classification, the Jaccard similarity of the brand sets a system names across its iterations (500 possible iteration pairs per industry-system group; a pair involving an empty answer is dropped), and the sign agreement of pooled CMA over the pairs defined and non-zero in both halves. Brand-set sensitivity removes each of the ten brands in turn and the RS leader.

\emph{Criterion validity and size comparison.} RS is correlated with the sampled-brand size benchmark, using a percentile bootstrap over brands, per industry and pooled on within-industry ranks over the 47 brands with a comparable figure. Undefined brand-resampled Spearman coefficients are excluded with counts recorded; e-commerce loses 2 of 2,000 in each query set. The five industry orderings are descriptive; no population-level test is applied. Benchmark and recommendation Gini levels can be compared empirically, with their different inputs stated. Positive association is distinguished from zero in fintech and SaaS, while intervals cross zero in the other three industries. The heterogeneous audited, company, trade and estimated figures limit the criterion's interpretation.

\emph{Open-vocabulary reading.} For every vacuum query an adjudicating language model read all non-empty answers without any sample restriction and counted every brand named; the record per query holds the most-named brand and its count, whether it is in the sample, the best sampled brand and the number of distinct brands named, every share recomputed over the query's non-empty answers. The readings are model-produced; the status of the authors' check is recorded in Section 3.7 of the main text. One reading records five empty answers where the response index records four; the index is used.

\emph{Plan chronology.} The plan was frozen on 10 September 2026 at 01:15:54 UTC, after 950 answers had been collected and length and at-cap diagnostics inspected, before brand statistics. All deviations are dated that day: 1 and 2 at 02:14:49 UTC, 3 at 06:20:22, 4 at 06:35:41, 5 at 08:05:45 and 6 at 08:28:09. Deviation 5 replaces E2 rules after algebra and the earlier movable-query null exposed their weak discrimination; Deviation 6 restores fixed queries to the primary null and settles scope, missing evidence and inference. The final all-query reference is a subsequent recomputation. The original prefix and final whole-file hashes are printed in the main Data availability statement. These changes precede any real wave-2 brand statistic.

\emph{Cross-wave comparison.} The comparison follows the frozen analysis plan: correlation of RS between waves; the change in Gini, vacuum rate, high-maximum-inclusion rate and full agreement with a paired bootstrap over queries; the share of queries whose top brand is unchanged, overall and per system; the Jaccard similarity of each system's brand sets between waves; the empty-answer count, the at-cap share and the mean answer length per system and wave; every item reported on all 250 and the frozen 200 non-comparison queries, each with and without the 13 all-empty wave-1 queries, and, following Deviation 2 of the frozen analysis plan, with and without the answers that end at the 1,024-token cap.

\section{System configuration and the empty-answer record}\label{ssec:systems}

Table~\ref{tab:s-systems} gives the configuration detail behind Table 3 of the main text. Every call carried the system prompt ``You are a helpful assistant with broad knowledge of businesses and technology.'', temperature 0.3, an output cap of 1,024 tokens (\texttt{max\_completion\_tokens} for OpenAI, \texttt{max\_output\_tokens} for Gemini, \texttt{max\_tokens} for Perplexity), 3 s between calls and a checkpoint every 25 calls with resume by the key industry, query, system, iteration, in that call order. Wave 1 ran from 23 February 2026, 19:33 UTC, to 24 February 2026, 13:12 UTC, from a Google Colab notebook runtime whose region was not recorded (the notebook uses Colab paths and helpers; the raw file carries no host field), with Python 3.10 and unpinned clients, and stored the answer text only. Wave 2 ran in September 2026 from a workstation on one residential connection in northern Europe (macOS 15.5, Python 3.13.7, openai 2.21.0, google-genai 2.2.0) and also stores the identifier the API returns, the token counts and the seconds per call. The Gemini client changed from \texttt{google.generativeai} to \texttt{google.genai} between waves with the same identifier and settings. An API call carries no client location beyond its source address.

\begin{table}[tbp]
\caption{The three systems as called in both waves. Wave 1 recorded no returned identifier and no token usage. Empty answers are answers returned with an empty body and no error flag; at-cap answers are wave-2 answers whose completion token count equals the 1,024-token cap (visible plus thinking tokens for Gemini).}\label{tab:s-systems}
\centering
\footnotesize
\setlength{\tabcolsep}{2pt}
\begin{tabularx}{\textwidth}{@{}>{\raggedright\arraybackslash}p{1.6cm}>{\raggedright\arraybackslash}p{1.9cm}>{\raggedright\arraybackslash}p{1.9cm}>{\raggedright\arraybackslash}p{1.2cm}YYY@{}}
\toprule
\textbf{System} & \textbf{Identifier requested} & \textbf{Identifier returned (wave 2)} & \textbf{Search} & \textbf{Client, wave 1 / wave 2} & \textbf{Empty answers, wave 1 / wave 2} & \textbf{At cap, wave 2} \\
\midrule
GPT-5.2 (OpenAI) & \texttt{gpt-5.2} & \texttt{gpt-5.2-}\allowbreak\texttt{2025-12-11} & none & openai, unpinned / openai 2.21.0 & 136 / 0 & 10.08\% (126/1,250) \\
Gemini 3 Flash (Google) & \texttt{gemini-3-}\allowbreak\texttt{flash-preview} & alias only & none & google.\-generative\-ai / google.\-genai 2.2.0 & 0 / 0 & 0\% \\
Perplexity sonar-pro & \texttt{sonar-pro} & alias only & built in, every call & openai client with the Perplexity base URL / same, 2.21.0 & 0 / 0 & 0\% \\
\bottomrule
\end{tabularx}
\end{table}

\emph{Empty answers in wave 1.} The 136 empty GPT-5.2 answers (Section 3.4 of the main text) fall by query type on comparison 83, problem-solution 31, exploration 17, direct 5 and trust 0, and by industry on health technology 64, fintech 33, SaaS 22, consulting 14 and e-commerce 3. The 43 queries with at least one empty answer are consulting 6, e-commerce 1, fintech 11, health technology 18 and SaaS 7; the 13 on which all five GPT-5.2 answers are empty are fintech 18, 39, 44; health technology 15, 18, 21, 23, 24, 31, 35; SaaS 15, 20, 22. Wave 2 completed on 10 September 2026 at 11:46:30 UTC with 3,750 answers, 750 complete query-system cells, zero transport errors and zero empty bodies; returned identifiers are those in Table~\ref{tab:s-systems}. All 1,250 September OpenAI answers record zero reasoning tokens; February has no usage fields, so a between-wave reasoning-token comparison is unavailable.

\emph{Answers at the output cap.} Equality with the 1,024-token cap flags possible truncation; the matcher scores the visible text. Deviation 2 requires each flag and every comparison with and without flagged answers. Gemini's thinking tokens count against the cap, so its at-cap check uses visible plus thinking tokens. Token counts were not stored in wave 1. The final flags identify 126 OpenAI answers (10.08\%), including 71 comparison answers, and none from Gemini or Perplexity. Between-wave visible-length comparisons are reported in Section S12; absent February token counts cannot be reconstructed from text.

\section{The size benchmark and inclusion map}\label{ssec:benchmark}\label{ssec:omitted}

\begin{figure}[t]
\centering
\includegraphics[width=\textwidth]{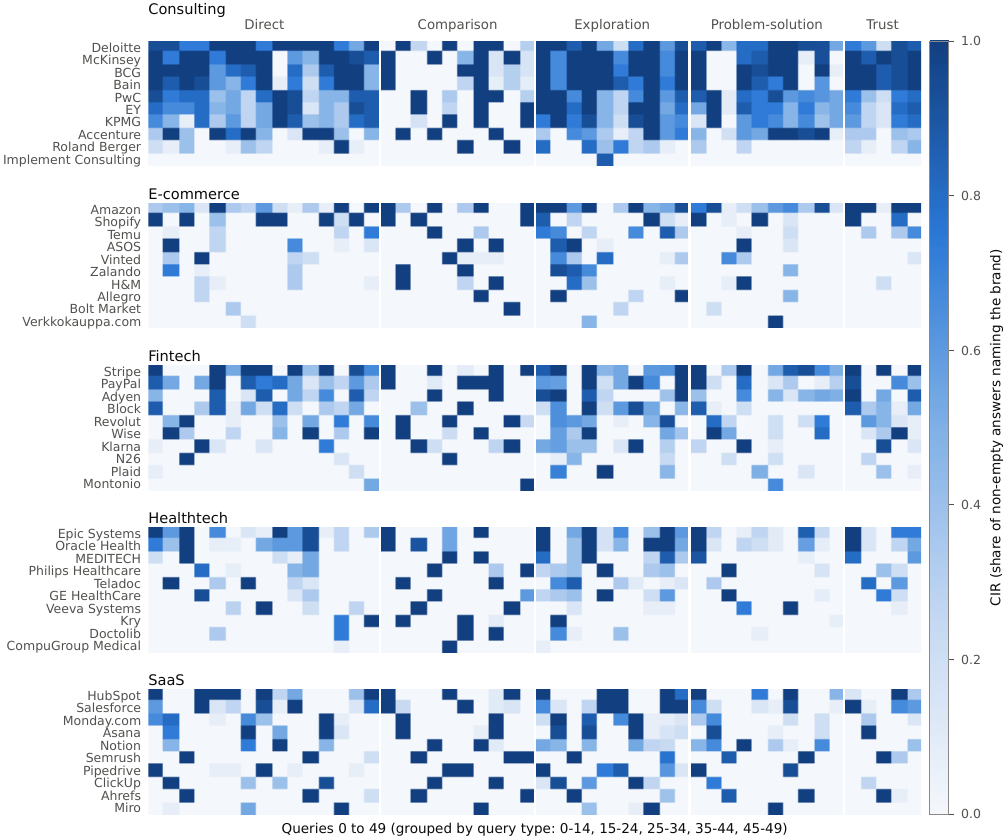}
\caption{Category Inclusion Rate by brand (rows, ordered by Recommendation Share) and query (columns, grouped by type); colour is the share of non-empty answers naming the brand.}
\label{fig:s-cir}
\end{figure}

Table~\ref{tab:s-brands} gives the size columns moved from the main brand table. Sources and basis flags are recorded brand by brand in the deposit; 29 figures are audited filings, and three brands lack a comparable figure. Figure~\ref{fig:s-cir} maps all 2,500 brand-query cells.

\begin{table}[p]
\caption{Brand sample with the external size marker (best comparable annual figure, USD billion, ECB 2025 average rates; Block net of gross bitcoin revenue, Amazon excluding AWS, Philips its health segments), its quality flag and headquarters countries.}\label{tab:s-brands}
\centering
\fontsize{7}{7.5}\selectfont
\setlength{\tabcolsep}{4pt}
\begin{tabular}{@{}lrllc@{}}
\toprule
\textbf{Brand} & \textbf{Size (USD bn)} & \textbf{Basis} & \textbf{Year} & \textbf{Quality} \\
\midrule
\multicolumn{5}{@{}l}{\textit{SaaS} (HQ: US, IL, SG, EE)} \\
Salesforce & 41.53 & revenue & FY2026 & audited \\
HubSpot & 3.13 & revenue & FY2025 & audited \\
Monday.com & 1.23 & revenue & FY2025 & audited \\
Asana & 0.79 & revenue & FY2026 & audited \\
Miro & 0.67 & ARR & 2025 & estimate \\
Notion & 0.50 & ARR & 2025 & estimate \\
Semrush & 0.44 & revenue & FY2025 & audited \\
ClickUp & 0.30 & ARR & 2025 & estimate \\
Pipedrive & 0.27 & revenue & FY2025 & audited \\
Ahrefs & 0.15 & ARR & 2024 & estimate \\
\midrule
\multicolumn{5}{@{}l}{\textit{Consulting} (HQ: US, UK, IE, NL, DE, DK)} \\
Deloitte & 70.50 & revenue & FY2025 & trade source \\
Accenture & 69.67 & revenue & FY2025 & audited \\
PwC & 56.90 & revenue & FY2025 & trade source \\
EY & 53.20 & revenue & FY2025 & trade source \\
KPMG & 39.80 & revenue & FY2025 & trade source \\
McKinsey & 16.00 & revenue & 2023 & estimate \\
BCG & 14.40 & revenue & 2025 & trade source \\
Bain & 7.50 & revenue & 2024 & estimate \\
Roland Berger & 1.14 & revenue & 2025 & trade source \\
Implement Consulting & 0.44 & revenue & FY2025 & audited \\
\midrule
\multicolumn{5}{@{}l}{\textit{Fintech} (HQ: US, UK, NL, SE, DE, EE)} \\
PayPal & 33.17 & net revenue & FY2025 & audited \\
Block & 15.69 & revenue, net of bitcoin & FY2025 & audited \\
Stripe & 6.80 & revenue & 2025 & estimate \\
Revolut & 5.96 & revenue & FY2025 & audited \\
Klarna & 3.51 & revenue & FY2025 & audited \\
Adyen & 2.67 & net revenue & FY2025 & audited \\
Wise & 2.50 & net revenue & FY2026 & audited \\
N26 & 0.57 & revenue & FY2025 & trade source \\
Plaid & 0.55 & ARR & 2025 & estimate \\
Montonio & $<$0.01 & revenue & FY2025 & trade source \\
\midrule
\multicolumn{5}{@{}l}{\textit{E-commerce} (HQ: US, CA, DE, SE, UK, LT, PL, CN, EE, FI)} \\
Amazon & 588.20 & revenue, ex AWS & FY2025 & audited \\
H\&M & 23.31 & revenue & FY2025 & audited \\
Zalando & 13.95 & revenue & FY2025 & audited \\
Shopify & 11.56 & revenue & FY2025 & audited \\
ASOS & 3.27 & revenue & FY2025 & audited \\
Allegro & 3.05 & revenue & FY2025 & audited \\
Vinted & 1.27 & revenue & FY2025 & audited \\
Verkkokauppa.com & 0.59 & revenue & FY2025 & audited \\
Temu & not used & GMV only & 2025 & estimate \\
Bolt Market & not used & no figure & FY2025 & none \\
\midrule
\multicolumn{5}{@{}l}{\textit{Health technology} (HQ: US, DE, FR, SE, NL)} \\
GE HealthCare & 20.63 & revenue & FY2025 & audited \\
Philips Healthcare & 15.38 & segment revenue & FY2025 & audited \\
Epic Systems & 6.70 & revenue & 2025 & estimate \\
Veeva Systems & 3.20 & revenue & FY2026 & audited \\
Teladoc & 2.53 & revenue & FY2025 & audited \\
CompuGroup Medical & 1.37 & revenue & FY2025 & audited \\
MEDITECH & 0.49 & revenue & FY2019 & audited \\
Doctolib & 0.39 & ARR & FY2024 & trade source \\
Kry & 0.25 & revenue & FY2024 & audited \\
Oracle Health & not used & percentage only & 2025 & trade source \\
\bottomrule
\end{tabular}
\end{table}

The sample of Table 1 in the main text is purposive. A frame drawn on the size marker would have added, among others: Atlassian, Microsoft Dynamics 365 and Zoho (SaaS); Capgemini, IBM Consulting and Oliver Wyman (consulting); Fiserv, Global Payments with Worldpay, and Nexi (fintech); eBay, Alibaba and Otto Group (e-commerce); Siemens Healthineers, Medtronic and Hims \& Hers (health technology). Nine candidates reach the top three when inserted separately alongside the sampled brands with measured figures; six do when all candidates enter jointly. These are ranks in the heterogeneous benchmark, without a population-ranking claim. The size benchmark of the sampled brands itself is documented brand by brand in the deposit, with the basis of each figure and its quality flag. Where a brand's as-filed figure and the figure used differ (Amazon net of AWS, Block net of gross bitcoin revenue, Philips's health segments), the as-filed figure is given as a sensitivity in the deposit's benchmark summary; the benchmark concentration statistics (Gini and HHI, \citealt{rhoades1993herfindahl}) are computed on the figures used. On those figures the within-sample size Gini and top-three share are 0.45 and 59.8 per cent (consulting, 10 brands), 0.83 and 96.9 (e-commerce, 8), 0.63 and 77.9 (fintech, 10), 0.61 and 83.8 (health technology, 9) and 0.81 and 93.6 (SaaS, 10).

\section{Correction record and agreement decomposition}\label{ssec:corrections}

The submitted method described multiple mentions and an index exceeding one, but its code counted each brand at most once per answer. The deposited inclusion-rate file instead used a one-system denominator of five: its maximum was 3.0, with 800 of 2,500 cells above one, inconsistent with its share and vacuum files. The per-industry Gini and agreement entries of submitted Tables 2 and 5 do not reproduce from that deposit, although the overall means do. Undefined conditional probabilities were replaced by zero in submitted Table 3, creating positive asymmetry mechanically. On the corrected mentions this convention still gives 4,738 asymmetric and 3,174 co-mentioned records at the 0.10 analysis threshold (ratio 1.49:1); keeping undefined cells undefined gives 0.28:1.

Substring aliases produced 14,005 presence rows, including EY inside ordinary words; the guarded matcher gives 10,463 (9,831 own-industry). EY falls from 2,912 answers to 437 and from RS 0.955 to 0.588, rank six in consulting. Its corrected top three is Deloitte, McKinsey and BCG; other industry leader sets remain unchanged. Of 157 classified submitted quantities, 12 flip, 13 otherwise change and 132 remain unchanged; the CSV also contains 12 unclassified trace rows, which are outside that denominator.

The query claim also changes: 49 comparison and six other prompts name individual sampled brands; two other prompts name only brand groups (Big 4 versus MBB, and Big 4). The frozen 200-query companion excludes comparisons but retains six prompts naming individual sampled brands; the other 194 include Big 4, SAP and Slack prompts. No comparison query was a vacuum empirically in wave 1. Finally, the 136 empty GPT-5.2 answers were scored as answers naming nobody. Treating them as missing changes the mean Gini from 0.302 to 0.303, vacuums from 21 to 19 and high maximum inclusion from 190 to 204 queries. Full agreement is 143/250 under both conventions and 143/237 under both when restricted to queries each system answered.

Table~\ref{tab:s-decomposition} reports both sequential orders and Shapley contributions. R restricts top brands to the query's industry; T uses lowest mean first-mention ordinal rank, then original brand-list order; M installs the guarded matcher; S counts no-top cases as disagreement and keeps every query in the denominator. The historical intermediates have varying denominators. T ranks mentions among all matched brands: restricting that intermediate rank to the industry's brands would give 150/243, not the reported 152/243. The final result remains 143/250. All 16 subsets of R, T, M and S determine the Shapley values over 24 orders; contributions sum to 15.6 percentage points and describe a computational correction.

\begin{table}[tbp]
\caption{Agreement decomposition; increments and Shapley values are percentage points.}\label{tab:s-decomposition}
\centering\footnotesize
\begin{tabular}{@{}llrrr@{}}
\toprule Order & Changes applied & Count / denominator & Agreement (\%) & Increment \\
\midrule
RTMS & none & 104/250 & 41.6000 & baseline \\
RTMS & R & 147/243 & 60.4938 & +18.8938 \\
RTMS & RT & 152/243 & 62.5514 & +2.0576 \\
RTMS & RTM & 156/239 & 65.2720 & +2.7205 \\
RTMS & RTMS & 143/250 & 57.2000 & -8.0720 \\
MRTS & none & 104/250 & 41.6000 & baseline \\
MRTS & M & 145/240 & 60.4167 & +18.8167 \\
MRTS & RM & 153/239 & 64.0167 & +3.6001 \\
MRTS & RTM & 156/239 & 65.2720 & +1.2552 \\
MRTS & RTMS & 143/250 & 57.2000 & -8.0720 \\
\midrule
\multicolumn{2}{l}{Shapley R / T / M / S} & \multicolumn{3}{l}{+9.6324 / +2.4177 / +9.2008 / $-5.6509$} \\
\bottomrule
\end{tabular}
\end{table}

\section{All query-type concentration estimates and intervals}\label{ssec:intervals}

Table~\ref{tab:s-intervals} retains every industry and mean query-type Gini, HHI, top-three share and fixed-15 Gini estimate. G is conditional Gini, H is normalised HHI, T is top-three share, and F is fixed-15 Gini. CI is BCa except P (percentile fallback); B marks BCa, U an uncomputed interval. Percentile G uses the retained linear-interpolation column; other percentile columns and every fallback use the inverted empirical CDF. These are distinct quantile conventions. Mean HHI and top-three intervals were not computed. The deposited concentration CSV records bias correction, acceleration, tied-replicate share and each fallback reason.

Sixteen concentration metric intervals across five source rows fail to bracket their point estimate; all remain reported. Mean comparison Gini has a clipped BCa bias correction: point 0.230728, fallback [0.309599, 0.454236], retained linear-percentile interval [0.310347, 0.454245]. Its fixed-15 point is 0.238192 with fallback [0.315995, 0.458313]. No non-bracketing rule filters a published estimate. Elsewhere, the SaaS pair ratio at threshold 0.20 has one infinite bootstrap replicate out of 2,000 because its denominator is zero; the full-distribution fallback is [0, 0.333333]. The pooled fixed-15 pair ratio at 0.20 falls back to [0, 0.108108] because BCa has zero width while replicates vary. No reported fallback endpoint in this run is infinite.

\clearpage
{\scriptsize
\setlength{\tabcolsep}{3pt}
\begin{longtable}{@{}p{3.2cm}crp{2.6cm}p{2.6cm}c@{}}
\caption{All query-type concentration estimates: reported and percentile 95\% intervals.}\label{tab:s-intervals}\\
\toprule Industry / type & Metric & Point & CI & Percentile & Method \\
\midrule\endfirsthead
\multicolumn{6}{l}{Table \thetable\ continued}\\
\toprule Industry / type & Metric & Point & CI & Percentile & Method \\
\midrule\endhead
\bottomrule\endfoot
Consulting / direct & G & 0.2479 & [0.2240, 0.2619] & [0.2344, 0.2771] & B \\
Consulting / direct & H & 1219.4 & [1180.8, 1243.2] & [1197.6, 1272.4] & B \\
Consulting / direct & T & 0.4312 & [0.4122, 0.4488] & [0.4159, 0.4552] & B \\
Consulting / direct & F & 0.2479 & [0.2240, 0.2619] & [0.2342, 0.2771] & B \\
Consulting / comparison & G & 0.2190 & [0.1946, 0.2133] & [0.2289, 0.4655] & B \\
Consulting / comparison & H & 1174.1 & [1151.7, 1167.2] & [1188.3, 1716.9] & B \\
Consulting / comparison & T & 0.4219 & [0.3970, 0.4193] & [0.4245, 0.6321] & B \\
Consulting / comparison & F & 0.2212 & [0.1924, 0.2158] & [0.2288, 0.4722] & B \\
Consulting / exploration & G & 0.2064 & [0.1756, 0.2358] & [0.1805, 0.2392] & B \\
Consulting / exploration & H & 1156.7 & [1102.9, 1201.7] & [1114.7, 1224.8] & B \\
Consulting / exploration & T & 0.4068 & [0.3836, 0.4424] & [0.3838, 0.4426] & B \\
Consulting / exploration & F & 0.2064 & [0.1756, 0.2358] & [0.1804, 0.2392] & B \\
Consulting / problem solution & G & 0.2529 & [0.2251, 0.2788] & [0.2380, 0.3633] & B \\
Consulting / problem solution & H & 1251.2 & [1213.8, 1303.1] & [1225.1, 1417.8] & B \\
Consulting / problem solution & T & 0.4283 & [0.4053, 0.4669] & [0.4122, 0.5268] & B \\
Consulting / problem solution & F & 0.2513 & [0.2244, 0.2762] & [0.2363, 0.3632] & B \\
Consulting / trust & G & 0.3136 & [0.2454, 0.4342] & [0.2453, 0.4342] & B \\
Consulting / trust & H & 1313.4 & [1220.4, 1685.0] & [1215.5, 1628.7] & B \\
Consulting / trust & T & 0.5082 & [0.4303, 0.6594] & [0.4302, 0.6400] & B \\
Consulting / trust & F & 0.3136 & [0.2454, 0.4342] & [0.2430, 0.4342] & B \\
E-commerce / direct & G & 0.5379 & [0.4006, 0.6562] & [0.4382, 0.7231] & B \\
E-commerce / direct & H & 2177.7 & [1535.6, 3226.2] & [1640.0, 3706.8] & B \\
E-commerce / direct & T & 0.7203 & [0.5662, 0.8395] & [0.6039, 0.9068] & B \\
E-commerce / direct & F & 0.5379 & [0.4006, 0.6562] & [0.4372, 0.7230] & B \\
E-commerce / comparison & G & 0.3542 & [0.2482, 0.3825] & [0.3277, 0.6429] & B \\
E-commerce / comparison & H & 1440.8 & [1212.5, 1523.8] & [1362.8, 2686.2] & B \\
E-commerce / comparison & T & 0.5379 & [0.4334, 0.5741] & [0.5019, 0.8052] & B \\
E-commerce / comparison & F & 0.3443 & [0.2482, 0.3633] & [0.3270, 0.6352] & B \\
E-commerce / exploration & G & 0.3732 & [0.3103, 0.4271] & [0.3455, 0.6097] & B \\
E-commerce / exploration & H & 1554.9 & [1378.2, 1958.5] & [1411.8, 2442.8] & B \\
E-commerce / exploration & T & 0.5437 & [0.4873, 0.6101] & [0.5116, 0.7870] & B \\
E-commerce / exploration & F & 0.3732 & [0.3103, 0.4271] & [0.3454, 0.6097] & B \\
E-commerce / problem solution & G & 0.4833 & [0.3384, 0.5854] & [0.3982, 0.7778] & B \\
E-commerce / problem solution & H & 2124.6 & [1448.1, 3215.8] & [1601.6, 4503.3] & B \\
E-commerce / problem solution & T & 0.6510 & [0.5214, 0.7638] & [0.5746, 0.9778] & B \\
E-commerce / problem solution & F & 0.4833 & [0.3384, 0.5854] & [0.3980, 0.7778] & B \\
E-commerce / trust & G & 0.7513 & [0.6875, 0.7805] & [0.7156, 0.8041] & B \\
E-commerce / trust & H & 3808.4 & [3148.5, 4687.5] & [3290.5, 4837.6] & B \\
E-commerce / trust & T & 0.9558 & [0.8406, 1.0000] & [0.8816, 1.0000] & B \\
E-commerce / trust & F & 0.7513 & [0.6875, 0.7805] & [0.7156, 0.8041] & B \\
Fintech / direct & G & 0.3974 & [0.2938, 0.4736] & [0.3333, 0.5270] & B \\
Fintech / direct & H & 1498.1 & [1276.5, 1724.3] & [1353.3, 1935.1] & B \\
Fintech / direct & T & 0.5513 & [0.4626, 0.6240] & [0.4980, 0.6852] & B \\
Fintech / direct & F & 0.3974 & [0.2938, 0.4736] & [0.3333, 0.5270] & B \\
Fintech / comparison & G & 0.3398 & [0.2631, 0.3616] & [0.3242, 0.5898] & B \\
Fintech / comparison & H & 1360.6 & [1227.2, 1388.6] & [1341.5, 2271.0] & B \\
Fintech / comparison & T & 0.5309 & [0.4518, 0.5727] & [0.4949, 0.7521] & B \\
Fintech / comparison & F & 0.3410 & [0.2629, 0.3511] & [0.3287, 0.5900] & B \\
Fintech / exploration & G & 0.3192 & [0.2205, 0.4104] & [0.2488, 0.4602] & B \\
Fintech / exploration & H & 1318.0 & [1168.0, 1505.3] & [1215.5, 1673.6] & B \\
Fintech / exploration & T & 0.4915 & [0.4087, 0.5682] & [0.4328, 0.6243] & B \\
Fintech / exploration & F & 0.3180 & [0.2182, 0.4039] & [0.2471, 0.4595] & B \\
Fintech / problem solution & G & 0.3981 & [0.2470, 0.4754] & [0.3274, 0.5698] & B \\
Fintech / problem solution & H & 1568.1 & [1217.0, 1872.2] & [1353.3, 2229.3] & B \\
Fintech / problem solution & T & 0.5707 & [0.4380, 0.6397] & [0.5124, 0.7276] & B \\
Fintech / problem solution & F & 0.4000 & [0.2463, 0.4697] & [0.3369, 0.5682] & B \\
Fintech / trust & G & 0.3785 & [0.3045, 0.4709] & [0.3064, 0.5696] & B \\
Fintech / trust & H & 1438.6 & [1293.1, 1715.4] & [1306.0, 2225.2] & B \\
Fintech / trust & T & 0.5374 & [0.4706, 0.6316] & [0.4746, 0.7163] & B \\
Fintech / trust & F & 0.3785 & [0.3045, 0.4709] & [0.3064, 0.5694] & B \\
Health technology / direct & G & 0.3916 & [0.2626, 0.4582] & [0.3247, 0.5841] & B \\
Health technology / direct & H & 1594.3 & [1241.6, 1935.3] & [1341.0, 2283.1] & B \\
Health technology / direct & T & 0.5961 & [0.4569, 0.6912] & [0.5094, 0.7613] & B \\
Health technology / direct & F & 0.3940 & [0.2630, 0.4614] & [0.3260, 0.5853] & B \\
Health technology / comparison & G & 0.1089 & [0.0741, 0.0797] & [0.1725, 0.5376] & B \\
Health technology / comparison & H & 1044.5 & [1026.1, 1026.9] & [1101.7, 2020.4] & B \\
Health technology / comparison & T & 0.3664 & [0.3517, 0.3540] & [0.4102, 0.7006] & B \\
Health technology / comparison & F & 0.1322 & [0.1049, 0.1094] & [0.1736, 0.5432] & B \\
Health technology / exploration & G & 0.4203 & [0.2626, 0.5459] & [0.2977, 0.5875] & B \\
Health technology / exploration & H & 1558.5 & [1260.1, 2011.5] & [1298.8, 2252.7] & B \\
Health technology / exploration & T & 0.5895 & [0.4441, 0.7477] & [0.4656, 0.7833] & B \\
Health technology / exploration & F & 0.4236 & [0.2637, 0.5554] & [0.2946, 0.5905] & B \\
Health technology / problem solution & G & 0.4921 & [0.4026, 0.5592] & [0.4446, 0.7092] & B \\
Health technology / problem solution & H & 1783.2 & [1522.5, 2027.8] & [1641.4, 3353.5] & B \\
Health technology / problem solution & T & 0.6510 & [0.5195, 0.7744] & [0.5707, 0.9270] & B \\
Health technology / problem solution & F & 0.4926 & [0.3933, 0.5638] & [0.4394, 0.7124] & B \\
Health technology / trust & G & 0.5403 & [0.4833, 0.6225] & [0.4837, 0.6880] & B \\
Health technology / trust & H & 1961.5 & [1758.9, 2375.6] & [1774.7, 2993.9] & B \\
Health technology / trust & T & 0.6642 & [0.6016, 0.7808] & [0.6019, 0.9231] & B \\
Health technology / trust & F & 0.5403 & [0.4833, 0.6225] & [0.4837, 0.6880] & B \\
SaaS / direct & G & 0.2444 & [0.1102, 0.3434] & [0.1680, 0.5193] & B \\
SaaS / direct & H & 1232.5 & [1051.5, 1536.7] & [1096.5, 2119.0] & B \\
SaaS / direct & T & 0.4807 & [0.3778, 0.6030] & [0.4008, 0.6975] & B \\
SaaS / direct & F & 0.2444 & [0.1102, 0.3434] & [0.1680, 0.5193] & B \\
SaaS / comparison & G & 0.1317 & [0.1063, 0.1249] & [0.1607, 0.5377] & B \\
SaaS / comparison & H & 1066.6 & [1051.8, 1063.1] & [1091.3, 2057.1] & B \\
SaaS / comparison & T & 0.3944 & [0.3688, 0.3918] & [0.3977, 0.7034] & B \\
SaaS / comparison & F & 0.1523 & [0.1120, 0.1273] & [0.1682, 0.5411] & B \\
SaaS / exploration & G & 0.2309 & [0.0918, 0.3535] & [0.1303, 0.4726] & B \\
SaaS / exploration & H & 1165.2 & [1020.9, 1368.2] & [1054.6, 1780.3] & B \\
SaaS / exploration & T & 0.4540 & [0.3571, 0.5741] & [0.3816, 0.6733] & B \\
SaaS / exploration & F & 0.2307 & [0.0844, 0.3517] & [0.1289, 0.4736] & B \\
SaaS / problem solution & G & 0.2357 & [0.1006, 0.3015] & [0.1751, 0.5795] & B \\
SaaS / problem solution & H & 1184.6 & [1033.9, 1294.1] & [1106.7, 2236.8] & B \\
SaaS / problem solution & T & 0.4585 & [0.3698, 0.5276] & [0.4112, 0.7411] & B \\
SaaS / problem solution & F & 0.2357 & [0.1006, 0.3015] & [0.1742, 0.5795] & B \\
SaaS / trust & G & 0.4696 & [0.3678, 0.5838] & [0.3962, 0.7540] & B \\
SaaS / trust & H & 1722.2 & [1465.1, 2226.4] & [1493.0, 3958.1] & B \\
SaaS / trust & T & 0.6160 & [0.5041, 0.7350] & [0.5338, 0.9400] & B \\
SaaS / trust & F & 0.4696 & [0.3678, 0.5838] & [0.3962, 0.7540] & B \\
Mean / direct & G & 0.3638 & [0.3141, 0.3783] & [0.3501, 0.4607] & B \\
Mean / direct & H & 1544.4 & not computed & not computed & U \\
Mean / direct & T & 0.5559 & not computed & not computed & U \\
Mean / direct & F & 0.3643 & [0.3139, 0.3786] & [0.3506, 0.4612] & B \\
Mean / comparison & G & 0.2307 & [0.3096, 0.4542] & [0.3103, 0.4542] & P \\
Mean / comparison & H & 1217.3 & not computed & not computed & U \\
Mean / comparison & T & 0.4503 & not computed & not computed & U \\
Mean / comparison & F & 0.2382 & [0.3160, 0.4583] & [0.3160, 0.4583] & P \\
Mean / exploration & G & 0.3100 & [0.2634, 0.3303] & [0.2892, 0.4054] & B \\
Mean / exploration & H & 1350.7 & not computed & not computed & U \\
Mean / exploration & T & 0.4971 & not computed & not computed & U \\
Mean / exploration & F & 0.3104 & [0.2642, 0.3323] & [0.2890, 0.4071] & B \\
Mean / problem solution & G & 0.3724 & [0.3281, 0.3635] & [0.3804, 0.5131] & B \\
Mean / problem solution & H & 1582.3 & not computed & not computed & U \\
Mean / problem solution & T & 0.5519 & not computed & not computed & U \\
Mean / problem solution & F & 0.3726 & [0.3271, 0.3646] & [0.3792, 0.5141] & B \\
Mean / trust & G & 0.4907 & [0.4462, 0.5040] & [0.4774, 0.5917] & B \\
Mean / trust & H & 2048.8 & not computed & not computed & U \\
Mean / trust & T & 0.6563 & not computed & not computed & U \\
Mean / trust & F & 0.4907 & [0.4462, 0.5040] & [0.4773, 0.5917] & B \\
\end{longtable}
}

\section{Stability, scoring conventions and sensitivities}\label{ssec:stability}

\begin{figure}[t]
\centering
\includegraphics[width=\textwidth]{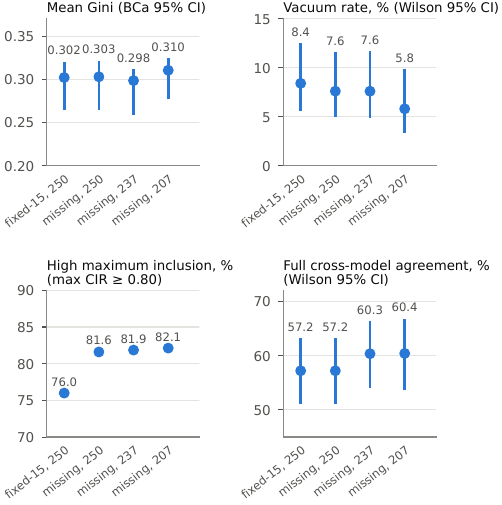}
\caption{Headline statistics under the two scoring conventions and the two empties sensitivities, all industries pooled: fixed-15 (250 queries), missing (250), missing without the 13 all-empty queries (237) and without the 43 affected queries (207). Mean Gini 0.302, 0.303, 0.298, 0.310; vacuum rate 8.4, 7.6, 7.6, 5.8 per cent; high maximum inclusion rate 76.0, 81.6, 81.9, 82.1 per cent; full agreement 57.2, 57.2, 60.3, 60.4 per cent. The top brand per system is the argmax of naming counts and is the same under both conventions, so the first two agreement values coincide.}
\label{fig:s-conventions}
\end{figure}

Table~\ref{tab:s-stability} collects the reliability and sensitivity checks that Section 4.7 of the main text summarises, and Figure~\ref{fig:s-conventions} shows the four headline statistics under the two scoring conventions and the two empties sensitivities. Per industry, the split-half Pearson correlation of RS is 0.999 (consulting), 0.998 (e-commerce), 0.997 (fintech), 0.997 (health technology) and 0.988 (SaaS); the vacuum kappa is 0.91 in e-commerce, 1.00 in fintech and 0.93 in health technology, and near zero in consulting and SaaS, which hold one and two vacuums at 96 per cent raw agreement. The within-system CMA halves hold three and two answers and are the coarsest possible split, with sign agreement 71 to 78 per cent pooled. The rank-weighted RS of Table 5 correlates with plain RS at Spearman 0.92 (consulting), 0.99 (e-commerce), 0.84 (fintech), 0.81 (health technology) and 0.88 (SaaS) over the ten brands; it keeps the leader in four industries and puts McKinsey ahead of Deloitte in consulting.

\begin{table}[tbp]
\caption{Stability and sensitivity. Split halves are iterations $\{1, 3, 5\}$ against $\{2, 4\}$. Drop-affected removes the 43 queries with at least one empty GPT-5.2 answer; drop-all-empty removes the 13 queries on which all five GPT-5.2 answers are empty. Gini values are the unweighted mean over industries with BCa intervals; rates are pooled.}\label{tab:s-stability}
\centering
\footnotesize
\setlength{\tabcolsep}{3pt}
\begin{tabularx}{\textwidth}{@{}Ycc@{}}
\toprule
Statistic & Missing convention & Fixed-15 \\
\midrule
Split-half RS, Pearson $r$ / Spearman $\rho$ (50 brands) & 0.998 / 0.995 & 0.998 / 0.995 \\
Split-half vacuum classification, Cohen's $\kappa$ (raw agreement) & 0.81 (97.6\%) & 0.89 (98.4\%) \\
Iteration Jaccard of brand sets, GPT-5.2 / Gemini / Perplexity & 0.84 / 0.84 / 0.79 & 0.81 / 0.84 / 0.79 \\
Split-half pooled CMA, sign agreement / Spearman & 84.0\% / 0.642 & 85.6\% / 0.674 \\
Matcher precision / recall on the adjudicated gold sample & 300/300 / 278/282 & \\
\midrule
Query set & Gini [95\% CI]; vacuums; high maximum inclusion; full agreement & \\
\midrule
All 250 queries, fixed-15 & 0.30 [0.26, 0.32]; 21 (8.4\%); 76.0\%; 57.2\% & \\
All 250 queries, missing & 0.30 [0.26, 0.32]; 19 (7.6\%); 81.6\%; 57.2\% & \\
Drop-all-empty (237) & 0.30 [0.26, 0.31]; 18 (7.6\%); 81.9\%; 60.3\% & \\
Drop-affected (207) & 0.31 [0.28, 0.32]; 12 (5.8\%); 82.1\%; 60.4\% & \\
Non-comparison (200) & 0.33 [0.29, 0.35]; 19 (9.5\%); 77.0\%; 56.0\% & \\
Non-comparison, drop-affected (182) & 0.34 [0.29, 0.36]; 12 (6.6\%); 79.7\%; 58.2\% & \\
\midrule
Leave-one-brand-out Gini range per industry & 0.05 to 0.12 & 0.05 to 0.12 \\
Gini change when the leader is dropped & $-0.10$ (e-commerce) to $+0.01$ (consulting) & \\
E-commerce Gini without the alias ``bolt food'' & 0.433 (with: 0.421); vacuums 7 and 7 & \\
\bottomrule
\end{tabularx}
\end{table}

Every submitted-hypothesis verdict is the same under both conventions and on both drop sets, and scoring conventions leave the top brand of each retained query unchanged (Table~\ref{tab:s-stability}, Figure~\ref{fig:s-conventions}). Single brand removals can swap adjacent industries in the concentration ordering: e-commerce without Amazon (0.33) sits below fintech (0.34), consulting without Implement Consulting (0.15) below SaaS (0.19), and fintech without Montonio (0.29) below health technology (0.33). The brand whose removal moves the Gini most is the smallest brand in three industries (Implement Consulting, Montonio, CompuGroup Medical) and the leader in two (Amazon, HubSpot).

On all 250 queries under the conditional estimand, the maximum industry split-half Gini gap is 0.012850425 (at most 0.013); it is 0.019097034 on the 200-query companion, and 0.015747380 and 0.016199541 under fixed-15 respectively. Undefined halves remain undefined and are excluded with their counts recorded. The pooled conditional halves have no undefined query; this does not imply every system answered in each half. Drop-all-empty leaves partial-empty queries, so it need not equal fixed-15: mean Gini is 0.298363 conditional and 0.299452 fixed-15, with 194 versus 190 high-maximum-inclusion queries. Removing all 43 affected queries (and the corresponding companion set) makes the conventions equal.

\section{Query-type regression by industry}\label{ssec:glm}

Table~\ref{tab:s-glm} gives the per-industry rows behind the pooled odds ratios of Table 10 in the main text. Each industry model has 50 query clusters; a query-type term is identified from the 10 queries of that type (5 for trust), so the per-industry rows are read for direction only and the trust rows are flagged.

\begin{table}[tbp]
\caption{Per-industry logistic models of brand presence per non-empty answer: odds ratios with 95\% intervals relative to direct queries and to GPT-5.2, brand fixed effects, standard errors clustered by query. Trust terms rest on five clusters and are flagged with an asterisk.}\label{tab:s-glm}
\centering
\footnotesize
\setlength{\tabcolsep}{3pt}
\begin{tabular}{@{}llcccc@{}}
\toprule
Industry & Term & Odds ratio [95\% CI] & $p$ & Fixed-15 [95\% CI] & $p$ \\
\midrule
Consulting & Comparison & 0.25 [0.15, 0.43] & $<0.001$ & 0.23 [0.13, 0.39] & $<0.001$ \\
 & Exploration & 1.69 [0.93, 3.07] & 0.085 & 1.68 [0.93, 3.04] & 0.085 \\
 & Problem-solution & 0.69 [0.37, 1.31] & 0.260 & 0.62 [0.31, 1.22] & 0.168 \\
 & Trust* & 0.94 [0.46, 1.91] & 0.860 & 0.94 [0.46, 1.90] & 0.860 \\
 & Gemini 3 Flash & 0.62 [0.44, 0.88] & 0.007 & 0.72 [0.51, 1.00] & 0.050 \\
 & Perplexity sonar-pro & 0.64 [0.46, 0.90] & 0.010 & 0.74 [0.53, 1.03] & 0.076 \\
\midrule
E-commerce & Comparison & 1.51 [1.00, 2.29] & 0.052 & 1.46 [0.96, 2.20] & 0.073 \\
 & Exploration & 2.20 [1.19, 4.08] & 0.012 & 2.20 [1.19, 4.07] & 0.012 \\
 & Problem-solution & 0.90 [0.52, 1.56] & 0.712 & 0.90 [0.52, 1.56] & 0.712 \\
 & Trust* & 1.13 [0.60, 2.10] & 0.709 & 1.13 [0.60, 2.10] & 0.709 \\
 & Gemini 3 Flash & 0.64 [0.52, 0.78] & $<0.001$ & 0.65 [0.53, 0.80] & $<0.001$ \\
 & Perplexity sonar-pro & 0.75 [0.62, 0.91] & 0.004 & 0.77 [0.63, 0.93] & 0.008 \\
\midrule
Fintech & Comparison & 0.79 [0.58, 1.07] & 0.128 & 0.64 [0.49, 0.84] & 0.001 \\
 & Exploration & 1.53 [0.88, 2.66] & 0.129 & 1.38 [0.83, 2.30] & 0.219 \\
 & Problem-solution & 0.80 [0.51, 1.25] & 0.319 & 0.70 [0.42, 1.18] & 0.181 \\
 & Trust* & 1.19 [0.65, 2.19] & 0.566 & 1.19 [0.66, 2.16] & 0.566 \\
 & Gemini 3 Flash & 0.58 [0.48, 0.70] & $<0.001$ & 0.73 [0.58, 0.91] & 0.006 \\
 & Perplexity sonar-pro & 0.57 [0.45, 0.73] & $<0.001$ & 0.71 [0.54, 0.95] & 0.020 \\
\midrule
Health technology & Comparison & 1.53 [1.01, 2.33] & 0.045 & 0.93 [0.59, 1.46] & 0.742 \\
 & Exploration & 1.90 [1.17, 3.08] & 0.010 & 1.74 [0.99, 3.04] & 0.052 \\
 & Problem-solution & 0.56 [0.29, 1.08] & 0.084 & 0.49 [0.26, 0.91] & 0.025 \\
 & Trust* & 1.09 [0.62, 1.92] & 0.770 & 1.13 [0.64, 2.00] & 0.670 \\
 & Gemini 3 Flash & 0.48 [0.36, 0.63] & $<0.001$ & 0.76 [0.55, 1.06] & 0.102 \\
 & Perplexity sonar-pro & 0.47 [0.34, 0.64] & $<0.001$ & 0.74 [0.51, 1.08] & 0.116 \\
\midrule
SaaS & Comparison & 0.99 [0.73, 1.34] & 0.954 & 0.81 [0.60, 1.09] & 0.162 \\
 & Exploration & 1.59 [1.10, 2.30] & 0.014 & 1.55 [1.09, 2.22] & 0.016 \\
 & Problem-solution & 0.80 [0.48, 1.34] & 0.397 & 0.80 [0.48, 1.34] & 0.397 \\
 & Trust* & 0.70 [0.49, 1.00] & 0.052 & 0.70 [0.49, 1.00] & 0.052 \\
 & Gemini 3 Flash & 0.81 [0.69, 0.95] & 0.011 & 0.92 [0.77, 1.11] & 0.406 \\
 & Perplexity sonar-pro & 0.86 [0.74, 1.00] & 0.057 & 0.98 [0.82, 1.17] & 0.813 \\
\bottomrule
\end{tabular}
\end{table}

Every query-type concentration estimate and interval, including comparison and problem-solution, is retained in Section~\ref{ssec:intervals}. The mean maximum CIR per query is 0.90 on direct, 1.00 on comparison, 0.91 on exploration, 0.77 on problem-solution and 0.89 on trust prompts, and the vacuum rate by type is 9.3, 0.0, 4.0, 18.0 and 4.0 per cent. The industry leader gives way to another top-three brand on some prompt types: McKinsey leads consulting on comparison and trust prompts and BCG on exploration prompts; Shopify leads e-commerce on direct prompts; PayPal ties Stripe on fintech comparison prompts; Oracle Health leads health technology on exploration prompts; Salesforce leads SaaS on exploration and trust prompts; in every case the second brand is within 0.18 and in all but one within 0.03, on subsets of 5 to 15 queries.

\section{Co-Mention Asymmetry at every threshold}\label{ssec:cma}

Table~\ref{tab:s-cma} gives the records and pairs counts of Table 8 in the main text at the four thresholds 0.05, 0.10, 0.15 and 0.20, and Table~\ref{tab:s-zero} shows the effect of the submitted version's zero-substitution convention on the corrected mention table; Figure~\ref{fig:s-listlength} (Section~\ref{ssec:procedures}) plots the list-length control. Under fixed-15 and the stratified null 84 pairs are significant (0 asymmetric, 84 co-mentioned) against 53 (0 and 53) under the missing convention. The median number of queries on which a pair is defined is 13 (range 1 to 36); pairs defined on fewer than ten queries are flagged in the deposit. The sign of the pooled CMA agrees between the two conventions in 430 of the pairs defined in both. On the 200 non-comparison queries the pooled records ratio at 0.10 is 0.28:1 [0.21, 0.36] and at 0.20 0.24:1 [0.18, 0.32].

\begin{table}[tbp]
\caption{Co-Mention Asymmetry by threshold, all 50 queries, missing convention. Records are (pair, query) combinations; asymmetric = CMA above the threshold, co-mentioned = CMA below its negative; ratio = asymmetric : co-mentioned with a BCa 95\% interval over queries; significant pairs = ordered pairs with nominal, unadjusted $p < 0.05$ under the system-stratified permutation null, asymmetric / co-mentioned.}\label{tab:s-cma}
\centering
\footnotesize
\setlength{\tabcolsep}{2pt}
\begin{tabularx}{\textwidth}{@{}>{\raggedright\arraybackslash}p{2.0cm}>{\centering\arraybackslash}p{1.1cm}YY>{\raggedright\arraybackslash}p{2.6cm}YYY@{}}
\toprule
Industry & Cut-off & Rec.\ asym. & Rec.\ co-m. & Ratio [95\% CI] & Pairs asym. & Pairs co-m. & Sig.\ pairs \\
\midrule
Consulting & 0.05 & 240 & 932 & 0.26:1 [0.16, 0.44] & 0 & 51 & 0 / 24 \\
 & 0.10 & 193 & 881 & 0.22:1 [0.13, 0.36] & 0 & 44 & 0 / 24 \\
 & 0.15 & 174 & 835 & 0.21:1 [0.12, 0.35] & 0 & 33 & 0 / 23 \\
 & 0.20 & 148 & 781 & 0.19:1 [0.10, 0.33] & 0 & 30 & 0 / 23 \\
E-commerce & 0.05 & 88 & 128 & 0.69:1 [0.38, 1.10] & 11 & 34 & 0 / 0 \\
 & 0.10 & 67 & 116 & 0.58:1 [0.30, 1.03] & 4 & 23 & 0 / 0 \\
 & 0.15 & 59 & 110 & 0.54:1 [0.27, 1.01] & 1 & 13 & 0 / 0 \\
 & 0.20 & 48 & 94 & 0.51:1 [0.26, 0.93] & 0 & 7 & 0 / 0 \\
Fintech & 0.05 & 144 & 390 & 0.37:1 [0.23, 0.61] & 7 & 53 & 0 / 9 \\
 & 0.10 & 115 & 361 & 0.32:1 [0.20, 0.56] & 4 & 35 & 0 / 7 \\
 & 0.15 & 98 & 329 & 0.30:1 [0.17, 0.53] & 2 & 20 & 0 / 4 \\
 & 0.20 & 78 & 288 & 0.27:1 [0.16, 0.48] & 2 & 15 & 0 / 3 \\
Health technology & 0.05 & 66 & 282 & 0.23:1 [0.12, 0.42] & 4 & 50 & 0 / 14 \\
 & 0.10 & 45 & 260 & 0.17:1 [0.08, 0.32] & 0 & 39 & 0 / 12 \\
 & 0.15 & 36 & 238 & 0.15:1 [0.06, 0.28] & 0 & 21 & 0 / 9 \\
 & 0.20 & 30 & 221 & 0.14:1 [0.05, 0.26] & 0 & 16 & 0 / 7 \\
SaaS & 0.05 & 90 & 168 & 0.54:1 [0.29, 0.85] & 4 & 32 & 0 / 6 \\
 & 0.10 & 67 & 151 & 0.44:1 [0.24, 0.72] & 2 & 20 & 0 / 6 \\
 & 0.15 & 54 & 139 & 0.39:1 [0.19, 0.66] & 0 & 14 & 0 / 5 \\
 & 0.20 & 49 & 127 & 0.39:1 [0.19, 0.68] & 0 & 6 & 0 / 4 \\
\midrule
Pooled & 0.05 & 628 & 1900 & 0.33:1 [0.26, 0.43] & 26 & 220 & 0 / 53 \\
 & 0.10 & 487 & 1769 & 0.28:1 [0.21, 0.36] & 10 & 161 & 0 / 49 \\
 & 0.15 & 421 & 1651 & 0.25:1 [0.19, 0.35] & 3 & 101 & 0 / 41 \\
 & 0.20 & 353 & 1511 & 0.23:1 [0.17, 0.32] & 2 & 74 & 0 / 37 \\
\bottomrule
\end{tabularx}
\end{table}

\begin{table}[tbp]
\caption{The submitted zero-substitution convention on corrected mentions, analysed at $|\mathrm{CMA}| > 0.10$ under fixed-15. Its historical storage filter was 0.05; the last two columns keep undefined cells undefined at the same 0.10 analysis threshold.}\label{tab:s-zero}
\centering
\footnotesize
\setlength{\tabcolsep}{3pt}
\begin{tabularx}{\textwidth}{@{}>{\raggedright\arraybackslash}p{2.6cm}YYYYY@{}}
\toprule
Industry & Rec.\ asym.\ (zero subst.) & Rec.\ co-m.\ (zero subst.) & Ratio (zero subst.) & Ratio, defined only, fixed-15 & Ratio, defined only, missing \\
\midrule
Consulting & 901 & 1741 & 0.52:1 & 0.20:1 & 0.22:1 \\
E-commerce & 873 & 198 & 4.41:1 & 0.57:1 & 0.58:1 \\
Fintech & 1138 & 559 & 2.04:1 & 0.25:1 & 0.32:1 \\
Health technology & 856 & 359 & 2.38:1 & 0.15:1 & 0.17:1 \\
SaaS & 970 & 317 & 3.06:1 & 0.30:1 & 0.44:1 \\
\midrule
Pooled & 4738 & 3174 & 1.49:1 & 0.23:1 & 0.28:1 \\
\bottomrule
\end{tabularx}
\end{table}

The five most negative pooled pairs are PwC and EY ($-0.80$), Semrush and Ahrefs ($-0.77$), KPMG and EY ($-0.74$), Bain and BCG ($-0.73$) and Philips Healthcare and GE HealthCare ($-0.69$), each with a stratified permutation $p$ below 0.005; the most positive pooled pairs (Wise and Montonio, N26 and Montonio, $+0.30$) rest on two defined queries and have $p = 1$. The secondary list-length check has two industry-stratum mean differences above 0.01: fintech-long (0.013492) and health-technology-long (0.013539), so the maximum is at most 0.014. Broad pooled strata do not identify the cause of the primary co-mention result. Split-half sign agreement of pooled CMA is 84.0 per cent over the 338 pairs non-zero in both halves; between systems the sign agreement of pooled CMA is 55.8 per cent (GPT-5.2 and Gemini), 49.4 per cent (GPT-5.2 and Perplexity) and 60.7 per cent (Gemini and Perplexity). The leader carries the most negative mean CMA of its industry only in health technology (Epic Systems, $-0.27$); per industry the Spearman correlation between RS and mean CMA as brand $A$ is $-0.321$ (consulting), $-0.188$ (e-commerce), $-0.818$ (fintech, $p = 0.004$), $-0.939$ (health technology, $p < 0.001$) and $-0.358$ (SaaS).

The within-query E2 family has 450 tests in each convention and set. Conditional all-query and non-comparison results each have 58 BH-significant pairs and 43 with their sign reproduced in both halves; 38 have negative observed CMA and five positive. Relative to their null means, 31 of the 43 are more co-mentioned and 12 less. These are two different classifications: the sign of observed CMA and its direction of departure from its own null. Fixed-15 has 60 significant / 50 departing pairs on all queries and 59 / 49 on the companion. E2 is not supported in all four combinations. The all-query fixed-15 rho is inside its null band, but the tail and pair component is not supported, so the composite remains not supported. The 120-query movable-only sensitivity has a conditional ratio of 0.329499 against [0.208280, 0.239187]; it is separate from the all-query primary ratio 0.275297 against [0.175676, 0.201302]. Under fixed-15/all, rho is $-0.612121$ against [$-0.633939$, $-0.517576$] and the ratio 0.233186 against [0.148026, 0.169901]; on fixed-15/non-comparison, rho is $-0.527273$ against [$-0.510303$, $-0.367273$] and the ratio 0.256786 against [0.161964, 0.186167]. These departures are associations; no intervention identifies a causal displacement process.

\section{Open-vocabulary reading of the vacuum queries}\label{ssec:vacuums}

Table~\ref{tab:s-vacuums} gives, for each of the 19 vacuum queries, the brand an adjudicating language model found most often when it read every non-empty answer without the sample restriction, its count, the number of distinct brands named and whether any brand reaches a majority; Section 4.3 of the main text gives the summary counts. Three queries remain strict open-vocabulary vacuums: consulting 37 and health technology 5 and 40. Health technology 44 has Azure in 7/14 answers, CVI exactly 0.50, so it is a boundary case. Four queries lack a strict majority; this separate count includes that boundary. The models named between 11 and 72 distinct brands per query (median 29). Two fintech queries (39 and 44) that are vacuums under the fixed-15 convention only were also read: Plaid and Stripe lead them at 5 of 10 non-empty answers, and no brand reaches a majority.

\begin{sidewaystableorg}[p]
\caption{The 19 vacuum queries (CVI $> 0.50$) read without the sample restriction. Counts are over non-empty answers; both brand columns use the open-vocabulary reader consistently, with matcher conflicts disclosed below; author check pending. Type: D direct, E exploration, P problem-solution, T trust.}\label{tab:s-vacuums}
\centering
\footnotesize
\setlength{\tabcolsep}{2pt}
\begin{tabularx}{\linewidth}{@{}lcYc>{\raggedright\arraybackslash}p{2.3cm}>{\raggedright\arraybackslash}p{2.8cm}cc@{}}
\toprule
Industry & Idx & Query & Type & Best sampled brand (count) & Most-named brand (count) & Distinct & Majority \\
\midrule
Health technology & 11 & Best EHR for small medical practices & D & Epic Systems (1/15) & Tebra (Kareo) (13/15) & 23 & yes \\
Health technology & 44 & Launch a digital health service in the Nordics & P & Kry (1/14) & Microsoft Azure (7/14) & 28 & no \\
Health technology & 40 & Monitor patients remotely & P & none (0/11) & VitalConnect (5/11) & 20 & no \\
Health technology & 31 & Which health tech startups are emerging & E & Teladoc (1/10) & Ambience Healthcare (10/10) & 65 & yes \\
E-commerce & 44 & Buy luxury fashion online & P & Amazon (2/15) & Net-a-Porter (14/15) & 50 & yes \\
E-commerce & 47 & Fashion retailer with the best reputation & T & H\&M (3/15) & Nordstrom (14/15) & 45 & yes \\
Health technology & 43 & Radiology AI software & P & GE HealthCare (2/13) & Aidoc (13/13) & 45 & yes \\
E-commerce & 6 & Marketplace for electronics in the Nordics & D & Amazon (4/15) & Elkj{\o}p / Elgiganten / Gigantti (12/15) & 24 & yes \\
Health technology & 39 & Online appointment booking for patients & P & Epic Systems (4/15) & SimplePractice (15/15) & 36 & yes \\
Health technology & 13 & Health tech for pharmaceutical companies & D & Veeva Systems (3/11) & Insilico Medicine (9/11) & 24 & yes \\
Health technology & 5 & Health technology for life sciences & D & Veeva Systems (4/14) & Google DeepMind AlphaFold (6/14) & 72 & no \\
E-commerce & 5 & Best quick commerce delivery app & D & Amazon (5/15) & Instacart (15/15) & 28 & yes \\
E-commerce & 10 & Marketplace for sustainable shopping & D & Amazon (5/15) & EarthHero (14/15) & 39 & yes \\
E-commerce & 30 & Quick commerce services available & E & Amazon (5/15) & Gopuff (12/15) & 51 & yes \\
E-commerce & 42 & Find sustainable products online & P & Amazon (6/15) & EarthHero (15/15) & 51 & yes \\
Health technology & 36 & Offer telehealth to my patients & P & Teladoc (5/15) & Doxy.me (15/15) & 29 & yes \\
SaaS & 13 & Best CRM for real estate agents & D & HubSpot (6/15) & Follow Up Boss (15/15) & 24 & yes \\
Consulting & 37 & Consultant for an SAP implementation & P & Deloitte (5/11) & Deloitte (5/11) & 11 & no \\
SaaS & 44 & Build landing pages without a developer & P & HubSpot (6/15) & Unbounce (15/15) & 21 & yes \\
\bottomrule
\end{tabularx}
\end{sidewaystableorg}

The sample-artefact cases lie in sub-categories absent from the brand list, such as quick commerce, sustainable retail, telehealth tooling, radiology AI and real-estate CRM. Two best-sampled-brand readings conflict with the matcher: health technology 43 is Epic Systems 3/13 under the matcher and GE HealthCare 2/13 under the reader; health technology 40 is Epic Systems 1/11 under the matcher and no sampled brand 0/11 under the reader. The main vacuum table uses the matcher, while Table~\ref{tab:s-vacuums} uses the reader in both brand columns. Both conflicts remain unresolved pending author adjudication.

\section{The submitted hypotheses by industry}\label{ssec:hypotheses}

Tables~\ref{tab:s-pooled-hypotheses} and~\ref{tab:s-hypotheses} give the pooled and per-industry verdicts, on all queries and on the non-comparison companion. Every verdict is the same under the fixed-15 convention. No hypothesis is supported in any scope. H4 uses one clustering pooled over industries (one clustering over all answers) and has no per-industry row.

\begin{table}[tbp]
\caption{The five hypotheses of the submitted version under one interval rule, all 250 queries; cut-offs set before analysis and not registered; NS not supported; H4 carries a ten-seed range in place of an interval.}\label{tab:s-pooled-hypotheses}
\centering
\footnotesize
\setlength{\tabcolsep}{3pt}
\begin{tabularx}{\textwidth}{@{}lYYc@{}}
\toprule
H & Cut-off & Observed, pooled [95\% CI] & Verdict \\
\midrule
H1 & Gini $> 0.60$ & 0.30 [0.26, 0.32] & NS \\
H2 & Records ratio $> 3$:1 at $|\mathrm{CMA}| > 0.10$ & 0.28:1 [0.21, 0.36] & NS \\
H3 & Vacuums $> 20\%$ of queries & 7.6\% [4.9, 11.6] & NS \\
H4 & Emergent clusters $> 15\%$ & 0.0\% at every seed (submitted 4.2\%) & NS \\
H5 & Full agreement $< 50\%$ & 57.2\% [51.0, 63.2] & NS \\
\bottomrule
\end{tabularx}
\end{table}

\begin{table}[tbp]
\caption{The five hypotheses of the submitted version under one interval rule, per industry. Verdicts: NS not supported, Inc.\ inconclusive (interval includes the cut). Point estimates with 95\% intervals where the verdict is inconclusive; H4 is one clustering over all industries and carries a range over ten seeds.}\label{tab:s-hypotheses}
\centering
\footnotesize
\setlength{\tabcolsep}{2.5pt}
\begin{tabularx}{\textwidth}{@{}lYYYYYY@{}}
\toprule
H & Scope & Consulting & E-commerce & Fintech & Health technology & SaaS \\
\midrule
H1 Gini $> 0.60$ & all & NS (0.23) & NS (0.42) & NS (0.34) & NS (0.33) & NS (0.19) \\
 & non-comparison & NS (0.23) & NS (0.46) & NS (0.36) & NS (0.41) & NS (0.21) \\
H2 ratio $> 3$:1 & all & NS (0.22:1) & NS (0.58:1) & NS (0.32:1) & NS (0.17:1) & NS (0.44:1) \\
 & non-comparison & NS (0.23:1) & NS (0.58:1) & NS (0.32:1) & NS (0.18:1) & NS (0.42:1) \\
H3 vacuums $> 20\%$ & all & NS (2.0\%) & Inc.\ (14.0\% [7.0, 26.2]) & NS (0.0\%) & Inc.\ (18.0\% [9.8, 30.8]) & NS (4.0\%) \\
 & non-comparison & NS (2.5\%) & Inc.\ (17.5\% [8.7, 31.9]) & NS (0.0\%) & Inc.\ (22.5\% [12.3, 37.5]) & NS (5.0\%) \\
H4 clusters $> 15\%$ & all (pooled) & \multicolumn{5}{l}{NS (0.0\% at every one of ten seeds; submitted 4.2\%)} \\
H5 agreement $< 50\%$ & all & Inc.\ (50.0\% [36.6, 63.4]) & NS (64.0\% [50.1, 75.9]) & Inc.\ (62.0\% [48.2, 74.1]) & Inc.\ (48.0\% [34.8, 61.5]) & Inc.\ (62.0\% [48.2, 74.1]) \\
 & non-comparison & Inc.\ (47.5\% [32.9, 62.5]) & Inc.\ (60.0\% [44.6, 73.7]) & Inc.\ (60.0\% [44.6, 73.7]) & Inc.\ (50.0\% [35.2, 64.8]) & Inc.\ (62.5\% [47.0, 75.8]) \\
\bottomrule
\end{tabularx}
\end{table}

\emph{The emergent-category clustering re-run.} The submitted version clustered the stored answer embeddings (BGE-M3, 1,024 dimensions) with UMAP (15 neighbours, 10 components, cosine) and HDBSCAN (minimum cluster size 20, minimum samples 5) under BERTopic, anchored each cluster to the query nearest its mean embedding, and called a cluster emergent when that similarity fell below 0.60; it reported 3 of 71 clusters (4.2 per cent), all three with empty keyword sets. The re-run uses the same embeddings and parameters over ten UMAP seeds (0 to 9), first on all 3,750 answers and then without the 136 empty answers. With the empties left in, every seed returns exactly three emergent clusters of sizes 64, 38 and 34, holding the 136 empty answers and nothing else: the empty string embeds to one vector, so the empties form clusters with no words and one shared nearest query. Without the empties, the 3,614 answers form 68 to 76 clusters per seed (median 71.5; outlier share 13.3 to 16.8 per cent), and no cluster falls below 0.60 at any seed, or below a cut of 0.66 set from the design itself as the fifth percentile of each query's similarity to its nearest other query; the lowest cluster-to-query similarity anywhere is 0.68 against a median of 0.77. Cluster brand labels recomputed with the guarded matcher place EY in the top three of four to five consulting clusters per seed and of no cluster outside consulting; under the substring matcher EY had appeared in the top three of 61 of 71 clusters across all five industries. The pinned environment and per-seed outputs are in the deposit.

H4 is not separately clustered on the non-comparison companion; that row is n/a. Its all-query zero-emergence result applies to the stored wave-1 answer embeddings. Query embeddings and new encodings use pinned BGE-M3 snapshot \texttt{5617a9f6\allowbreak 1b028005\allowbreak a4858fda\allowbreak c845db40\allowbreak 6aefb181}. The original notebook did not record the answer encoder revision or complete runtime. In a separate 292-answer wave-1 diagnostic, stored and new vectors had median cosine similarity 1.000 and minimum 0.972421; seed 0 produced four clusters from stored vectors and five from new ones, with zero emergent clusters in both. The notebook selected CUDA if available, otherwise CPU, but actual device and package provenance are absent. The cause remains unresolved; it cannot be assigned to device precision. Current-tokenizer lengths in that subset are at most 1,202 against the current 8,192-token limit. Cache identity and row-alignment checks prevent mismatched reuse but do not resolve historical comparability. This limits direct cross-wave clustering comparisons. All ten September seeds completed on 3,750 answers: 73 to 77 clusters, zero emergent under the 0.60 rule at every seed, and one unanchored cluster under the 0.664 benchmark at seed 0 (zero at the other nine). The lowest nearest-query similarity is 0.6509. These results do not resolve the historical encoder limitation; Section S12 reports each seed. No separate at-cap or non-comparison clustering was prescribed or run.

\section{September replication and cross-wave results}\label{ssec:wave2}

The actual September wave was finalised on 11 September 2026 under the frozen plan and its six recorded deviations. The original and finalised response files have identical bytes: 3,750 answers, no transport error, no empty body. All 250 queries and the frozen non-comparison companion of 200 are retained. Removing the 13 February queries with no OpenAI answer gives 237 and 196 queries. The separate sensitivity drops 126 September OpenAI answers at the output cap. Every estimate below retains its stated population and denominator. OpenAI response-level mean length rises from 2,785 to 2,957 characters, while its matched query-level change is 1 [\ensuremath{-}24, 28] characters: February has 136 missing responses and 13 unmatched queries, so the populations and weighting differ. Gemini records a mean 551 thoughts tokens in September; February recorded no token usage. Full per-industry, query-type, threshold, fixed-15, interval-diagnostic and pair-level outputs are retained in the computational inventory for the subsequent deposit; they are not yet in the published version 2.0.

\begingroup\footnotesize\setlength{\tabcolsep}{3pt}
\begin{longtable}{@{}>{\raggedright\arraybackslash}p{2.4cm}>{\raggedright\arraybackslash}p{1cm}>{\raggedright\arraybackslash}p{1cm}>{\raggedright\arraybackslash}p{2.2cm}>{\raggedright\arraybackslash}p{2.4cm}>{\raggedright\arraybackslash}p{2.4cm}@{}}
\caption{All non-empty answers. Changes are September minus February; rate changes are percentage points. Intervals are paired percentile bootstraps over queries; pooled Gini averages the industry replicates. RS correlations use 50 brands.}\label{tab:s-wave2-change}\\
\toprule
Query set & RS $\rho$ & RS $r$ & $\Delta$ Gini [95\% CI] & $\Delta$ vacuum [95\% CI] & $\Delta$ high inclusion [95\% CI] \\
\midrule
\endfirsthead
\toprule
Query set & RS $\rho$ & RS $r$ & $\Delta$ Gini [95\% CI] & $\Delta$ vacuum [95\% CI] & $\Delta$ high inclusion [95\% CI] \\
\midrule
\endhead
All 250 & 0.994 & 0.996 & -0.006 [-0.014, 0.005] & 0.0 [-2.4, 2.4] & 0.8 [-2.4, 4.0] \\
Non-comparison 200 & 0.994 & 0.996 & -0.007 [-0.016, 0.006] & 0.0 [-2.5, 2.5] & 1.0 [-3.0, 5.0] \\
All minus 13 (237) & 0.993 & 0.996 & -0.007 [-0.015, 0.005] & 0.0 [-2.1, 2.5] & 0.8 [-2.5, 3.8] \\
Non-comparison minus 13 (196) & 0.993 & 0.996 & -0.007 [-0.016, 0.006] & 0.0 [-2.6, 3.1] & 1.0 [-3.1, 5.1] \\
\bottomrule
\end{longtable}
\endgroup

\begingroup\footnotesize\setlength{\tabcolsep}{3pt}
\begin{longtable}{@{}>{\raggedright\arraybackslash}p{2.8cm}>{\raggedright\arraybackslash}p{3.1cm}>{\raggedright\arraybackslash}p{2.2cm}>{\raggedright\arraybackslash}p{3.3cm}@{}}
\caption{All non-empty answers. Agreement change uses every query, in percentage points; unchanged-top shares are percentages. System order is OpenAI, Gemini, Perplexity.}\label{tab:s-wave2-top}\\
\toprule
Query set & $\Delta$ agreement [95\% CI] & Same pooled top & Same top per system \\
\midrule
\endfirsthead
\toprule
Query set & $\Delta$ agreement [95\% CI] & Same pooled top & Same top per system \\
\midrule
\endhead
All 250 & 3.6 [-1.6, 8.8] & 88.8 & 86.4 / 91.6 / 73.6 \\
Non-comparison 200 & 1.0 [-5.0, 7.0] & 86.5 & 88.0 / 90.5 / 70.0 \\
All minus 13 (237) & 1.3 [-4.2, 6.8] & 88.2 & 91.1 / 91.6 / 72.6 \\
Non-comparison minus 13 (196) & 0.5 [-5.6, 6.6] & 86.2 & 89.8 / 90.3 / 69.9 \\
\bottomrule
\end{longtable}
\endgroup

\begingroup\footnotesize\setlength{\tabcolsep}{3pt}
\begin{longtable}{@{}>{\raggedright\arraybackslash}p{2.4cm}>{\raggedright\arraybackslash}p{1cm}>{\raggedright\arraybackslash}p{1cm}>{\raggedright\arraybackslash}p{2.2cm}>{\raggedright\arraybackslash}p{2.4cm}>{\raggedright\arraybackslash}p{2.4cm}@{}}
\caption{Excluding September answers at the output cap. Changes are September minus February; rate changes are percentage points. Intervals are paired percentile bootstraps over queries; pooled Gini averages the industry replicates. RS correlations use 50 brands.}\label{tab:s-wave2-change_nocap}\\
\toprule
Query set & RS $\rho$ & RS $r$ & $\Delta$ Gini [95\% CI] & $\Delta$ vacuum [95\% CI] & $\Delta$ high inclusion [95\% CI] \\
\midrule
\endfirsthead
\toprule
Query set & RS $\rho$ & RS $r$ & $\Delta$ Gini [95\% CI] & $\Delta$ vacuum [95\% CI] & $\Delta$ high inclusion [95\% CI] \\
\midrule
\endhead
All 250 & 0.994 & 0.996 & -0.007 [-0.015, 0.004] & 0.0 [-2.4, 2.4] & 0.8 [-2.4, 4.0] \\
Non-comparison 200 & 0.994 & 0.996 & -0.008 [-0.017, 0.006] & 0.0 [-2.5, 2.5] & 1.0 [-3.0, 5.0] \\
All minus 13 (237) & 0.994 & 0.996 & -0.008 [-0.016, 0.005] & 0.0 [-2.1, 2.5] & 0.8 [-2.5, 3.8] \\
Non-comparison minus 13 (196) & 0.993 & 0.996 & -0.008 [-0.016, 0.006] & 0.0 [-2.6, 3.1] & 1.0 [-3.1, 5.1] \\
\bottomrule
\end{longtable}
\endgroup

\begingroup\footnotesize\setlength{\tabcolsep}{3pt}
\begin{longtable}{@{}>{\raggedright\arraybackslash}p{2.8cm}>{\raggedright\arraybackslash}p{3.1cm}>{\raggedright\arraybackslash}p{2.2cm}>{\raggedright\arraybackslash}p{3.3cm}@{}}
\caption{Excluding September answers at the output cap. Agreement change uses every query, in percentage points; unchanged-top shares are percentages. System order is OpenAI, Gemini, Perplexity.}\label{tab:s-wave2-top_nocap}\\
\toprule
Query set & $\Delta$ agreement [95\% CI] & Same pooled top & Same top per system \\
\midrule
\endfirsthead
\toprule
Query set & $\Delta$ agreement [95\% CI] & Same pooled top & Same top per system \\
\midrule
\endhead
All 250 & 1.2 [-4.0, 6.4] & 88.4 & 90.0 / 91.6 / 73.6 \\
Non-comparison 200 & 0.0 [-6.0, 6.0] & 86.0 & 90.0 / 90.5 / 70.0 \\
All minus 13 (237) & 0.4 [-4.6, 5.9] & 88.2 & 91.1 / 91.6 / 72.6 \\
Non-comparison minus 13 (196) & 0.0 [-6.1, 6.1] & 86.2 & 90.3 / 90.3 / 69.9 \\
\bottomrule
\end{longtable}
\endgroup

\begingroup\footnotesize\setlength{\tabcolsep}{3pt}
\begin{longtable}{@{}>{\raggedright\arraybackslash}p{2cm}>{\raggedright\arraybackslash}p{2.2cm}>{\raggedright\arraybackslash}p{1cm}>{\raggedright\arraybackslash}p{1cm}>{\raggedright\arraybackslash}p{2.5cm}>{\raggedright\arraybackslash}p{2.5cm}@{}}
\caption{Per-industry RS correlations and leaders, all non-empty answers. Each correlation uses the ten sampled brands.}\label{tab:s-wave2-industries}\\
\toprule
Industry & Query set & RS $\rho$ & RS $r$ & February leader & September leader \\
\midrule
\endfirsthead
\toprule
Industry & Query set & RS $\rho$ & RS $r$ & February leader & September leader \\
\midrule
\endhead
Consulting & All 250 & 1.000 & 0.996 & Deloitte & Deloitte \\
Consulting & Non-comparison 200 & 1.000 & 0.996 & Deloitte & Deloitte \\
E-commerce & All 250 & 0.939 & 0.997 & Amazon & Amazon \\
E-commerce & Non-comparison 200 & 0.976 & 0.997 & Amazon & Amazon \\
Fintech & All 250 & 0.952 & 0.989 & Stripe & Stripe \\
Fintech & Non-comparison 200 & 0.964 & 0.987 & Stripe & Stripe \\
Health technology & All 250 & 0.988 & 0.996 & Epic Systems & Epic Systems \\
Health technology & Non-comparison 200 & 0.952 & 0.996 & Epic Systems & Epic Systems \\
SaaS & All 250 & 0.976 & 0.991 & HubSpot & HubSpot \\
SaaS & Non-comparison 200 & 0.964 & 0.992 & HubSpot & HubSpot \\
\bottomrule
\end{longtable}
\endgroup

\begingroup\footnotesize\setlength{\tabcolsep}{3pt}
\begin{longtable}{@{}>{\raggedright\arraybackslash}p{2.2cm}>{\raggedright\arraybackslash}p{2.4cm}>{\raggedright\arraybackslash}p{2cm}>{\raggedright\arraybackslash}p{3.8cm}>{\raggedright\arraybackslash}p{1cm}@{}}
\caption{E1 descriptive size-benchmark correlations, September with and without at-cap answers. The same frozen benchmark excludes three incomparable size markers (47 usable brands overall).}\label{tab:s-wave2-size}\\
\toprule
Wave & Query set & Industry & Size--RS rho [95\% CI] & Brands \\
\midrule
\endfirsthead
\toprule
Wave & Query set & Industry & Size--RS rho [95\% CI] & Brands \\
\midrule
\endhead
September & All 250 & Consulting & 0.418 [-0.358, 0.923] & 10 \\
September & All 250 & E-commerce & 0.571 [-0.259, 1.000] & 8 \\
September & All 250 & Fintech & 0.758 [0.208, 0.987] & 10 \\
September & All 250 & Health technology & 0.533 [-0.158, 0.878] & 9 \\
September & All 250 & SaaS & 0.745 [0.148, 1.000] & 10 \\
September & Non-comparison 200 & Consulting & 0.418 [-0.364, 0.925] & 10 \\
September & Non-comparison 200 & E-commerce & 0.524 [-0.538, 1.000] & 8 \\
September & Non-comparison 200 & Fintech & 0.721 [0.119, 0.962] & 10 \\
September & Non-comparison 200 & Health technology & 0.550 [-0.119, 0.930] & 9 \\
September & Non-comparison 200 & SaaS & 0.709 [-0.006, 0.987] & 10 \\
September, uncapped & All 250 & Consulting & 0.418 [-0.355, 0.925] & 10 \\
September, uncapped & All 250 & E-commerce & 0.571 [-0.231, 1.000] & 8 \\
September, uncapped & All 250 & Fintech & 0.758 [0.202, 0.987] & 10 \\
September, uncapped & All 250 & Health technology & 0.533 [-0.217, 0.862] & 9 \\
September, uncapped & All 250 & SaaS & 0.745 [0.132, 0.987] & 10 \\
September, uncapped & Non-comparison 200 & Consulting & 0.418 [-0.585, 0.925] & 10 \\
September, uncapped & Non-comparison 200 & E-commerce & 0.524 [-0.333, 1.000] & 8 \\
September, uncapped & Non-comparison 200 & Fintech & 0.758 [0.159, 0.987] & 10 \\
September, uncapped & Non-comparison 200 & Health technology & 0.550 [-0.179, 0.927] & 9 \\
September, uncapped & Non-comparison 200 & SaaS & 0.709 [0.044, 1.000] & 10 \\
\bottomrule
\end{longtable}
\endgroup

\begingroup\footnotesize\setlength{\tabcolsep}{3pt}
\begin{longtable}{@{}>{\raggedright\arraybackslash}p{2cm}>{\raggedright\arraybackslash}p{2.4cm}>{\raggedright\arraybackslash}p{2.3cm}>{\raggedright\arraybackslash}p{4.7cm}@{}}
\caption{E1 industry concentration orderings, reported descriptively beside the unchanged size benchmark. Five industries do not support a general rank-order test.}\label{tab:s-wave2-ordering}\\
\toprule
Wave & Query set & Statistic & Observed orderings \\
\midrule
\endfirsthead
\toprule
Wave & Query set & Statistic & Observed orderings \\
\midrule
\endhead
September & All 250 & 5 industries (gini) & recommendation ecommerce > fintech > healthtech > consulting > saas; size benchmark ecommerce > saas > fintech > healthtech > consulting; test: none: five purposively chosen industries \\
September & All 250 & 5 industries (top3\_share) & recommendation ecommerce > fintech > healthtech > saas > consulting; size benchmark ecommerce > saas > healthtech > fintech > consulting; test: none: five purposively chosen industries \\
September & All 250 & 5 industries (hhi) & recommendation ecommerce > fintech > healthtech > consulting > saas; size benchmark ecommerce > saas > fintech > healthtech > consulting; test: none: five purposively chosen industries \\
September & Non-comparison 200 & 5 industries (gini) & recommendation ecommerce > healthtech > fintech > consulting > saas; size benchmark ecommerce > saas > fintech > healthtech > consulting; test: none: five purposively chosen industries \\
September & Non-comparison 200 & 5 industries (top3\_share) & recommendation ecommerce > healthtech > fintech > saas > consulting; size benchmark ecommerce > saas > healthtech > fintech > consulting; test: none: five purposively chosen industries \\
September & Non-comparison 200 & 5 industries (hhi) & recommendation ecommerce > healthtech > fintech > consulting > saas; size benchmark ecommerce > saas > fintech > healthtech > consulting; test: none: five purposively chosen industries \\
September, uncapped & All 250 & 5 industries (gini) & recommendation ecommerce > fintech > healthtech > consulting > saas; size benchmark ecommerce > saas > fintech > healthtech > consulting; test: none: five purposively chosen industries \\
September, uncapped & All 250 & 5 industries (top3\_share) & recommendation ecommerce > fintech > healthtech > saas > consulting; size benchmark ecommerce > saas > healthtech > fintech > consulting; test: none: five purposively chosen industries \\
September, uncapped & All 250 & 5 industries (hhi) & recommendation ecommerce > fintech > healthtech > consulting > saas; size benchmark ecommerce > saas > fintech > healthtech > consulting; test: none: five purposively chosen industries \\
September, uncapped & Non-comparison 200 & 5 industries (gini) & recommendation ecommerce > healthtech > fintech > consulting > saas; size benchmark ecommerce > saas > fintech > healthtech > consulting; test: none: five purposively chosen industries \\
September, uncapped & Non-comparison 200 & 5 industries (top3\_share) & recommendation ecommerce > healthtech > fintech > saas > consulting; size benchmark ecommerce > saas > healthtech > fintech > consulting; test: none: five purposively chosen industries \\
September, uncapped & Non-comparison 200 & 5 industries (hhi) & recommendation ecommerce > healthtech > fintech > consulting > saas; size benchmark ecommerce > saas > fintech > healthtech > consulting; test: none: five purposively chosen industries \\
\bottomrule
\end{longtable}
\endgroup

\begingroup\footnotesize\setlength{\tabcolsep}{3pt}
\begin{longtable}{@{}>{\raggedright\arraybackslash}p{2.2cm}>{\raggedright\arraybackslash}p{2.3cm}>{\raggedright\arraybackslash}p{1.2cm}>{\raggedright\arraybackslash}p{3cm}>{\raggedright\arraybackslash}p{2.9cm}@{}}
\caption{Pooled expectation components in both query sets. E1 is descriptive. E2 intervals are fixed-margins null bands; E5-a is the first paired agreement difference, with the second difference below; E5-b is answered-query agreement with its Wilson interval. The component rules and null references remain those of the frozen deviations. Unavailable evidence is not a successful test.}\label{tab:s-wave2-expectations}\\
\toprule
Wave & Query set & Part & Estimate [interval or null band] & Component verdict \\
\midrule
\endfirsthead
\toprule
Wave & Query set & Part & Estimate [interval or null band] & Component verdict \\
\midrule
\endhead
February & All 250 & E1-a & 0.635 [0.335, 0.820] & descriptive (Deviation 4; no verdict) \\
February & Non-comparison 200 & E1-a & 0.609 [0.299, 0.795] & descriptive (Deviation 4; no verdict) \\
February & All 250 & E2-a' & -0.525 [-0.513, -0.362] & not supported \\
February & All 250 & E2-b' & 0.275 [0.176, 0.201] & not supported \\
February & Non-comparison 200 & E2-a' & -0.515 [-0.474, -0.321] & not supported \\
February & Non-comparison 200 & E2-b' & 0.279 [0.177, 0.203] & not supported \\
February & All 250 & E5-a & 0.141 [0.083, 0.199] & supported \\
February & All 250 & E5-b & 0.603 [0.540, 0.664] & supported \\
February & Non-comparison 200 & E5-a & 0.130 [0.061, 0.198] & supported \\
February & Non-comparison 200 & E5-b & 0.571 [0.501, 0.639] & supported \\
September & All 250 & E1-a & 0.616 [0.315, 0.790] & descriptive (Deviation 4; no verdict) \\
September & Non-comparison 200 & E1-a & 0.597 [0.273, 0.787] & descriptive (Deviation 4; no verdict) \\
September & All 250 & E2-a' & -0.455 [-0.627, -0.428] & supported \\
September & All 250 & E2-b' & 0.220 [0.144, 0.166] & not supported \\
September & Non-comparison 200 & E2-a' & -0.430 [-0.646, -0.435] & not supported \\
September & Non-comparison 200 & E2-b' & 0.225 [0.147, 0.170] & not supported \\
September & All 250 & E5-a & 0.120 [0.064, 0.184] & supported \\
September & All 250 & E5-b & 0.608 [0.546, 0.666] & supported \\
September & Non-comparison 200 & E5-a & 0.110 [0.040, 0.185] & supported \\
September & Non-comparison 200 & E5-b & 0.570 [0.501, 0.637] & supported \\
September, uncapped & All 250 & E1-a & 0.616 [0.315, 0.790] & descriptive (Deviation 4; no verdict) \\
September, uncapped & Non-comparison 200 & E1-a & 0.606 [0.288, 0.796] & descriptive (Deviation 4; no verdict) \\
September, uncapped & All 250 & E2-a' & -0.447 [-0.627, -0.418] & supported \\
September, uncapped & All 250 & E2-b' & 0.232 [0.151, 0.174] & not supported \\
September, uncapped & Non-comparison 200 & E2-a' & -0.389 [-0.629, -0.418] & not supported \\
September, uncapped & Non-comparison 200 & E2-b' & 0.235 [0.153, 0.177] & not supported \\
September, uncapped & All 250 & E5-a & 0.112 [0.052, 0.177] & supported \\
September, uncapped & All 250 & E5-b & 0.606 [0.543, 0.665] & supported \\
September, uncapped & Non-comparison 200 & E5-a & 0.097 [0.025, 0.170] & supported \\
September, uncapped & Non-comparison 200 & E5-b & 0.566 [0.496, 0.633] & supported \\
\bottomrule
\end{longtable}
\endgroup

\begingroup\footnotesize\setlength{\tabcolsep}{3pt}
\begin{longtable}{@{}>{\raggedright\arraybackslash}p{2.2cm}>{\raggedright\arraybackslash}p{2.3cm}>{\raggedright\arraybackslash}p{2.7cm}>{\raggedright\arraybackslash}p{2.7cm}>{\raggedright\arraybackslash}p{1.7cm}@{}}
\caption{Agreement differences under the answered-query denominator of each pair. First: Gemini--OpenAI minus Gemini--Perplexity. Second: Gemini--OpenAI minus OpenAI--Perplexity. Values are proportions with paired percentile intervals; both intervals must be entirely positive for the pair-ordering component to be supported.}\label{tab:s-wave2-pairdiff}\\
\toprule
Wave & Query set & First difference [95\% CI] & Second difference [95\% CI] & E5-a \\
\midrule
\endfirsthead
\toprule
Wave & Query set & First difference [95\% CI] & Second difference [95\% CI] & E5-a \\
\midrule
\endhead
February & All 250 & 0.141 [0.083, 0.199] & 0.110 [0.047, 0.171] & supported \\
February & Non-comparison 200 & 0.130 [0.061, 0.198] & 0.107 [0.035, 0.179] & supported \\
September & All 250 & 0.120 [0.064, 0.184] & 0.120 [0.060, 0.180] & supported \\
September & Non-comparison 200 & 0.110 [0.040, 0.185] & 0.115 [0.045, 0.185] & supported \\
September, uncapped & All 250 & 0.112 [0.052, 0.177] & 0.120 [0.058, 0.183] & supported \\
September, uncapped & Non-comparison 200 & 0.097 [0.025, 0.170] & 0.106 [0.036, 0.178] & supported \\
\bottomrule
\end{longtable}
\endgroup

\begingroup\footnotesize\setlength{\tabcolsep}{3pt}
\begin{longtable}{@{}>{\raggedright\arraybackslash}p{1.9cm}>{\raggedright\arraybackslash}p{2cm}>{\raggedright\arraybackslash}p{1.6cm}>{\raggedright\arraybackslash}p{2.8cm}>{\raggedright\arraybackslash}p{1.6cm}>{\raggedright\arraybackslash}p{2.1cm}@{}}
\caption{Expectation composites, departing co-mention pairs and the query-independent agreement null. E2 and E5 each require both components; any unavailable required evidence yields an inconclusive result. These completed rows have no unavailable required component.}\label{tab:s-wave2-composite}\\
\toprule
Wave & Query set & E2 & E2 pair evidence & E5 & E5 Monte Carlo null \\
\midrule
\endfirsthead
\toprule
Wave & Query set & E2 & E2 pair evidence & E5 & E5 Monte Carlo null \\
\midrule
\endhead
February & All 250 & not supported & ratio position above; pairs tested 450, significant after correction 58, departing 43; queries 250, of which 130 admit no trade (point-mass null) & supported & Monte Carlo null 6.1\% [3.4, 9.3]; 143 agreeing queries \\
February & Non-comparison 200 & not supported & ratio position above; pairs tested 450, significant after correction 58, departing 43; queries 200, of which 83 admit no trade (point-mass null) & supported & Monte Carlo null 8.1\% [4.6, 12.2]; 112 agreeing queries \\
September & All 250 & not supported & ratio position above; pairs tested 441, significant after correction 31, departing 31; queries 250, of which 132 admit no trade (point-mass null) & supported & Monte Carlo null 5.7\% [3.2, 8.8]; 152 agreeing queries \\
September & Non-comparison 200 & not supported & ratio position above; pairs tested 441, significant after correction 30, departing 30; queries 200, of which 86 admit no trade (point-mass null) & supported & Monte Carlo null 7.5\% [4.0, 11.5]; 114 agreeing queries \\
September, uncapped & All 250 & not supported & ratio position above; pairs tested 441, significant after correction 29, departing 29; queries 250, of which 135 admit no trade (point-mass null) & supported & Monte Carlo null 5.6\% [2.9, 8.7]; 146 agreeing queries \\
September, uncapped & Non-comparison 200 & not supported & ratio position above; pairs tested 441, significant after correction 28, departing 28; queries 200, of which 89 admit no trade (point-mass null) & supported & Monte Carlo null 7.3\% [4.0, 11.1]; 112 agreeing queries \\
\bottomrule
\end{longtable}
\endgroup

\begingroup\footnotesize\setlength{\tabcolsep}{3pt}
\begin{longtable}{@{}>{\raggedright\arraybackslash}p{2.4cm}>{\raggedright\arraybackslash}p{3.6cm}>{\raggedright\arraybackslash}p{2.8cm}>{\raggedright\arraybackslash}p{1.2cm}>{\raggedright\arraybackslash}p{1.4cm}@{}}
\caption{Pooled September logistic models with brand fixed effects and standard errors clustered by industry--query (250 clusters). Query-type reference is direct and system reference is OpenAI. Rows count eligible answer-by-brand observations.}\label{tab:s-wave2-glm}\\
\toprule
Answers & Term & Odds ratio [95\% CI] & p & Rows \\
\midrule
\endfirsthead
\toprule
Answers & Term & Odds ratio [95\% CI] & p & Rows \\
\midrule
\endhead
September & qtype=comparison & 0.78 [0.61, 1.00] & 0.0462 & 37500 \\
September & qtype=exploration & 2.00 [1.56, 2.55] & 0.0000 & 37500 \\
September & qtype=problem\_solution & 0.88 [0.68, 1.13] & 0.3155 & 37500 \\
September & qtype=trust & 0.95 [0.71, 1.27] & 0.7357 & 37500 \\
September & model=gemini & 0.66 [0.59, 0.73] & 0.0000 & 37500 \\
September & model=perplexity & 0.57 [0.51, 0.64] & 0.0000 & 37500 \\
September, uncapped & qtype=comparison & 0.77 [0.59, 1.00] & 0.0486 & 36240 \\
September, uncapped & qtype=exploration & 2.00 [1.57, 2.55] & 0.0000 & 36240 \\
September, uncapped & qtype=problem\_solution & 0.90 [0.70, 1.15] & 0.3957 & 36240 \\
September, uncapped & qtype=trust & 0.95 [0.71, 1.27] & 0.7224 & 36240 \\
September, uncapped & model=gemini & 0.65 [0.58, 0.73] & 0.0000 & 36240 \\
September, uncapped & model=perplexity & 0.56 [0.50, 0.64] & 0.0000 & 36240 \\
\bottomrule
\end{longtable}
\endgroup

\begingroup\footnotesize\setlength{\tabcolsep}{3pt}
\begin{longtable}{@{}>{\raggedright\arraybackslash}p{2.4cm}>{\raggedright\arraybackslash}p{3cm}>{\raggedright\arraybackslash}p{2cm}>{\raggedright\arraybackslash}p{2cm}>{\raggedright\arraybackslash}p{2cm}@{}}
\caption{Full agreement as percentage / denominator: D1 every query, D2 every system returned an answer, D3 every system has a sampled top brand.}\label{tab:s-wave2-denominators}\\
\toprule
Wave & Query set & D1 & D2 & D3 \\
\midrule
\endfirsthead
\toprule
Wave & Query set & D1 & D2 & D3 \\
\midrule
\endhead
February & All 250 & 57.2 / 250 & 60.3 / 237 & 67.8 / 211 \\
February & Non-comparison 200 & 56.0 / 200 & 57.1 / 196 & 65.9 / 170 \\
February & All minus 13 (237) & 60.3 / 237 & 60.3 / 237 & 67.8 / 211 \\
February & Non-comparison minus 13 (196) & 57.1 / 196 & 57.1 / 196 & 65.9 / 170 \\
September & All 250 & 60.8 / 250 & 60.8 / 250 & 68.2 / 223 \\
September & Non-comparison 200 & 57.0 / 200 & 57.0 / 200 & 65.9 / 173 \\
September & All minus 13 (237) & 61.6 / 237 & 61.6 / 237 & 68.5 / 213 \\
September & Non-comparison minus 13 (196) & 57.7 / 196 & 57.7 / 196 & 65.7 / 172 \\
September, uncapped & All 250 & 58.4 / 250 & 60.6 / 241 & 68.2 / 214 \\
September, uncapped & Non-comparison 200 & 56.0 / 200 & 56.6 / 198 & 65.5 / 171 \\
September, uncapped & All minus 13 (237) & 60.8 / 237 & 61.0 / 236 & 68.2 / 211 \\
September, uncapped & Non-comparison minus 13 (196) & 57.1 / 196 & 57.1 / 196 & 65.5 / 171 \\
\bottomrule
\end{longtable}
\endgroup

\begingroup\footnotesize\setlength{\tabcolsep}{3pt}
\begin{longtable}{@{}>{\raggedright\arraybackslash}p{1.4cm}>{\raggedright\arraybackslash}p{2.5cm}>{\raggedright\arraybackslash}p{1.5cm}>{\raggedright\arraybackslash}p{1.4cm}>{\raggedright\arraybackslash}p{1cm}>{\raggedright\arraybackslash}p{1.2cm}>{\raggedright\arraybackslash}p{2.4cm}@{}}
\caption{Cross-wave sampled-brand-set Jaccard, pooled by system. Cells with no answer in either wave are excluded. Two empty brand sets score one; the last column excludes those sets.}\label{tab:s-wave2-jaccard}\\
\toprule
Answers & Query set & System & Scored / total & Mean & Both empty & Mean excluding empty \\
\midrule
\endfirsthead
\toprule
Answers & Query set & System & Scored / total & Mean & Both empty & Mean excluding empty \\
\midrule
\endhead
All answers & All 250 & openai & 237/250 & 0.886 & 3 & 0.885 \\
All answers & Non-comparison 200 & openai & 196/200 & 0.870 & 3 & 0.868 \\
All answers & All minus 13 (237) & openai & 237/237 & 0.886 & 3 & 0.885 \\
All answers & Non-comparison minus 13 (196) & openai & 196/196 & 0.870 & 3 & 0.868 \\
All answers & All 250 & gemini & 250/250 & 0.882 & 8 & 0.878 \\
All answers & Non-comparison 200 & gemini & 200/200 & 0.865 & 8 & 0.859 \\
All answers & All minus 13 (237) & gemini & 237/237 & 0.881 & 7 & 0.878 \\
All answers & Non-comparison minus 13 (196) & gemini & 196/196 & 0.867 & 7 & 0.862 \\
All answers & All 250 & perplexity & 250/250 & 0.672 & 12 & 0.655 \\
All answers & Non-comparison 200 & perplexity & 200/200 & 0.635 & 12 & 0.612 \\
All answers & All minus 13 (237) & perplexity & 237/237 & 0.664 & 10 & 0.649 \\
All answers & Non-comparison minus 13 (196) & perplexity & 196/196 & 0.633 & 10 & 0.613 \\
Uncapped & All 250 & openai & 236/250 & 0.888 & 4 & 0.886 \\
Uncapped & Non-comparison 200 & openai & 196/200 & 0.875 & 4 & 0.873 \\
Uncapped & All minus 13 (237) & openai & 236/237 & 0.888 & 4 & 0.886 \\
Uncapped & Non-comparison minus 13 (196) & openai & 196/196 & 0.875 & 4 & 0.873 \\
Uncapped & All 250 & gemini & 250/250 & 0.882 & 8 & 0.878 \\
Uncapped & Non-comparison 200 & gemini & 200/200 & 0.865 & 8 & 0.859 \\
Uncapped & All minus 13 (237) & gemini & 237/237 & 0.881 & 7 & 0.878 \\
Uncapped & Non-comparison minus 13 (196) & gemini & 196/196 & 0.867 & 7 & 0.862 \\
Uncapped & All 250 & perplexity & 250/250 & 0.672 & 12 & 0.655 \\
Uncapped & Non-comparison 200 & perplexity & 200/200 & 0.635 & 12 & 0.612 \\
Uncapped & All minus 13 (237) & perplexity & 237/237 & 0.664 & 10 & 0.649 \\
Uncapped & Non-comparison minus 13 (196) & perplexity & 196/196 & 0.633 & 10 & 0.613 \\
\bottomrule
\end{longtable}
\endgroup

\begingroup\footnotesize\setlength{\tabcolsep}{3pt}
\begin{longtable}{@{}>{\raggedright\arraybackslash}p{1.5cm}>{\raggedright\arraybackslash}p{1.2cm}>{\raggedright\arraybackslash}p{1.6cm}>{\raggedright\arraybackslash}p{1.2cm}>{\raggedright\arraybackslash}p{1.6cm}>{\raggedright\arraybackslash}p{1cm}>{\raggedright\arraybackslash}p{3.3cm}@{}}
\caption{Answer lengths in characters among non-empty answers. Counts and means are response-level; differences and their intervals use paired query-level means.}\label{tab:s-wave2-length}\\
\toprule
System & Feb. n & Feb. mean & Sept. n & Sept. mean & Ratio & Paired difference [95\% CI] \\
\midrule
\endfirsthead
\toprule
System & Feb. n & Feb. mean & Sept. n & Sept. mean & Ratio & Paired difference [95\% CI] \\
\midrule
\endhead
openai & 1114 & 2784.5 & 1250 & 2957.2 & 1.062 & 1.0 [-24.3, 28.4] \\
gemini & 1250 & 2047.8 & 1250 & 2064.5 & 1.008 & 16.7 [-4.6, 38.0] \\
perplexity & 1250 & 2343.4 & 1250 & 1550.5 & 0.662 & -792.9 [-846.0, -740.9] \\
\bottomrule
\end{longtable}
\endgroup

\begingroup\footnotesize\setlength{\tabcolsep}{3pt}
\begin{longtable}{@{}>{\raggedright\arraybackslash}p{1cm}>{\raggedright\arraybackslash}p{1.5cm}>{\raggedright\arraybackslash}p{2cm}>{\raggedright\arraybackslash}p{2cm}>{\raggedright\arraybackslash}p{2cm}>{\raggedright\arraybackslash}p{2.9cm}@{}}
\caption{All ten actual September H4 seeds: 3,750 non-empty answers, pinned BGE-M3 on CPU, with unchanged UMAP and HDBSCAN parameters. Emergent uses similarity below 0.60; unanchored uses the frozen query-embedding benchmark. Historical wave-1 answer-encoder provenance remains unresolved, limiting direct cluster comparisons.}\label{tab:s-wave2-h4}\\
\toprule
Seed & Clusters & Outliers (\%) & Emergent & Unanchored & Minimum similarity \\
\midrule
\endfirsthead
\toprule
Seed & Clusters & Outliers (\%) & Emergent & Unanchored & Minimum similarity \\
\midrule
\endhead
0 & 76 & 18.8 & 0 & 1 & 0.651 \\
1 & 77 & 18.5 & 0 & 0 & 0.686 \\
2 & 75 & 17.6 & 0 & 0 & 0.716 \\
3 & 75 & 19.2 & 0 & 0 & 0.682 \\
4 & 76 & 18.0 & 0 & 0 & 0.710 \\
5 & 77 & 18.1 & 0 & 0 & 0.710 \\
6 & 75 & 17.5 & 0 & 0 & 0.666 \\
7 & 73 & 19.9 & 0 & 0 & 0.710 \\
8 & 77 & 17.6 & 0 & 0 & 0.666 \\
9 & 75 & 20.0 & 0 & 0 & 0.666 \\
\bottomrule
\end{longtable}
\endgroup

\begingroup\footnotesize\setlength{\tabcolsep}{3pt}
\begin{longtable}{@{}>{\raggedright\arraybackslash}p{1.3cm}>{\raggedright\arraybackslash}p{2.8cm}>{\raggedright\arraybackslash}p{5cm}>{\raggedright\arraybackslash}p{2.3cm}@{}}
\caption{Submitted hypothesis cut-offs applied to September, pooled. Full industry rows and fixed-15 counterparts remain in the computational tables; H4 is one all-query clustering.}\label{tab:s-wave2-hypotheses}\\
\toprule
H & Query set & Observed [interval or seed range] & Verdict \\
\midrule
\endfirsthead
\toprule
H & Query set & Observed [interval or seed range] & Verdict \\
\midrule
\endhead
H1 & All 250 & Gini = 0.30 [0.26, 0.31] & Not supported \\
H2 & All 250 & records 0.22:1 [0.16, 0.28] (381 asymmetric / 1731 co-mentioned); pooled pairs 0.06:1 & Not supported \\
H3 & All 250 & 7.6\% (19/250) [4.9, 11.6] & Not supported \\
H4 & All 250 & 0.0\% / 0.0\% / 0.0\% emergent under the 0.60 rule (min / median / max over 10 seeds; 0 clusters flagged per seed); 73 to 77 clusters (median 75.5); benchmark rule cut 0.664: 0 to 1 unanchored per seed & Not supported \\
H5 & All 250 & 60.8\% (152/250) [54.6, 66.6] & Not supported \\
H1 & Non-comparison 200 & Gini = 0.33 [0.28, 0.35] & Not supported \\
H2 & Non-comparison 200 & records 0.23:1 [0.17, 0.29] (376 asymmetric / 1671 co-mentioned); pooled pairs 0.06:1 & Not supported \\
H3 & Non-comparison 200 & 9.5\% (19/200) [6.2, 14.4] & Not supported \\
H4 & Non-comparison 200 & not clustered separately (the clustering pools the answers of all 250 queries) & n/a (not clustered separately) \\
H5 & Non-comparison 200 & 57.0\% (114/200) [50.1, 63.7] & Not supported \\
\bottomrule
\end{longtable}
\endgroup

\bibliography{references-revised}